\documentclass[a4paper,11pt]{article}
\usepackage[english]{babel}
\usepackage{enumitem}
\usepackage{amsmath}
\usepackage{amssymb,xcolor, appendix}
\usepackage{tikz}
\usepackage{tikz-network}
\usepackage{tikz}
\tikzset{
	treenode/.style = {shape=circle, rounded corners,
		draw, align=center,
		top color=white, bottom color=blue!20},
	root/.style     = {treenode, font=\Large, bottom color=red!30},
	env/.style      = {treenode, font=\ttfamily\normalsize}
}
\usepackage{bbm}
\usepackage{soul}
\usepackage{diagbox}
\usepackage{eurosym}
\usepackage[left=2.5cm, right=2.5cm, top=2.5cm, bottom=2.5cm]{geometry}
\usepackage{amsfonts}
\usepackage{lscape}
\usepackage{graphicx, float}
\usepackage{xcolor}
\usepackage{multirow}
\usepackage{subfloat}
\usepackage{caption, subcaption}
\usepackage{array}
\usepackage{mathrsfs}
\usepackage{rotating}
\usepackage{setspace} 
\usepackage{titling}
\allowdisplaybreaks
\newtheorem{theorem}{Theorem}

\newtheorem{lemma}{Lemma}

\newtheorem{proposition}{Proposition}

\newtheorem{assumption}{Assumption}
\newtheorem{assumptionalt}{Assumption}[assumption]
\newenvironment{assumptionp}[1]{
	\renewcommand\theassumptionalt{#1}
	\assumptionalt
}{\endassumptionalt}

\newtheorem{assump}{Assumption}

\newlist{condenum}{enumerate}{1} % 'condenum': a new, enumerate-like list env.
\setlist[condenum]{label=\bfseries Condition \arabic*.,
	ref=\arabic*, wide}
\usepackage[affil-it]{authblk}
\usepackage[colorlinks=true,linkcolor=blue]{hyperref}
\usepackage[square,sort&compress]{natbib}
\AtBeginDocument{%
	\hypersetup{citecolor=blue}}

\title{\Large{Partial Identification of Distributional Treatment Effects in Panel Data using Copula Equality Assumptions}\\ (\textit{First Draft}\footnote[1]{Please do not quote without the permission of the authors. An earlier version of this paper was presented at the 2024 Annual Conference of the International Association for Applied Econometrics (IAAE) and the 2024 International Panel Data Conference (IPDC).} )}
\date{October 31, 2024}

\author{Heshani Madigasekara\footnote[2]{Monash University. Email: \textbf{\texttt{Heshani.Madigasekara@monash.edu}}.} ,  D. S. Poskitt\footnote[3]{Monash University. Email: \textbf{\texttt{Donald.Poskitt@monash.edu}}.} , Lina Zhang\footnote[4]{University of Amsterdam and Tinbergen Institute. Email: \textbf{\texttt{l.zhang5@uva.nl}}.} , \textit{\&} Xueyan Zhao\footnote[5]{Monash University. Email: \textbf{\texttt{Xueyan.Zhao@monash.edu}}.}}

\usepackage[skip=10pt plus1pt, indent=30pt]{parskip}

\begin{document}
	% \setstretch{1.5}
	\maketitle

	%Jayen jayama wewa!!!!!!!!

	\begin{abstract} 
		This paper aims to partially identify the distributional treatment effects (DTEs) that depend on the unknown joint distribution of treated and untreated potential outcomes. We construct the DTE bounds using panel data and allow individuals to switch between the treated and untreated states more than once over time. Individuals are grouped based on their past treatment history, and DTEs are allowed to be heterogeneous across different groups. We provide two alternative {group-wise copula equality assumptions} to bound the unknown joint and the DTEs, both of which leverage information from the past observations. Testability of these two assumptions are also discussed, and test results are presented. We apply this method to study the treatment effect heterogeneity of exercising on the adults’ body weight. These results demonstrate that our method improves the identification power of the DTE bounds compared to the existing methods.
		
	\end{abstract}

	\section{Introduction} \label{Section Introduction}
	
	In program evaluation literature, the causal effect is the central measure of interest to estimate. Most studies in this regard, have focused on summary measures that depend only on the marginal distributions of potential outcomes. These include the mean of the treatment effect of interest such as the average treatment effect (ATE), %marginal treatment effect (MTE), 
	local average treatment effect (LATE), the average treatment effect on the treated (ATT), etc. (e.g., \cite{flores2018average}, \cite{heckman1996identification}, \cite{angrist1995identification} and \cite{heckman2005structural}). However, these summary parameters only provide limited information about the whole distribution of the causal effect for a particular group. For example, average effect parameters become inadequate if one is interested in the {fraction} of people that are negatively affected by an intervention regardless of a positive {average} treatment effect or whether a program will result in extremely large benefits but only for a low fraction of people. This study focuses on estimating the bounds for partially identified \textit{distributional} treatment effects (DTEs) for individual treatment groups using panel data.

	Estimating the mean of a treatment effect requires only the marginal distributions of the treated and untreated potential outcomes. However, identifying the distributional treatment effects depends on the joint distribution of the two potential outcomes and is a more challenging task. 
	The fundamental reason for this difficulty is that, in program evaluation using observational data, it is impossible to simultaneously observe both treated and untreated potential outcomes for the same individual at a given time point. One can observe either the treated or the untreated potential outcome for an individual, but \textit{never both}. In the literature, this issue is known as the fundamental problem of causal inference (\cite{holland1986statistics}, \cite{flores2018average}). For example, one could identify the marginal distributions of the treated and untreated potential outcomes under ideal experimental data or standard assumptions such as selection only on the observables or random assignment of the treatment. However, the \textit{copula}, which links the marginals of the potential outcomes into the joint distribution, or the dependence structure between the potential outcomes required to identify the joint distribution, is unobserved due to the fundamental problem of causal inference. Thus, researchers either impose strong assumptions to point-identify the copula or bound such dependence with less restrictive but more plausible assumptions.

	%
	%\subsection{Literature Review} \label{Sec Literature Review}
	
	According to the discussion in \cite{callaway2021bounds}, the studies conducted in this area can be categorized into two main groups based on the assumptions they impose on this unknown dependence or copula structure. The first approach is to bound the joint distribution and its functionals without imposing any restrictions on this unobserved copula (for example, \cite{fan2009partial}, \cite{fan2010sharp} and \cite{russell2021sharp}). Most of the studies in this area have used two well-known classical bounds: (i) the Fr$ \acute{e} $chet-Hoeffiding (FH) bounds (\cite{hoeffding1940masstabinvariante} and \cite{frechet1951tableaux}) to bound the joint distribution of potential outcomes, and (ii)  Williamson and Downs (WD) bounds (\cite{williamson1990probabilistic}) to bound the distribution of the treatment effects. Both methods derive their resulting bounds by considering only the information on marginal distributions of the treated and untreated potential outcomes without placing restrictions on the unknown dependence structure. 
	Unfortunately, such bounds are likely to be wide and uninformative. As a result, imposing additional assumptions to obtain meaningful and more informative bounds on the joint distribution and DTEs has become desirable. 
	
	The other approach is to impose restrictions directly on the unknown copula, assuming the dependence is known. %, assuming it to be known.  
	According to \cite{callaway2021bounds} and \cite{frandsen2018testing}, rank invariance was the leading choice among many studies that fall into this category.\footnote{ \cite{doksum1974empirical} was the first to introduce the concept of rank invariance (i.e., also known as perfect positive dependence) in estimating the distributional impacts of the treatment.} Since it assumes the position or rank of an individual in the distribution of treated potential outcomes is the same as that individual's rank in the distribution of untreated potential outcomes, it is considered a very strong assumption that imposes stringent restrictions on the heterogeneity of the treatment effects. In particular, it allows constant treatment effects, making it inapplicable in many real-world scenarios (see discussion in \cite{frandsen2018testing}, \cite{chernozhukov2019inference} and \cite{frandsen2021partial}).

	Due to the implausibility of these severe point-identifying assumptions, researchers have focused on developing tighter bounds on the DTEs by imposing less strong and plausible assumptions. \cite{manski1997} used the Monotone treatment response (MTR) assumption to bound the distributional treatment effects. Simply, it states that, for each individual, an increase in the treatment variable will either increase the outcome or leave it unchanged. \cite{frandsen2021partial} derive sharp bounds on the DTEs by relying on the assumption that potential outcomes are mutually stochastically increasing (a positive dependence). Both studies use cross-sectional data and provide substantially tighter bounds, yet they can be implausible in some situations due to their nature.

	This paper exploits the additional information provided in panel data and individual history to gain identification of the treatment effect distribution. \cite{callaway2018quantile} and \cite{callaway2019quantile} have exploited the use of panel data and have worked on point-identifying the quantile treatment effects accounting for the dependence between the treated and untreated potential outcomes under a Difference-in-Difference (DiD) setting. Unlike these studies, our study is conducted in a non-DiD setting, allowing for general treatment patterns.
	
	The method developed in this paper follows the existing work in \cite{callaway2021bounds}. \cite{callaway2021bounds} bounds the DTE parameters using WD bounds (\cite{williamson1990probabilistic}) with the ``copula stability assumption" (CSA). His approach requires at least three periods of panels, and he considers a special case where all individuals are \textit{untreated} in the first two periods and they can choose to be treated or untreated only in the last period (i.e., (0, 0, 0) or (0, 0, 1)). Rather than imposing restrictions directly on the unknown dependence of the treated and untreated potential outcomes in the last period, the CSA assumes that the dependence or the copula of the untreated potential outcomes of adjacent periods, be invariant over time for the treated group (i.e., (0, 0, 1)) (for more details, see the discussion in Section \ref{Chap2_Subsec CSA}). Even though he has considered only this treated group (0, 0, 1), an analogous assumption can be made if one is interested in identifying the DTEs for the case (1, 1, 0).

	% \subsection{Objectives of the Current Study} \label{Sec objectives current study}
	
	The current study complements and extends \cite{callaway2021bounds} by also considering panel data with three time periods but focusing on DTEs for other groups, allowing for more general treatment patterns where individuals can switch between the treated and untreated states unrestrictedly across all three time periods. This results in eight treatment path groups. Consequently, the CSA in \cite{callaway2021bounds} cannot be similarly imposed on these other treatment groups except for the $(0,0,1)$ and $(1,1,0)$ treatment groups. Hence, we explore two alternative group-wise copula equality (GCE) assumptions to bound the joint distribution of treated and untreated potential outcomes, as well as the distributional treatment effect parameters for all of these groups. 
	
	Specifically, as the CSA of \cite{callaway2021bounds}, neither of these alternative GCE assumptions imposes restrictions directly on the unknown copula but limits its possibilities by restricting the unobserved copula of the potential outcomes in adjacent periods—leading to bounds instead of point-identifying the DTEs. These GCE assumptions follow different techniques to identify the target unobserved copula, both leveraging information from a similar treatment group as the target group. In other words, given the target group, the first assumption equates its unobserved copula with an observed copula from the past, while the second equates it with an observed copula between the same two adjacent periods, both considering another similar treatment group. Then, as in existing works (e.g., \cite{callaway2021bounds}, \cite{frandsen2021partial}), we apply the FH bounds to bound the joint distribution and WD bounds to bound the distribution of the treatment effects, but in conjunction with either of these key copula assumptions.

	Relative to the conventional point identifying assumptions on the dependence structure of the potential outcomes, these are relatively more plausible and weaker restrictions but can still result in informative bounds for the DTE parameters. 
	Additionally, they yield tighter identified sets for the DTE parameters compared to the existing methods, which impose no restrictions on the copula structure of the potential outcomes. However, for this task, we require access to a rich panel, as the techniques developed here require a sufficient number of observations in a matching group. %However, for this task, we require access to a rich panel as the techniques developed here use the information from another similar group to identify the unknown group-wise estimates. 

	Two types of testing methods—parametric and non-parametric—are discussed to assess the testability of the two copula assumptions using historical observations from the panel. The results of these tests are presented, which, on the whole, fail to reject the two copula equality assumptions. Monte Carlo simulations are undertaken with alternative designs, and the results show that the DTE bounds are well estimated in finite samples. Point-wise 95\% confidence regions are estimated to account for finite sample variations using the numerical bootstrap method proposed by \cite{hong2018numerical}. Finally, we present an empirical example to study the distributional treatment effects of exercises on individuals' BMI, using data from the Household, Income and Labour Dynamics in Australia (HILDA) survey (\cite{HILDA_2015}). 
	
	This study mainly presents the identification of the distributional treatment effect parameters for one particular treatment group: those who are treated in all three periods. Assumptions and the approach to bound the DTEs for all the other seven groups are discussed in the Supplementary Appendix under Section D. %\ref{Sec other groups}. 
	Therefore, our approach facilitates bounding the DTEs for all treated and untreated groups, and consequently allows for the identification of DTEs for the entire population.

	The structure of this paper is organized as follows: In Section \ref{Sec Model Setup}, we discuss the general model setup. Section \ref{Chap2_Sec DTE Bounds} covers the approach developed to partially identify the DTEs, including the testability and plausibility of the key assumptions in practice. Section \ref{Chap2_Estimation and Inference} describes the estimation and inference methods used in this study. Estimation is conducted using distribution and quantile regression, while inference is performed using the numerical bootstrap method proposed by \cite{hong2018numerical} (following similar approaches to those in \cite{callaway2021bounds}). Numerical illustrations and Monte Carlo simulations under different designs are provided in Section \ref{Chap2_Numerical Illustration and Sim}. An empirical illustration is presented in Section \ref{Chap2_Section Application}. All proofs and additional details required to understand the developed method are included in the Supplementary Appendix.\footnote{The Supplementary Appendix for this paper can be provided upon request.} %\ref{Chap2_Appendix}.

	\section{Model Setup}\label{Sec Model Setup}
	%Model Setup and the Parameters of Interest
	Our approach requires at least three periods of panels and aims to partially identify group-wise distributional treatment effects by allowing for more general treatment patterns.
	In this study, we focus on a baseline model with only three periods of panels, where $ s $ denotes a generic time period such that $ s \in \{t-2,\, t-1,\, t\} $. For each individual $ i $ at a time period $ s $, we observe $ \{Y_{is}, D_{is}\}$, where $D_{is}\in\{0,1\}$ is a binary treatment variable and $ Y_{is} $ is the observed outcome. %\textcolor{red}{(Lina: X variable is never used throughout the study. So, I deleted them.)} 
	Let,
	$$Y_{is}=D_{is}Y_{1is}+(1-D_{is})Y_{0is} \quad  \text{for } s \in \{t-2,\, t-1,\, t\},$$ 
	where $Y_{1is}$ and	$Y_{0is}$ are the treated and untreated potential outcomes, respectively.\footnote{The results in this study can be easily extended to incorporate covariates, which will be discussed later in Section \ref{Chap2_Section Application}.}

	Denote $\boldsymbol{Y}_{it}=(Y_{it-2},Y_{it-1},Y_{it})'$ and $\boldsymbol{D}_{it}=(D_{it-2},,D_{it-1},D_{it})'$ as the set of values that we observe in all three periods. %The vector $ \mathbf{Y}_{it} $ represents the observed outcomes of an individual, and $ \mathbf{X}_{it} $ contains the covariate values for that individual $i$ in that three time periods. 
	The vector $ \boldsymbol{D}_{it} $ represents the treatment statuses of an individual $i$ from period $ t-2 $ to $ t $, and is referred to as the treatment pattern at period $t$. 
	As we focus on group-wise DTEs, let us first introduce the grouping concept used to group the individuals in this study. 
	\vspace{-0.2cm}
	\subsection{Groups} As we allow for general treatment patterns, the vector $ \boldsymbol{D}_{it}  $ can exhibit different patterns, and we group individuals according to their treatment patterns at period $t$. Because we consider three time periods and a binary treatment, there are eight possible treatment patterns, which we define as the eight groups:
	\begin{align*}
		\boldsymbol{D}_{it}=(0,0,1)\quad\quad \boldsymbol{D}_{it}=(0,1,1) \quad\quad
		\boldsymbol{D}_{it}=(1,0,1)\quad\quad\boldsymbol{D}_{it}=(1,1,1)\\
		\boldsymbol{D}_{it}=(0,0,0)\quad\quad\boldsymbol{D}_{it}=(0,1,0)\quad\quad
		\boldsymbol{D}_{it}=(1,0,0)\quad\quad\boldsymbol{D}_{it}=(1,1,0) .
	\end{align*}
	In this study, we provide the opportunity to identify the DTEs for all the eight cases mentioned above. 
	
	It is important to introduce the groups for two reasons. Firstly, in many empirical applications, the motivation of the researcher can be different. For example, one might be interested in the DTE of those who remained in a program throughout the study (i.e., $ \boldsymbol{D}_{it}= (1,1,1)$) or of those who remained untreated and switched to be treated at the last period (i.e., $ \boldsymbol{D}_{it}= (0,0,1)$). %, by introducing two alternative copula assumptions to recover the missing dependence structure between the potential outcomes.
	Secondly, grouping individuals based on their treatment patterns allows us to control the unobserved heterogeneity at the group level and to study the DTEs by groups.

	\subsection{Parameters of Interest}

	We are interested in two group-wise distributional treatment effects parameters: the distribution of the treatment effect (DoTE) and the quantile of the treatment effect (QoTE), at period $t$ for a given treatment group $ \boldsymbol{D}_{t} $.\footnote{%\textcolor{magenta}
		{Throughout the study, we omit the subscript $i$ for notation simplicity.}} 
	
	The DoTE for a group $ \boldsymbol{D}_{t} $ is the cumulative distribution function of the treatment effect at period $t$, which provides the fraction of individuals in group $ \boldsymbol{D}_{t} $ with a treatment effect less than any given value $ \delta $:
	\begin{equation*}
		DoTE_{ \boldsymbol{D}_{t}} (\delta)= P(Y_{1t} - Y_{0t} \leq \delta \mid \boldsymbol{D}_{t}), \quad -\infty < \delta < +\infty.
	\end{equation*}
	%where $ d_{s} \in \{0,1\}$ for $ s \in \{t-2, t-1\} $, with $ d_{t}=1 $, as they are treated in the last period $ t $.
	The DoTE can be estimated and plotted for different values of $ \delta $. 
	For example, if we are interested in the fraction of individuals who are worse off by the treatment, this can be identified by $ {DoTE_{ \boldsymbol{D}_{t}}(\delta = 0)}$ or for some negative value of $ \delta$. On the other hand, by estimating $ 1 - DoTE_{ \boldsymbol{D}_{t}}(0) $, we can identify the fraction of individuals who have benefited from the treatment, and so on.
	
	The second parameter of interest, the QoTE for a given group $\boldsymbol{D}_{t}$ at period $t$, can be obtained by simply inverting the DoTE as below:
	\begin{equation*}
		QoTE_{ \boldsymbol{D}_{t}}(\tau) = \inf\{\delta: DoTE_{ \boldsymbol{D}_{t}} (\delta)\geq \tau \mid \boldsymbol{D}_{t}\}, \quad \tau \in (0,1).
	\end{equation*}
	As the DoTE, the QoTE can also be estimated and plotted for different values of $\tau$.
	For example, for continuously distributed outcomes, the $ QoTE_{ \boldsymbol{D}_{t}}(0.05) $ is simply the $ 5^{th} $ percentile of the treatment effect for individuals in group $ \boldsymbol{D}_{t} $, which identifies the treatment effect of the individuals who are in the lower tail of the distribution. Meanwhile, the treatment effect of those who have achieved the highest benefits can be obtained by considering the higher percentiles, such as the $ 95^{th} $ percentile (i.e., $ QoTE_{ \boldsymbol{D}_{t}}(0.95) $). Another important measure is the median effect of the treatment, which can be identified by $ QoTE_{ \boldsymbol{D}_{t}}(0.5) $. Thus, both group-wise DoTE and QoTE parameters provide more detailed information on the individual-level heterogeneity of the treatment effects compared to the widely used average treatment effect parameters.

	\subsection{General Model Assumptions}	
	%We next introduce some important assumptions that will be maintained throughout the study and are general to all eight possible groups $\boldsymbol{D}_{t}$. These assumptions are also employed by \citet{callaway2021bounds}.
	We next introduce the first three model assumptions that are general to all eight possible groups, $\boldsymbol{D}_{t}$. These assumptions are also employed by \citet{callaway2021bounds}.
	\begin{assumptionp}{1} \label{Chap2_ass1}
		$\{\boldsymbol{Y}_{t},\boldsymbol{D}_{t}\}$ are independently and identically distributed across all individuals $i$. 
		%$\{\boldsymbol{Y}_{it},\boldsymbol{D}_{it},\boldsymbol{X}_{it}\}:= \{(Y_{it-2},Y_{it-1},Y_{it})',(D_{it-1},D_{it-2},D_{it})',(X_{it-2},X_{it-2},X_{it})'\}$ are independently and identically distributed across all individuals. 
	\end{assumptionp}
	Assumption \ref{Chap2_ass1} does not restrict the time series dependence among the observations for each individual $i$.

	\begin{assumptionp}{2}\label{Chap2_ass2} $F_{Y_{1t}|\boldsymbol{D}_{t}}$ and $F_{Y_{0t}|\boldsymbol{D}_{t}}$ are both identified.
	\end{assumptionp}
	Assumption \ref{Chap2_ass2} assumes that the marginal distributions of both  treated potential outcome $F_{Y_{1t}|\boldsymbol{D}_{t}}$ and the untreated potential outcome $F_{Y_{0t}|\boldsymbol{D}_{t}}$, for individuals in a group $ \boldsymbol{D}_{t}$, are identified. 
	
	%\textcolor{magenta}
	{For example, consider the treatment group $\boldsymbol{D}_{t} = (1,1,1)$. For this group,} the marginal distribution $F_{Y_{1t}|\boldsymbol{D}_{t} = (1,1,1)}$ can be identified directly from the data, as the treated outcome of an individual in this group  $\boldsymbol{D}_{t} = (1,1,1)$ at period $ t $ is observable. However, since their untreated potential outcome at period $t$ is \textit{unobserved}, assuming the point identification of the counterfactual marginal $F_{Y_{0t}|\boldsymbol{D}_{t} = (1,1,1)}$ needs further illustration. 
	In settings where the treatment assignment is random, one could identify this counterfactual marginal with no additional assumptions. However, in settings with observational data, some additional identification assumptions will be required.
	Here, we follow \cite{callaway2021bounds} and invoke the Change-in-Changes (CiC) method introduced by \cite{athey2006identification}.
	%\textcolor{magenta}
	%{One possible way is to utilize a result of the Change-in-Changes (CiC) method introduced by \cite{athey2006identification}. %(see Appendix Section \ref{Appendix Sec CiC} for more details). 
		For example, under this setting, we identify the counterfactual marginal $F_{Y_{0t}|\boldsymbol{D}_{t} = (1,1,1)}$ for group $\boldsymbol{D}_{t} = (1,1,1)$ by using the observed information of the distributions of its control group $\boldsymbol{D}_{t} = (1,1,0)$, as given below in Equation \eqref{Chap2_CiC_marginal}.\footnote{%\textcolor{magenta}
			{The CiC method can be used to identify the counterfactual marginal for any given group, $\boldsymbol{D}_{t}$. The counterfactual marginal of a treated group at period $t$ (e.g., $\boldsymbol{D}_{t} = (1,0,1)$), is identified using information from its untreated group at period $t$ with the same treatment history (e.g., $\boldsymbol{D}_{t} = (1,0,0)$), and vice versa.}}
		\begin{equation}
			F_{Y_{0t}|\boldsymbol{D}_{t} = (1,1,1)}(y)=F_{Y_{1t-1}|\boldsymbol{D}_{t} = (1,1,1)}\left(F^{-1}_{Y_{1t-1}|\boldsymbol{D}_{t} = (1,1,0)}\left(F_{Y_{0t}|\boldsymbol{D}_{t} = (1,1,0)}(y)\right)\right).\label{Chap2_CiC_marginal}
		\end{equation}
		As pointed out by \cite{callaway2021bounds}, the CiC method imposes relatively less restrictive assumptions and is more flexible in addressing the unobserved heterogeneity, compared to the other possible methods that can be used to identify the counterfactual marginal.\footnote{For example, the Panel DID method of \cite{callaway2019quantile}, Quantile Difference in Differences of \cite{athey2006identification}, or the selection on observables. 
		} 

		\begin{assumptionp}{3}\label{Chap2_ass4}
			%\textcolor{magenta}
			{$ Y_{dt}$, $Y_{dt-1} $, and $Y_{dt-2} $ conditional on $\boldsymbol{D}_{t} $ are continuously distributed. }%conditional on the treatment pattern $\boldsymbol{D}_{t} = (1, 1,1)$.
		\end{assumptionp}
		
		%Assumption \ref{Chap2_ass4} is beneficial for utilizing the copula assumptions imposed in this study. 
		%\textcolor{magenta}
		{We use copulas that describe the dependence structure between the potential outcomes in the identification process of the DTEs. A copula is a continuous function that maps the uniform margins to a joint distribution. When the outcomes are continuously distributed, it enables a one-to-one correspondence between the joint distribution and the copula of the potential outcomes, ensuring that each joint distribution corresponds to a unique copula, and vice versa.} Further, Assumption \ref{Chap2_ass4} allows the quantile functions (i.e., QoTE) derived by taking the inverse of the distribution functions (i.e., DoTE) to be well-defined.

		These three model assumptions will be maintained throughout the study. However, we require additional assumptions to identify the distributional treatment effects as they depend on the joint distribution of the potential outcomes. The following section presents the two key alternative GCE assumptions employed in this study to identify these DTEs separately under two model setups.
		
		\section[GCE Assumptions and DTE Bounds]{Group-wise Copula Equality  Assumptions  and Distributional Treatment Effect Bounds}	\label{Chap2_Sec DTE Bounds}
		%\subsection{GCE Assumptions and DTE bounds}
		
		In this section, we introduce the fourth assumption of our model: Assumption 4. We present our GCE assumptions under two alternative versions of Assumption 4, each discussed separately under two model setups, which we refer to as Model 1 and Model 2 in Sections \ref{Chap2_Subsec Model setup 1} and \ref{Subsec Model setup 2}, respectively. Furthermore, we discuss the plausibility and testability of these alternative assumptions in Sections \ref{Chap2_SubChap2_Sec plausibility} and \ref{Chap2_Subsec Testability}.
		
		%\textcolor{magenta}
		{Assumptions \ref{Chap2_ass1} to \ref{Chap2_ass4} alone will not be sufficient as} the challenge in bounding the DoTE and QoTE arises from the unknown dependence structure between the two potential outcomes. According to \cite{sklar1959fonctions}, the joint distribution of the treated and untreated potential outcomes can be expressed in terms of its marginal distributions and the copula function as below:
		\begin{align}\label{Chap2_Eq sklar formula}
			F_{Y_{0t}, Y_{1t}|\boldsymbol{D}_{t}}(y_{0},y_{1}) &= C_{Y_{0t}, Y_{1t}|\boldsymbol{D}_{t} }\left(F_{Y_{0t}|\boldsymbol{D}_{t}}(y_{0}),\, F_{Y_{1t}|\boldsymbol{D}_{t}}(y_{1})\right),
		\end{align}
		where $ C_{Y_{0t}, Y_{1t}|\boldsymbol{D}_{t}}\left(\cdot, \, \cdot\right): [0,1]^{2} \longrightarrow [0,1] $ denotes the unknown copula that captures the dependence structure between the two potential outcomes. This copula is unknown because given a group $\boldsymbol{D}_{t}$, at the last period $t$, either the treated or the untreated potential outcomes can be observed, and never both. %\st{For example, the untreated potential outcome, $ Y_{0t} $ at period $ t $,  is \textit{unobserved} for group $\boldsymbol{D}_{t} = (1,1,1)$.}}
	
	A common approach in the literature to partially identify the joint distribution is the use of the Fr$ \acute{e} $chet-Hoeffiding bounds \citep{frechet1951tableaux,hoeffding1940masstabinvariante}, where its upper and lower bounds are constructed  %, namely, comonotonicity and countermonotonicity.\footnote{In probability theory, \textit{comonotonicity} refers to the perfect positive dependence between the components of a random vector. This means, that the components in the random vector, can be represented as increasing functions of a single random variable. In two dimensions (i.e., for two components) it is also possible to consider perfect negative dependence, which is called \textit{countermonotonicity}.} 
	under perfect positive and perfect negative dependence between the two variables, respectively. Therefore, the  Fr$ \acute{e} $chet-Hoeffiding bounds define a collection of joint distributions without placing any restrictions on variables' dependence, but using only the information of the marginal distributions (see Supplementary Appendix Section E %\ref{Chap2 Appendix Sec FHandWD}
	 for more details). Consequently, the Fr$ \acute{e} $chet-Hoeffiding bounds often result in relatively wide bounds \citep{heckman1997instrumental,callaway2021bounds,frandsen2021partial}. %Due to this, even though, these bounds rule out that there is a common treatment effect for all individuals, they preclude meaningful economic inferences when analysing the individual-level treatment effect heterogeneity.
	Another common approach is to impose restrictions directly on the unknown copula of $(Y_{0t}, Y_{1t})$. It is often considered a stringent approach as it may not be easy to verify these restrictions in practice. For example, one of the most widely used assumptions is the rank invariance assumption, which assumes a perfect positive dependence between the treated and untreated potential outcomes. %\textcolor{magenta}{\st{for the treated individuals.}} 
	This restricts individuals, to have the same rank in the marginal distributions of the two potential outcomes, which may not hold in some applications.

	In this study, we aim at an intermediate solution, achieved by combining restrictions on the unknown copula of potential outcomes in the adjacent periods $t-1$ and $t$, and a conditional version of Fr$ \acute{e} $chet-Hoeffiding bounds on the joint distribution of $(Y_{0t}, Y_{1t})$ given the potential outcome in the previous period $t-1$. %\textcolor{magenta}
	{To elaborate further, consider the treatment group $\boldsymbol{D}_{t} = (1,1,1)$}. In particular, similar to \eqref{Chap2_Eq sklar formula}, we can get,
	\begin{align}\label{Chap2_Eq sklar conditional_formula}
		&F_{Y_{0t}, Y_{1t}|Y_{1t-1},\boldsymbol{D}_{t} = (1,1,1)}(y_{0},y_{1}\mid y^{\prime}) \nonumber\\
		&= C_{Y_{0t}, Y_{1t}|Y_{1t-1},\boldsymbol{D}_{t} = (1,1,1)}\left(F_{Y_{0t}|Y_{1t-1},\boldsymbol{D}_{t} = (1,1,1)}(y_{0}\mid y^{\prime}),\, F_{Y_{1t}|Y_{1t-1},\boldsymbol{D}_{t} = (1,1,1)}(y_{1}\mid y^{\prime})\mid y^{\prime}\right),
	\end{align}
	where the marginal distribution $F_{Y_{1t}|Y_{1t-1},\boldsymbol{D}_{t} = (1,1,1)}(y_{1}\mid y^{\prime})$ is directly identifiable from the observed data, while the other marginal distribution $F_{Y_{0t}|Y_{1t-1},\boldsymbol{D}_{t} = (1,1,1)}(y_{0}\mid y^{\prime})$ is not. In addition,
	the copula $C_{Y_{0t}, Y_{1t}|Y_{1t-1},\boldsymbol{D}_{t} = (1,1,1)}(\cdot,\cdot\mid y^{\prime})$ can be bounded using a conditional version of Fr$ \acute{e} $chet-Hoeffiding bounds and the two marginals as:
	\begin{align}\label{Chap2_conditional_FH_bounds1}
		F_{Y_{1 t}, Y_{0 t} \mid Y_{1 t-1},\boldsymbol{D}_{t}=(1,1,1)}^{L}\left(y_{1}, y_{0} \mid y^{\prime}\right) &\leq F_{Y_{1 t}, Y_{0 t} \mid Y_{1 t-1}, \boldsymbol{D}_{t}=(1,1,1)}\left(y_{1}, y_{0} \mid y^{\prime}\right) \nonumber\\&\qquad \qquad\leq F_{Y_{1 t}, Y_{0 t} \mid Y_{1 t-1},\boldsymbol{D}_{t}=(1,1,1)}^{U}\left(y_{1}, y_{0} \mid y^{\prime}\right),
	\end{align}
	where
	\begin{align}\label{Chap2_conditional_FH_lower_bounds}
		&F_{Y_{1 t}, Y_{0 t} \mid Y_{1 t-1}, \boldsymbol{D}_{t}=(1,1,1)}^{L}\left(y_{1}, y_{0} \mid y^{\prime}\right)\nonumber\\
		&\qquad=\max \left\{F_{Y_{1 t} \mid Y_{1 t-1}, \boldsymbol{D}_{t}=(1,1,1)}\left(y_{1} \mid y^{\prime}\right)+F_{Y_{0 t} \mid Y_{1 t-1}, \boldsymbol{D}_{t}=(1,1,1)}\left(y_{0} \mid y^{\prime}\right)-1,0\right\}, \\
		\label{conditional_FH_upper_bounds}
		&F_{Y_{1 t}, Y_{0 t} \mid Y_{1 t-1}, \boldsymbol{D}_{t}=(1,1,1)}^{U}\left(y_{1}, y_{0} \mid y^{\prime}\right)\nonumber\\
		&\qquad=\min \left\{F_{Y_{1 t} \mid Y_{1 t-1}, \boldsymbol{D}_{t}=(1,1,1)}\left(y_{1} \mid y^{\prime}\right), F_{Y_{0 t} \mid Y_{1 t-1}, \boldsymbol{D}_{t}=(1,1,1)}\left(y_{0} \mid y^{\prime}\right)\right\}.
	\end{align}
	Given \eqref{Chap2_conditional_FH_bounds1}, \eqref{Chap2_conditional_FH_lower_bounds}, and \eqref{conditional_FH_upper_bounds}, we can see that the bounds for $F_{Y_{1 t}, Y_{0 t} \mid Y_{1 t-1}, \boldsymbol{D}_{t}=(1,1,1)}\left(y_{1}, y_{0} \mid y^{\prime}\right)$ depend %\textcolor{magenta}{\st{only}} 
	on the unknown conditional marginal $F_{Y_{0 t} \mid Y_{1 t-1}, \boldsymbol{D}_{t}=(1,1,1)}\left(y_{0} \mid y^{\prime}\right)$ \textminus which can be identified if the joint distribution $F_{Y_{0 t},Y_{1 t-1}| \boldsymbol{D}_{t}=(1,1,1)}\left(y_{0} , y^{\prime}\right)$ is known.
	The arguments of \cite{sklar1959fonctions} lead to: % by considering the dependence between the potential outcomes. 
	\begin{align}\label{Chap2_Eqn joint:Y0t&Y1t-1_intro}
		F_{Y_{0 t}, Y_{1 t-1} \mid \boldsymbol{D}_{t}= (1,1,1)}\left(y_{0}, y^{\prime}\right)=C_{Y_{0 t}, Y_{1 t-1} \mid \boldsymbol{D}_{t}= (1,1,1)}\left(F_{Y_{0 t} \mid \boldsymbol{D}_{t}= (1,1,1)}\left(y_{0}\right), F_{Y_{1t-1} \mid \boldsymbol{D}_{t}= (1,1,1)}\left(y^{\prime}\right)\right).
	\end{align}
	Since $F_{Y_{0 t} \mid \boldsymbol{D}_{t}= (1,1,1)}$ (by Assumption \ref{Chap2_ass2}) and $F_{Y_{1t-1} \mid \boldsymbol{D}_{t}= (1,1,1)}$ are both identifiable, the only missing component in the bounds for $F_{Y_{1 t}, Y_{0 t} \mid Y_{1 t-1}, \boldsymbol{D}_{t}=(1,1,1)}\left(y_{1}, y_{0} \mid y^{\prime}\right)$ is the unknown copula  of potential outcomes across times,  $C_{Y_{0 t}, Y_{1 t-1} \mid \boldsymbol{D}_{t}= (1,1,1)}$.
	
	In the following subsections, we will introduce two group-wise copula equality (GCE) assumptions to identify $C_{Y_{0 t}, Y_{1 t-1} \mid \boldsymbol{D}_{t}= (1,1,1)}$, separately under the two model setups, Model 1 and Model 2. Once we obtain the lower and upper bounds for the conditional joint distribution of $F_{Y_{0t}, Y_{1t}|Y_{1t-1},\boldsymbol{D}_{t} = (1,1,1)}(y_{0},y_{1}\mid y^{\prime})$ in \eqref{Chap2_Eq sklar conditional_formula}, the bounds for the unconditional joint distribution $F_{Y_{0t}, Y_{1t}|\boldsymbol{D}_{t} = (1,1,1)}(y_{0},y_{1})$ can be obtained by integrating out $Y_{1t-1}$.
	%Model 1 utilizes the first alternative GCE assumption, while Model 2 utilizes the second alternative GCE assumption.	
	Even though we consider all eight treatment groups (i.e., both treated and untreated groups), to avoid repeating the same material in the main text, the methodology to bound the DTEs is illustrated for one particular group, $ \boldsymbol{D}_{t}= (1,1,1)$, where individuals remain treated throughout the three periods. The results for other groups are explained in Section D %\ref{Sec other groups} 
	of the Supplementary Appendix. Given this, we omit the subscript $\boldsymbol{D}_{t}$ in our target parameters, DoTE and QoTE to ease notation.

	\subsection{DTE bounds for Group $ \mathbf{\textit{D}}_{t}= (1,1,1)$ - Model 1 } \label{Chap2_Subsec Model setup 1}

	In this section, we introduce the first GCE assumption on the copula of $(Y_{1t-1}, Y_{0t})$ for group $\boldsymbol{D}_{t}= (1,1,1)$. We refer to models that satisfy the assumptions in this section as Model 1.

\begin{figure}[H]
		\caption{\centering Visualising the First Alternative GCE Assumption \ref{Chap2_ass3}}\label{Chap2_fig tree DA(I)}
		\begin{tikzpicture}[node distance=1.5cm, minimum size=0.1cm, inner sep=2, auto]
			%%%%%%%%%%%%%%%%%%%%% left figure  %%%%%%%%%%%%%%%%%%%%%
			\node [draw, circle, fill=blue!30] (G) {\textcolor{black}{\textbf{1}}};
			
			% New dotted node above G
			\node [draw, circle, dotted, line width=1pt, above = 0.35cm of G] (H) {\textbf{0}};
			\node [right of=H, xshift=-0.9cm, font=\large] {\textbf{{?}}};
			
			\node [draw, circle,  fill=gray!30, above left =1cm and 1.25cm of G] (C) {\textcolor{black}{\textbf{1}}};
			\node [draw, circle, fill=gray!30, above left =0.3cm and 1.25cm of C] (A) {\textcolor{black}{\textbf{1}}};
			\node [draw, circle, fill=gray!30, below left =0.3cm and 1.25cm of C] (B) {\textbf{0}}; % Adjust the distance as needed
			
			\node [draw, circle, fill=gray!30, below left =1cm and 1.25cm of G] (D) {\textbf{0}};
			\node [draw, circle, fill=gray!30, above left =0.3cm and 1.25cm of D] (E) {\textbf{1}};
			\node [draw, circle, fill=gray!30, below left =0.3cm and 1.25cm of D] (F) {\textbf{0}}; % Adjust the distance as needed
			
			% text 
			\node [left of=A, xshift=-0.2cm, font=\small] {$\boldsymbol{D}_t=(1,1,1)$};
			\node [left of=B, xshift=-0.2cm, font=\small] {$\boldsymbol{D}_t=(0,1,1)$};
			\node [left of=E, xshift=-0.2cm, font=\small] {$\boldsymbol{D}_t=(1,0,1)$};
			\node [left of=F, xshift=-0.2cm, font=\small] {$\boldsymbol{D}_t=(0,0,1)$};
			
			\node [above = 0.5cm of A, font=\small] (t1) {$(t-2)$};
			\node [right = 0cm and 0.5cm of t1, font=\small](t2) {$(t-1)$};
			\node [right = 0cm and 0.75cm of t2, font=\small] {$(t)$};
			
			% title
			\node [above = 0.3cm and 0cm of t2, font=\small] {\textbf{Groups with $D_{t}=1$}};
			
			% Arrows
			\draw [-, line width=1pt, dashed] (A) -- (C);
			\draw [-] (B) -- (C);     
			\draw [-, line width=1pt] (E) -- (D);
			\draw [-] (F) -- (D);
			\draw [-, line width=1pt] (D) -- (G);
			\draw [-, line width=1pt, dashed] (C) -- (G);
			
			% Curved arrows
			\draw [->, line width=1pt, dotted, bend left=70] (C) to node[midway, above=0.1cm] {\textcolor{black}{\tiny$C^{\ast}_{Y_{1t-1},Y_{0t}|\boldsymbol{D}_{t} = (1, 1, 1)}$}}  (H);
			\draw [->, line width=1pt, bend left=85] (E) to node[midway, above=0.1] {\textcolor{black}{\tiny$C_{Y_{1t-2},Y_{0t-1}|\boldsymbol{D}_{t} = (1, 0, 1)}$}}   (D);

			%%%%%%%%%%%%%%%%%%%%% right figure   %%%%%%%%%%%%%%%%%%%%%
			\node [draw, circle, fill=blue!30, right = 0 and 7.5cm of G] (Gr) {\textbf{0}};
			
			\node [draw, circle, fill=gray!30, above left =1cm and 1.25cm of Gr] (Cr) {\textbf{1}};
			\node [draw, circle, fill=gray!30, above left =0.3cm and 1.25cm of Cr] (Ar) {\textbf{1}};
			\node [draw, circle, fill=gray!30, below left =0.3cm and 1.25cm of Cr] (Br) {\textbf{0}}; % Adjust the distance as needed
			
			\node [draw, circle, fill=gray!30, below left =1cm and 1.25cm of Gr] (Dr) {\textbf{0}};
			\node [draw, circle, fill=gray!30, above left =0.3cm and 1.25cm of Dr] (Er) {\textbf{1}};
			\node [draw, circle, fill=gray!30, below left =0.3cm and 1.25cm of Dr] (Fr) {\textbf{0}}; % Adjust the distance as needed
			
			% text 
			\node [left of=Ar, xshift=-0.2cm, font=\small] {$\boldsymbol{D}_t=(1,1,0)$};
			\node [left of=Br, xshift=-0.2cm, font=\small] {$\boldsymbol{D}_t=(0,1,0)$};
			\node [left of=Er, xshift=-0.2cm, font=\small] {$\boldsymbol{D}_t=(1,0,0)$};
			\node [left of=Fr, xshift=-0.2cm, font=\small] {$\boldsymbol{D}_t=(0,0,0)$};
			
			\node [above = 0.5cm of Ar, font=\small] (t1r) {$(t-2)$};
			\node [right = 0cm and 0.5cm of t1r, font=\small](t2r) {$(t-1)$};
			\node [right = 0cm and 0.75cm of t2r, font=\small] {$(t)$};
			
			% title
			\node [above = 0.3cm and 0cm of t2r, font=\small] {\textbf{Groups with $D_{t}=0$}};
			
			% Arrows
			\draw [-] (Ar) -- (Cr);
			\draw [-] (Br) -- (Cr);
			\draw [-] (Er) -- (Dr);
			\draw [-] (Fr) -- (Dr);
			\draw [-] (Dr) -- (Gr);
			\draw [-] (Cr) -- (Gr);
			
			% Curved arrows
			% Curved arrows
			%\draw [->, line width=1pt, bend left=70, color=red] (Cr) to node[midway, above=0.3cm] {\textcolor{black}{$C_{Y_{1t-1},Y_{0t}|\mathbf{D}_{t} = (1, 1, 0)}$}}  (Gr);
		\end{tikzpicture}\\\\
		\textit{Notes: The treatment path denoted by dashed lines represents the target group $ \boldsymbol{D}_{t} = (1, 1, 1) $, while the bolded path represents the copula recovery group $ \boldsymbol{D}_{t} = (1, 0, 1) $. The dotted arrow shows the target unobserved treatment switching pattern of group $ \boldsymbol{D}_{t} = (1, 1, 1) $ and its unobserved copula (left term in \ref{Chap2_ass3}), while the solid arrow represents the observed treatment switching pattern of the group $ \boldsymbol{D}_{t} = (1, 0, 1) $ and its copula (right term in \ref{Chap2_ass3}), which we used to recover the unknown copula.}
	\end{figure}
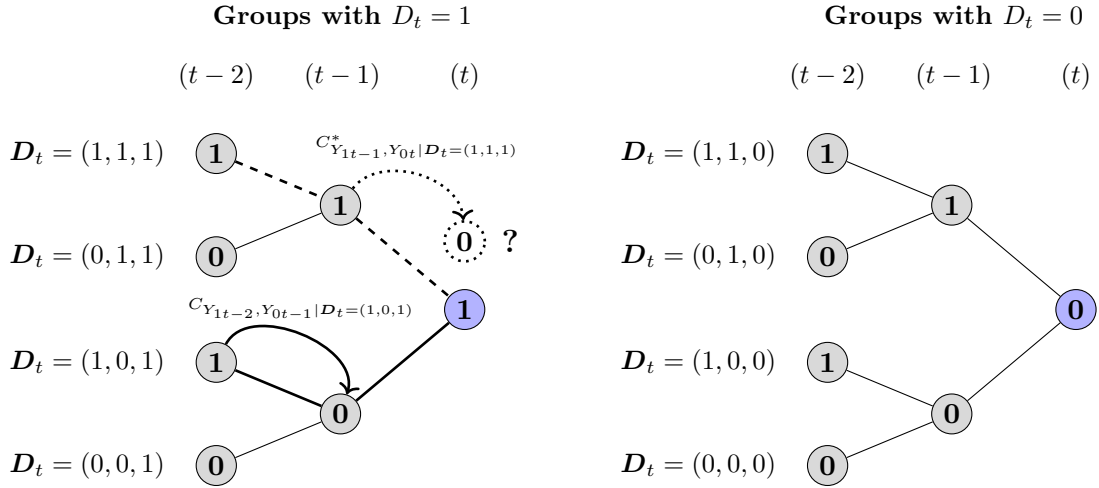

	\begin{assump}[GCE (I)]\label{Chap2_ass3} \textit{For all} $ (u, v) \in [0, 1]^{2} $, and for the treatment group $ \boldsymbol{D}_{t} = (1,1,1) $, we assume,
		\begin{align*}
			C^{\ast}_{Y_{1t-1},Y_{0t}|\boldsymbol{D}_{t} = (1, 1, 1)}(u, v) &= C_{Y_{1t-2},Y_{0t-1}|\boldsymbol{D}_{t} = (1, 0, 1)}(u, v),
		\end{align*}
		where the superscript ${\ast} $ denotes the ``unobserved" component.
	\end{assump}
	Assumption \ref{Chap2_ass3} utilizes information from another group of individuals who are treated at period $ t $, to recover the unobserved group-wise copula on the left-hand side (LHS) of the equality. Figure \ref{Chap2_fig tree DA(I)} represents all the eight possible treatment groups in a slightly different manner using a treatment diagram. We use this diagram to visualize and explain the logic employed under this first GCE assumption,  Assumption \ref{Chap2_ass3}. 
	
	In Figure \ref{Chap2_fig tree DA(I)}, the treatment path denoted by dashed lines represents the group of interest, $ \boldsymbol{D}_{t} = (1, 1, 1) $. The bolded treatment path represents the treatment pattern $ \boldsymbol{D}_{t} = (1, 0, 1) $, used to recover the unknown copula (i.e., the copula recovery group). The dotted arrow shows the unobserved copula of $ \boldsymbol{D}_{t} = (1, 1, 1) $.  %Hence, it's impossible to identify the dependence or the copula structure of the two potential outcomes, $Y_{1t-1}  $ and $ Y_{0t} $ for these treated individuals with $ \boldsymbol{D}_{t} = (1,1,1) $ (i.e., the LHS term on \ref{Chap2_ass3}), as only one of these variables is known. 
	In Assumption \ref{Chap2_ass3}, we assume that this unknown copula is equal to the known copula of the two potential outcomes $Y_{1t-2}  $ and $ Y_{0t-1} $ of an individual in the copula recover group $ \boldsymbol{D}_{t} = (1,0,1) $. The latter can be \textit{identified} as the potential outcomes $Y_{1t-2} $ and $ Y_{0t-1} $ are observed for individuals in $ \boldsymbol{D}_{t} = (1, 0, 1) $. This observed dependence is denoted by the solid arrow in Figure \ref{Chap2_fig tree DA(I)}.

	Under Assumption \ref{Chap2_ass3}, we match the treatment switching pattern that we are interested in, from period $ t-1 $ to $ t $ (i.e., dotted arrow), with an individual who also treated at $ t $ and has the same treatment switching pattern \textit{but} in different points in time, from period $ t-2 $ to $ t-1$ (i.e., solid arrow). Then, we assume that these two copula structures of the corresponding pair of outcomes are the same in the target group and copula recovery group.

	Next, we explain the identification procedure that is used in this study to partially identify the unknown joint distribution of the treated and untreated potential outcomes at period $t$; $F_{Y_{0 t}, Y_{1 t} \mid \boldsymbol{D}_{t}= (1,1,1)}$, and the target parameters, DoTE and QoTE.

	As explained above, we begin by identifying the joint distribution of the potential outcomes, $Y_{0 t}$ and $Y_{1 t-1}$ for the individuals with the treatment pattern $ \boldsymbol{D}_{t}= (1,1,1) $ (see Equation \eqref{Chap2_Eqn joint:Y0t&Y1t-1_intro}) \textminus which is also unknown. According to the arguments of \cite{sklar1959fonctions} and our first GCE Assumption \ref{Chap2_ass3}: % by considering the dependence between the potential outcomes. 
	\begin{align}\label{Chap2_Eqn joint:Y0t&Y1t-1}
		&F_{Y_{0 t}, Y_{1 t-1} \mid \boldsymbol{D}_{t}= (1,1,1)}\left(y_{0}, y^{\prime}_{1}\right)\nonumber\\
		&=C_{Y_{0 t}, Y_{1 t-1} \mid \boldsymbol{D}_{t}= (1,1,1)}\left(F_{Y_{0 t} \mid \boldsymbol{D}_{t}= (1,1,1)}\left(y_{0}\right), F_{Y_{1t-1} \mid \boldsymbol{D}_{t}= (1,1,1)}\left(y^{\prime}_{1}\right)\right) \qquad\text{(By \cite{sklar1959fonctions})} \nonumber\\
		&=C_{Y_{0 t-1}, Y_{1 t-2} \mid \boldsymbol{D}_{t}= (1,0,1)}\left(F_{Y_{0 t} \mid \boldsymbol{D}_{t}= (1,1,1)}\left(y_{0}\right), F_{Y_{1 t-1} \mid \boldsymbol{D}_{t}= (1,1,1)}\left(y^{\prime}_{1}\right)\right) \qquad\text{(By GCE \ref{Chap2_ass3})}\nonumber \\
		&=F_{Y_{0 t-1}, Y_{1 t-2} \mid \boldsymbol{D}_{t}= (1,0,1)}\left(F_{Y_{0 t-1} \mid \boldsymbol{D}_{t}= (1,0,1)}^{-1}  \left(F_{Y_{0 t} \mid \boldsymbol{D}_{t}= (1,1,1)}\left(y_{0}\right)\right),\right. \nonumber\\
		& \qquad \qquad \qquad \qquad \qquad \qquad \qquad \qquad \left. F_{Y_{1 t-2} \mid \boldsymbol{D}_{t}= (1,0,1)}^{-1}  \left(F_{Y_{1 t-1} \mid \boldsymbol{D}_{t}= (1,1,1)}\left(y^{\prime}_{1}\right)\right)\right).
	\end{align}
	In Equation \eqref{Chap2_Eqn joint:Y0t&Y1t-1}, all terms on the right-hand side (RHS) are identifiable either directly by observed data or indirectly by assumption.
	The unknown joint distribution of $ F_{Y_{0 t}, Y_{1 t-1} \mid \boldsymbol{D}_{t}= (1,1,1)} $ is a key factor in the identification of $F_{Y_{0t},Y_{1t} \mid \boldsymbol{D}_{t}= (1,1,1)} $ and its functionals.
	Next, utilising this preliminary result in Equation \eqref{Chap2_Eqn joint:Y0t&Y1t-1}, we identify the counterfactual marginal distribution of $ Y_{0t} $ conditional on the previous outcome $Y_{1t-1} $: $F_{Y_{0 t} \mid Y_{1 t-1}, \boldsymbol{D}_{t}= (1,1,1)}$. This result is given in Lemma \ref{Chap2_lemma1}. 
	
	\begin{lemma}[Distribution of $Y_{0t}$ Conditional  on $Y_{1 t-1}$ under Model 1]\label{Chap2_lemma1}
		Under Assumptions \ref{Chap2_ass1} to \ref{Chap2_ass3} and using the results in the Equation \eqref{Chap2_Eqn joint:Y0t&Y1t-1}, the counterfactual marginal distribution of $ Y_{0t} $ conditional on the previous outcome $ Y_{1t-1} $ for group $ \boldsymbol{D}_{t}= (1,1,1)  $, can be identified by, 
		\begin{align*}
			&F_{Y_{0 t} \mid Y_{1 t-1}, \boldsymbol{D}_{t}= (1,1,1)}\left(y_{0} \mid y^{\prime}_{1}\right)\\
			&=F_{Y_{0 t-1}\mid  Y_{1 t-2}, \boldsymbol{D}_{t}= (1,0,1)}\left(F_{Y_{0 t-1} \mid \boldsymbol{D}_{t}= (1,0,1)}^{-1}  \left(F_{Y_{0 t} \mid \boldsymbol{D}_{t}= (1,1,1)}\left(y_{0}\right)\right) \mid \right. \\
			& \qquad \qquad \qquad \qquad \qquad \qquad \qquad \qquad \left. F_{Y_{1 t-2} \mid \boldsymbol{D}_{t}= (1,0,1)}^{-1}  \left(F_{Y_{1 t-1} \mid \boldsymbol{D}_{t}= (1,1,1)}\left(y^{\prime}_{1}\right)\right)\right).
		\end{align*}
		(The proof of Lemma \ref{Chap2_lemma1} is provided in the Supplementary Appendix A)%\ref{Chap2_Appendix Proofs})
	\end{lemma}

	As we have identified the two conditional marginal distributions of the treated and untreated potential outcomes: (i) $ F_{Y_{1t}|Y_{1t-1},\boldsymbol{D}_{t}= (1,1,1)} $ using sample data, and (ii) $ F_{Y_{0t}|Y_{1t-1},\boldsymbol{D}_{t}= (1,1,1)} $ using Lemma \ref{Chap2_lemma1}. We can derive the joint distribution of $Y_{1 t}$ and $Y_{0 t}$, conditional on the outcome $Y_{1t-1}$ in the previous period, for the target group $ \boldsymbol{D}_{t}= (1,1,1) $, by simply applying a conditional version of the Fr$ \acute{e} $chet-Hoeffiding bounds. We provide this result under Lemma \ref{Chap2_lemma_condJoint}. 
	
	\begin{lemma}[Bounds on the Joint Distribution Conditional  on $Y_{1 t-1}$]\label{Chap2_lemma_condJoint} Under Assumptions \ref{Chap2_ass1} to \ref{Chap2_ass4} and the GCE Assumption \ref{Chap2_ass3}, the lower and upper bounds for the conditional joint distribution of $F_{Y_{1 t}, Y_{0 t} \mid Y_{1 t-1}, \boldsymbol{D}_{t}=(1,1,1)}\left(y_{1}, y_{0} \mid y^{\prime}\right) $ defined in Equations \eqref{Chap2_conditional_FH_lower_bounds} and \eqref{conditional_FH_upper_bounds} are identified, 
		where the expression of $F_{Y_{0t} \mid Y_{1t-1},\,\boldsymbol{D}_{t}=(1,1,1)} (y_{0}\mid y^{\prime})$ is given in Lemma \ref{Chap2_lemma1}.%\\\\
		%Proof: The proof of Lemma \ref{Chap2_lemma_condJoint} is provided in the Appendix.
	\end{lemma}
	
	Given the results in Lemma \ref{Chap2_lemma_condJoint}, we can proceed to derive the bounds on the (unconditional) joint distribution of potential outcomes $Y_{1 t}$ and $Y_{0 t}$ for the individuals in the target group $ \boldsymbol{D}_{t}=(1,1,1) $ at period $t$, as given below under Theorem \ref{Chap2_theorem_joint}.
	
	\begin{theorem}[Bounds on the Joint Distribution under Model 1]\label{Chap2_theorem_joint} Under Assumptions \ref{Chap2_ass1} to \ref{Chap2_ass3}, we have, %the bounds on the joint distribution of the potential outcomes for the treated individuals with the treatment pattern $ \boldsymbol{D}_{t}=(1,1,1) $ at period $ t $, are given by,
		\begin{align*}
			F_{Y_{1 t}, Y_{0 t} \mid \boldsymbol{D}_{t}=(1,1,1)}^{L}\left(y_{1}, y_{0}\right) \leq F_{Y_{1 t}, Y_{0 t} \mid \boldsymbol{D}_{t}=(1,1,1)}\left(y_{1}, y_{0}\right) \leq F_{Y_{1 t}, Y_{0 t} \mid \boldsymbol{D}_{t}=(1,1,1)}^{U}\left(y_{1}, y_{0}\right)
		\end{align*}
		where,
		\begin{align*}
			&F_{Y_{1 t}, Y_{0 t} \mid \boldsymbol{D}_{t}=(1,1,1)}^{L}\left(y_{1}, y_{0}\right)=\mathrm{E}\left[F_{Y_{1 t}, Y_{0 t} \mid Y_{1 t-1},\boldsymbol{D}_{t}=(1,1,1)}^{L}\left(y_{1}, y_{0} \mid Y_{1 t-1}\right) \mid \boldsymbol{D}_{t}=(1,1,1)\right] \\
			&F_{Y_{1 t}, Y_{0 t} \mid \boldsymbol{D}_{t}=(1,1,1)}^{U}\left(y_{1}, y_{0}\right)=\mathrm{E}\left[F_{Y_{1 t}, Y_{0 t} \mid Y_{1 t-1},\boldsymbol{D}_{t}=(1,1,1)}^{U}\left(y_{1}, y_{0} \mid Y_{1 t-1}\right) \mid \boldsymbol{D}_{t}=(1,1,1)\right],
		\end{align*}
		where $F_{Y_{1 t}, Y_{0 t} \mid Y_{1 t-1}, \boldsymbol{D}_{t}=(1,1,1)}^{L}$ and $F_{Y_{1 t}, Y_{0 t} \mid Y_{1 t-1}, \boldsymbol{D}_{t}=(1,1,1)}^{U}$ are given in  \eqref{Chap2_conditional_FH_lower_bounds} and \eqref{conditional_FH_upper_bounds}.%Lemma \ref{Chap2_lemma_condJoint}.%\\\\
		%	(Proof: The proof of Theorem \ref{Chap2_theorem_joint} is provided in the Appendix)
	\end{theorem}
	
	Next, we derive sharp bounds on the DoTE at period $t$ for group $ \boldsymbol{D}_{t}= (1,1,1) $, using a conditional version of the Williamson and Downs (WD) bounds proposed by \citet{williamson1990probabilistic}.\footnote{When there is no additional information besides the knowledge of marginal distributions, the bounds for distribution of the difference between two variables derived using WD inequality are proved to be sharp \citep{callaway2021bounds,fan2010sharp,williamson1990probabilistic}.}
	As we have identified the conditional counterfactual marginal distribution, $ F_{Y_{0 t} \mid Y_{1 t-1},\boldsymbol{D}_{t}= (1,1,1)}\left(y_{0} \mid y^{\prime}_{1}\right) $ (Lemma \ref{Chap2_lemma1}), using the WD bounds, we can obtain sharp bounds on the DoTE conditional on the previous outcome: ${F}_{Y_{1 t}-Y_{0 t} \mid Y_{1t-1},\,\boldsymbol{D}_{t}= (1,1,1)}$. This is given in Lemma \ref{Chap2_lemma_cond_TE}.

\begin{lemma}[DoTE Bounds Conditional  on $Y_{1 t-1}$ under Model 1]\label{Chap2_lemma_cond_TE} Under Assumptions \ref{Chap2_ass1} to \ref{Chap2_ass3} and using the results in Lemma \ref{Chap2_lemma1}, we have,%the bounds on the distribution of treatment effect for the individuals with the pattern $ \boldsymbol{D}_{t}= (1,1,1) $ at period $t$, conditional on the previous outcome $Y_{1 t-1}$, are given by, }
\begin{equation*}
	F_{Y_{1 t}-Y_{0 t} |Y_{1 t-1}, \boldsymbol{D}_{t}= (1,1,1)}^{L}\left(\delta | y^{\prime}_{1}\right) \leq F_{Y_{1 t}-Y_{0 t} |Y_{1 t-1}, \boldsymbol{D}_{t}= (1,1,1)}\left(\delta | y^{\prime}_{1}\right) \leq F_{Y_{1 t}-Y_{0 t} | Y_{1 t-1}, \boldsymbol{D}_{t}= (1,1,1)}^{U}\left(\delta | y^{\prime}_{1}\right),
\end{equation*}
where
\begin{align*}
	&F_{Y_{1 t}-Y_{0 t} \mid Y_{1 t-1}, \boldsymbol{D}_{t}= (1,1,1)}^{L}\left(\delta \mid y^{\prime}_{1}\right)\\
	&\qquad=\sup _{y} \max \left\{F_{Y_{1 t} \mid Y_{1 t-1}, \boldsymbol{D}_{t}= (1,1,1)}\left(y \mid y^{\prime}_{1}\right)-F_{Y_{0 t} \mid Y_{1 t-1}, \boldsymbol{D}_{t}= (1,1,1)}\left(y-\delta \mid y^{\prime}_{1}\right), 0\right\} \\
	&F_{Y_{1 t}-Y_{0 t} \mid Y_{1 t-1}, \boldsymbol{D}_{t}= (1,1,1)}^{U}\left(\delta \mid y^{\prime}_{1}\right)\\
	&\qquad=1+\inf _{y} \min \left\{F_{Y_{1 t} \mid Y_{1 t-1}, \boldsymbol{D}_{t}= (1,1,1)}\left(y \mid y^{\prime}_{1}\right)-F_{Y_{0 t} \mid Y_{1 t-1}, \boldsymbol{D}_{t}= (1,1,1)}\left(y-\delta \mid y^{\prime}_{1}\right), 0\right\},
\end{align*}
and $F_{Y_{0t} \mid Y_{1t-1},\, \boldsymbol{D}_{t}= (1,1,1)}  $ is given in Lemma \ref{Chap2_lemma1}.
%(Proof: The proof of Lemma \ref{Chap2_lemma_cond_TE} is provided in the Appendix)
\end{lemma}

The result given in Theorem \ref{Theorem_DoTE} provides the (unconditional) bounds on the DoTE for the individuals with the treatment pattern $ \boldsymbol{D}_{t}= (1,1,1)$ at period $ t $, using the results in Lemma \ref{Chap2_lemma_cond_TE}.
\begin{theorem}[DoTE Bounds under Model 1]\label{Theorem_DoTE} Under Assumptions \ref{Chap2_ass1} to \ref{Chap2_ass3}, we have,
$$
{DoTE}^{L}(\delta) \leq {DoTE}(\delta) \leq {DoTE}^{U}(\delta),
$$
where
\begin{align*}
{DoTE}^{L}(\delta)& =F_{Y_{1 t}-Y_{0 t} \mid \boldsymbol{D}_{t}= (1,1,1)}^{L}(\delta) \\ & =E_{Y_{1 t-1}}\left[F_{Y_{1 t}-Y_{0 t} \mid Y_{1 t-1}, \boldsymbol{D}_{t}= (1,1,1)}^{L}\left(\delta \mid Y_{1 t-1}\right) \mid \boldsymbol{D}_{t}= (1,1,1)\right] \\
{DoTE}^{U}(\delta)& =F_{Y_{1 t}-Y_{0 t} \mid \boldsymbol{D}_{t}= (1,1,1)}^{U}(\delta)\\ &=E_{Y_{1 t-1}}\left[F_{Y_{1 t}-Y_{0 t} \mid Y_{1 t-1}, \boldsymbol{D}_{t}= (1,1,1)}^{U}\left(\delta \mid Y_{1 t-1}\right) \mid \boldsymbol{D}_{t}= (1,1,1)\right],
\end{align*}
and $F_{Y_{1 t}-Y_{0 t} \mid Y_{1 t-1}, \boldsymbol{D}_{t}= (1,1,1)}^{L}$ and $F_{Y_{1 t}-Y_{0 t} \mid Y_{1 t-1}, \boldsymbol{D}_{t}= (1,1,1)}^{U}$ are given in Lemma \ref{Chap2_lemma_cond_TE}.
\end{theorem}

Once we identify the DoTE, we can use the result in Theorem \ref{Theorem_DoTE} to identify the QoTE at period $t$ for the target group $\boldsymbol{D_{t}} = (1,1,1)$. The bounds of the QoTE are obtained by simply inverting the bounds of the DoTE. 

\begin{theorem}[QoTE Bounds under Model 1]\label{Chap2_theorem_QoTE} Under Assumptions \ref{Chap2_ass1} to \ref{Chap2_ass3}, we have, 
\begin{align*}
{QoTE}^{L}(\tau) \leq {QoTE}(\tau) \leq {QoTE}^{U}(\tau),
\end{align*}
where
\begin{align*}
&{QoTE}^{L}(\tau)=\inf\left\{\delta: {DoTE}^{L}(\delta)\right\}\\
&{QoTE}^{U}(\tau)=\inf\left\{\delta: {DoTE}^{U}(\delta)\right\},
\end{align*}
and ${DoTE}^{L}(\delta)$ and ${DoTE}^{U}(\delta)$ are given in Theorem \ref{Theorem_DoTE}.
\end{theorem}

This completes the results under Model 1. The proofs of all lemmas and theorems in this subsection are provided in the Supplementary Appendix A. %\ref{Chap2_Appendix Proofs}. 

We, next discuss the approach to derive the bounds on the joint distribution of treated and untreated potential outcomes and the DTEs under Model 2 using an alternative GCE assumption.

\subsection{DTE Bounds for Group $\mathbf{\textit{D}}_{t} = (1,1,1) $ - Model 2 } \label{Subsec Model setup 2}

Model 2 utilizes the second GCE assumption to partially identify the target joint distribution and the DTEs. In this section, we still focus on the group $ \boldsymbol{D}_{t} = (1,1,1) $.

When bounding the target parameters, Model 2 requires the same Assumptions \ref{Chap2_ass1} to \ref{Chap2_ass4}, but a different GCE assumption given below under Assumption \ref{Chap2_ass3Model2}. The key difference here compared to Model 1 is that we utilize the group $\boldsymbol{D}_{t} = (1,1,0)$ as the copula recovery group for $ \boldsymbol{D}_{t} = (1,1,1) $, in which individuals are untreated in the last period $t$. 

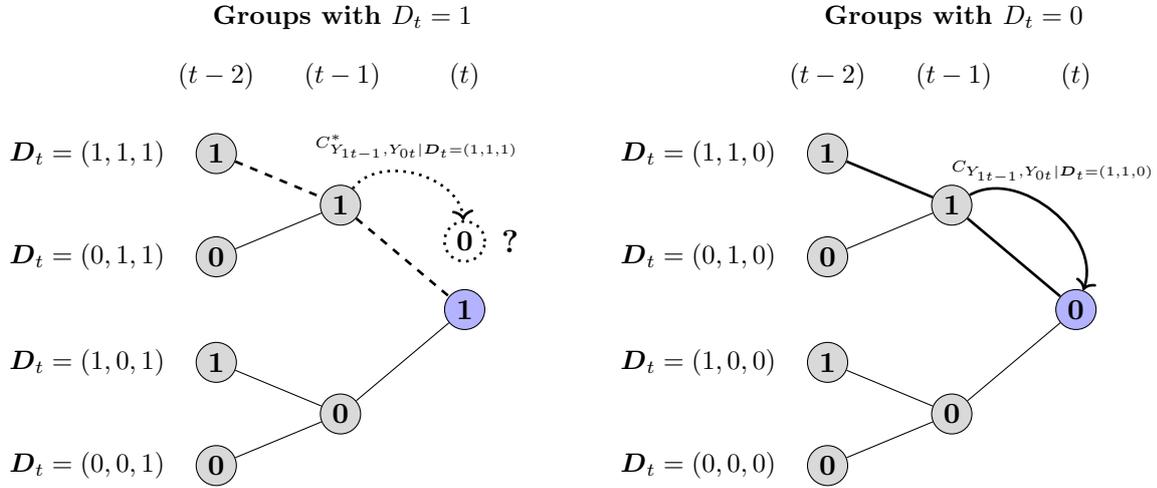
\begin{figure}[H]
\caption{\centering Visualising the Second Alternative GCE Assumption \ref{Chap2_ass3Model2}}\label{Chap2_fig tree DA(II)}
\begin{tikzpicture}[node distance=1.5cm, minimum size=0.1cm, inner sep=2, auto]
%%%%%%%%%%%%%%%%%%%%% left figure  %%%%%%%%%%%%%%%%%%%%%
\node [draw, circle, fill=blue!30] (G) {\textcolor{black}{\textbf{1}}};

% New dotted node above G
\node [draw, circle, dotted, line width=1pt, above = 0.35cm of G] (H) {\textbf{0}};
\node [right of=H, xshift=-0.9cm, font=\large] {\textbf{\textcolor{black}{?}}};

\node [draw, circle,  fill=gray!30, above left =1cm and 1.25cm of G] (C) {\textcolor{black}{\textbf{1}}};
\node [draw, circle, fill=gray!30, above left =0.3cm and 1.25cm of C] (A) {\textcolor{black}{\textbf{1}}};
\node [draw, circle, fill=gray!30, below left =0.3cm and 1.25cm of C] (B) {\textbf{0}}; % Adjust the distance as needed

\node [draw, circle, fill=gray!30, below left =1cm and 1.25cm of G] (D) {\textbf{0}};
\node [draw, circle, fill=gray!30, above left =0.3cm and 1.25cm of D] (E) {\textbf{1}};
\node [draw, circle, fill=gray!30, below left =0.3cm and 1.25cm of D] (F) {\textbf{0}}; % Adjust the distance as needed

% text 
\node [left of=A, xshift=-0.2cm, font=\small] {$\boldsymbol{D}_t=(1,1,1)$};
\node [left of=B, xshift=-0.2cm, font=\small] {$\boldsymbol{D}_t=(0,1,1)$};
\node [left of=E, xshift=-0.2cm, font=\small] {$\boldsymbol{D}_t=(1,0,1)$};
\node [left of=F, xshift=-0.2cm, font=\small] {$\boldsymbol{D}_t=(0,0,1)$};

\node [above = 0.5cm of A, font=\small] (t1) {$(t-2)$};
\node [right = 0cm and 0.5cm of t1, font=\small](t2) {$(t-1)$};
\node [right = 0cm and 0.75cm of t2, font=\small] {$(t)$};

% title
\node [above = 0.3cm and 0cm of t2, font=\small] {\textbf{Groups with $D_{t}=1$}};

% Arrows
\draw [-, line width=1pt, dashed] (A) -- (C);
\draw [-] (B) -- (C);     
\draw [-] (E) -- (D);
\draw [-] (F) -- (D);
\draw [-] (D) -- (G);
\draw [-, line width=1pt, dashed] (C) -- (G);

% Curved arrows
\draw [->, line width=1pt, dotted, bend left=70] (C) to node[midway, above=0.1cm] {\textcolor{black}{\tiny$C^{\ast}_{Y_{1t-1},Y_{0t}|\boldsymbol{D}_{t} = (1, 1, 1)}$}}  (H);
%\draw [->, line width=1pt, bend left=75, color=green] (E) to node[midway, above=0.1] {\textcolor{green}{$C_{Y_{1t-2},Y_{0t-1}|\mathbf{D}_{t} = (1, 0, 1)}$}}   (D);

%%%%%%%%%%%%%%%%%%%%% right figure   %%%%%%%%%%%%%%%%%%%%%
\node [draw, circle, fill=blue!30, right = 0 and 7.5cm of G] (Gr) {\textbf{0}};

\node [draw, circle, fill=gray!30, above left =1cm and 1.25cm of Gr] (Cr) {\textbf{1}};
\node [draw, circle, fill=gray!30, above left =0.3cm and 1.25cm of Cr] (Ar) {\textbf{1}};
\node [draw, circle, fill=gray!30, below left =0.3cm and 1.25cm of Cr] (Br) {\textbf{0}}; % Adjust the distance as needed

\node [draw, circle, fill=gray!30, below left =1cm and 1.25cm of Gr] (Dr) {\textbf{0}};
\node [draw, circle, fill=gray!30, above left =0.3cm and 1.25cm of Dr] (Er) {\textbf{1}};
\node [draw, circle, fill=gray!30, below left =0.3cm and 1.25cm of Dr] (Fr) {\textbf{0}}; % Adjust the distance as needed

% text 
\node [left of=Ar, xshift=-0.2cm, font=\small] {$\boldsymbol{D}_t=(1,1,0)$};
\node [left of=Br, xshift=-0.2cm, font=\small] {$\boldsymbol{D}_t=(0,1,0)$};
\node [left of=Er, xshift=-0.2cm, font=\small] {$\boldsymbol{D}_t=(1,0,0)$};
\node [left of=Fr, xshift=-0.2cm, font=\small] {$\boldsymbol{D}_t=(0,0,0)$};

\node [above = 0.5cm of Ar, font=\small] (t1r) {$(t-2)$};
\node [right = 0cm and 0.5cm of t1r, font=\small](t2r) {$(t-1)$};
\node [right = 0cm and 0.75cm of t2r, font=\small] {$(t)$};

% title
\node [above = 0.3cm and 0cm of t2r, font=\small] {\textbf{Groups with $D_{t}=0$}};

% Arrows
\draw [-, line width=1pt] (Ar) -- (Cr);
\draw [-] (Br) -- (Cr);
\draw [-] (Er) -- (Dr);
\draw [-] (Fr) -- (Dr);
\draw [-] (Dr) -- (Gr);
\draw [-, line width=1pt] (Cr) -- (Gr);

% Curved arrows
% Curved arrows
\draw [->, line width=1pt, bend left=70] (Cr) to node[midway, above=0.3cm] {\textcolor{black}{\tiny$C_{Y_{1t-1},Y_{0t}|\boldsymbol{D}_{t} = (1, 1, 0)}$}}  (Gr);
\end{tikzpicture}

\end{figure}

\begin{assump}[GCE (II)]\label{Chap2_ass3Model2}  For all $ (u, v) \in [0, 1]^{2} $, and for the treatment group $ \boldsymbol{D}_{t} = (1,1,1) $, we assume,% and for the target group $ \boldsymbol{D}_{t} = (1,1,1) $,% it is assumed that, 
\begin{align*}
C^{\ast}_{Y_{1t-1},Y_{0t}|\boldsymbol{D}_{t} = (1,1,1)}(u, v) &= C_{Y_{1t-1},Y_{0t}|\boldsymbol{D}_{t} = (1,1,0)}(u, v).
\end{align*}	
where the superscript ${\ast} $ denotes the ``unobserved" component.		
\end{assump}
Assumption \ref{Chap2_ass3Model2} provides an alternative technique to recover the same target unobserved copula for $ \boldsymbol{D}_{t} = (1, 1, 0) $, given in Assumption \ref{Chap2_ass3} (LHS). However, unlike Assumption \ref{Chap2_ass3}, which leverages information from another treated group at period $ t $, Assumption \ref{Chap2_ass3Model2} utilizes information from its untreated group at period $t$: $ \boldsymbol{D}_{t} = (1, 1, 0) $ which shares a similar treatment history. It assumes that the unknown copula between the two potential outcomes $ Y_{0t} $ and $ Y_{1t-1} $  for the target group $  \boldsymbol{D}_{t} = (1,1,1)  $, is equal to the copula of the potential outcomes in the same adjacent periods $ Y_{0t} $ and $ Y_{1t-1} $ for its untreated group $  \boldsymbol{D}_{t} = (1,1,0) $. Figure \ref{Chap2_fig tree DA(II)} visualizes the logic employed under Assumption \ref{Chap2_ass3Model2}.

As in Figure \ref{Chap2_fig tree DA(I)}, Figure \ref{Chap2_fig tree DA(II)} represent the treatment path of the target group $ \boldsymbol{D}_{t} = (1,1,1) $ using dashed lines, and the unknown copula with the dotted arrow. The bolded path represents the copula recovery group (or the untreated group) $ \boldsymbol{D}_{t} = (1, 1, 0) $, used under our second GCE assumption \ref{Chap2_ass3Model2} with its copula denoted by the solid arrow. 
Assumption \ref{Chap2_ass3Model2} assumes that the unknown copula (LHS) is the same as the identifiable copula of the treated potential outcome $Y_{1t-1}  $ at period $ t-1 $, and the untreated potential outcome $ Y_{0t} $ at period $ t $, of the untreated group $ \boldsymbol{D}_{t} = (1,1,0) $ (RHS).

Unlike in Model 1, which makes comparisons among the individuals who are treated at the last period $ t $ but across different points in time (i.e., ($ t-1 $ to $ t $) and ($ t-2 $ to $ t-1 $)), Model 2 makes comparisons in the same adjacent periods (i.e., $ t-1 $ to $ t $) but across individuals who are in two different groups in the last period $ t $ with a similar treatment history. 

Both Assumption \ref{Chap2_ass3Model2} and Assumption \ref{Chap2_ass3} acts similarly as the copula stability assumption (CSA) of \cite{callaway2021bounds}, as they do not impose restrictions directly on the unknown copula in Equation \eqref{Chap2_Eq sklar formula}. This makes them weaker and plausible than the other strong point-identifying assumptions.

In contrast to Model 1, here the same group $\boldsymbol{D}_{t} = (1,1,0)$ is used to identify both unobserved components: (i) the counterfactual marginal distribution for $\boldsymbol{D}_{t} = (1,1,1)$ under the CiC method (see Equation \eqref{Chap2_CiC_marginal}), and (ii) the unknown copula using Assumption \ref{Chap2_ass3Model2}. This can be advantageous when the available panel contains limited information. For instance, if a researcher aims to identify the DTEs for individuals with the pattern $\boldsymbol{D}_{t} = (1,1,1)$ but lacks observations for the group $\boldsymbol{D}_{t} = (1,0,1)$, Model 2 is preferable to Model 1. 

Next, we explain the derivations of the bounds on the joint distribution and the target parameters under Model 2 for group $ \boldsymbol{D}_{t}= (1,1,1) $. As in  Model  1, we first identify the joint distribution of $Y_{0 t}$ and $Y_{1 t-1}$ using Assumption \ref{Chap2_ass3Model2}. We have,
\begin{align}\label{Chap2_Eqn Model2JointY0t&Y1t-1}
&F_{Y_{0 t}, Y_{1 t-1} \mid \boldsymbol{D}_{t}= (1,1,1)}\left(y_{0}, y^{\prime}_{1}\right)\nonumber\\
&=C_{Y_{0 t}, Y_{1 t-1} \mid \boldsymbol{D}_{t}= (1,1,1)}\left(F_{Y_{0 t} \mid \boldsymbol{D}_{t}= (1,1,1)}\left(y_{0}\right), F_{Y_{1t-1} \mid \boldsymbol{D}_{t}= (1,1,1)}\left(y^{\prime}_{1}\right)\right)\nonumber \\
&=C_{Y_{0 t}, Y_{1 t-1} \mid \boldsymbol{D}_{t}= (1,1,0)}\left(F_{Y_{0 t} \mid \boldsymbol{D}_{t}= (1,1,1)}\left(y_{0}\right), F_{Y_{1 t-1} \mid \boldsymbol{D}_{t}= (1,1,1)}\left(y^{\prime}_{1}\right)\right)\nonumber \\
&=F_{Y_{0 t}, Y_{1 t-1} | \boldsymbol{D}_{t}= (1,1,0)}\left(F_{Y_{0 t} |\boldsymbol{D}_{t}= (1,1,0)}^{-1}  \left(F_{Y_{0 t}| \boldsymbol{D}_{t}= (1,1,1)}\left(y_{0}\right)\right), \right. \nonumber\\
& \qquad \qquad \qquad \qquad \qquad \qquad \qquad \qquad \left. F_{Y_{1 t-1} | \boldsymbol{D}_{t}= (1,1,0)}^{-1}  \left(F_{Y_{1 t-1} | \boldsymbol{D}_{t}= (1,1,1)}\left(y^{\prime}_{1}\right)\right)\right).
\end{align}

Using this result, the conditional counterfactual distribution of $ Y_{0t} $ given the outcome in the previous period is identified. Lemma \ref{Chap2_lemma1Model2} provides the resulting expression to identify this conditional counterfactual marginal distribution.

\begin{lemma}[Distribution of $Y_{0t}$ Conditional  on $Y_{1 t-1}$ under Model 2]\label{Chap2_lemma1Model2}
Under Assumptions \ref{Chap2_ass1} to \ref{Chap2_ass4} and \ref{Chap2_ass3Model2}, and the results in the Equation \eqref{Chap2_Eqn Model2JointY0t&Y1t-1}, we have,
\begin{align*}
&F_{Y_{0 t} \mid Y_{1 t-1},\boldsymbol{D}_{t}= (1,1,1)}\left(y_{0} \mid y^{\prime}_{1}\right)\\
&=F_{Y_{0 t}\mid  Y_{1 t-1}, \boldsymbol{D}_{t}= (1,1,0)}\left(F_{Y_{0 t} \mid \boldsymbol{D}_{t}= (1,1,0)}^{-1}  \left(F_{Y_{0 t} \mid \boldsymbol{D}_{t}= (1,1,1)}\left(y_{0}\right)\right) \mid  \right. \\
& \qquad \qquad \qquad \qquad \qquad \qquad \qquad \qquad \left.F_{Y_{1 t-1} \mid \boldsymbol{D}_{t}= (1,1,0)}^{-1}  \left(F_{Y_{1 t-1} \mid \boldsymbol{D}_{t}= (1,1,1)}\left(y^{\prime}_{1}\right)\right)\right).
\end{align*}
\end{lemma}

Once this unobserved conditional marginal distribution $ F_{Y_{0t}|Y_{1t-1},\boldsymbol{D}_{t}= (1,1,1)} (y_{0}\mid y^{\prime})  $ is identified, a similar procedure as used in Model 1 can be applied to identify the target joint and the DTE parameters of interest. That means we identify the bounds on the joint distribution of the treated and untreated potential outcomes and the target parameters (DoTE and QoTE) by applying the {Fr$ \acute{e} $chet-Hoeffiding} bounds and the WD bounds, respectively. 
To avoid repeating the same content, we leave all results and proofs of Model 2 to the Supplementary Appendix  Section B. %\ref{Appendix Sec M2 D(111)}.

In the following sections we discuss the plausibility and testability of our two key alternative GCE assumptions, \ref{Chap2_ass3} and \ref{Chap2_ass3Model2}.

\subsection{Plausibility of the GCE Assumptions}\label{Chap2_SubChap2_Sec plausibility}

This section discusses the plausibility of the two GCE assumptions imposed under Models 1 and 2 in other panel data models by considering the two-way fixed effects (TWFE) model. In particular, we provide some sufficient conditions under which the GCE assumptions hold. These additional conditions required for the GCE assumptions in the TWFE models are fairly weak, highlighting the applicability of our proposed methods in empirical studies.

The two-way fixed effects model can be considered the leading choice in panel data and cases with time-invariant unobserved heterogeneity. In general, the two-way fixed effects model can be defined as follows. Then, for $ s \in \{t-2, t-1,t\} $,
\begin{align} \label{Chap2_TWFEs model}
Y_{s} = \theta_{s} + \eta + \alpha D_{s} + V_{s}
\end{align}
where, $ Y_{s} $ is the outcome at a period $ s $ for an individual $ i $ with the treatment status $ D_{s} \in \{0,1\} $, $ \theta_{s} $ are the time fixed effects, $ \eta $ are the time-invariant unobserved heterogeneity that can be distributed differently for individuals in the treated groups and untreated groups, and $ V_{s} $ are the time-varying unobservables.

According to Equation \eqref{Chap2_TWFEs model}, the models for the treated and untreated potential outcomes can be expressed as below, under Equations \eqref{Chap2_treated TWFE} and \eqref{unChap2_treated TWFE}, respectively:
\begin{align}
Y_{1s} &= \theta_{s} + \eta + \alpha + V_{s} \label{Chap2_treated TWFE}\\ 
Y_{0s} &= \theta_{s} + \eta + V_{s} \label{unChap2_treated TWFE}
\end{align}
Next, we consider under what additional conditions the key copula assumptions hold in the two-way fixed effects model. Even though the additional conditions will be different for each of the eight treatment patterns, the procedure that needs to be followed to obtain the results under each case is the same. Therefore, the findings are illustrated only for the target group $ \boldsymbol{D}_{t} = (1,1,1)$ under Model 1, to avoid repeating the same material in the main study.

\subsubsection{For the Treatment Pattern $ \mathbf{\textit{D}}_{t} = (1,1,1)$ under Model 1}
In order to satisfy the first GCE Assumption \ref{Chap2_ass3} under Model 1 for the target group $ \boldsymbol{D}_{t} = (1,1,1)$, the following additional conditions given under Proposition \ref{Chap2_M1_prop_TWFE} are imposed.

\begin{proposition}\label{Chap2_M1_prop_TWFE}
In the two-way fixed effect model and under the additional condition given below, the GCE Assumption \ref{Chap2_ass3} in Model 1,  holds.
\begin{equation}\label{Chap2_M1_prop_TWFE Copula Eqn}
%C_{\eta_{i}+V_{it},\,\,\eta_{i}+\alpha+V_{it-1}\mid\boldsymbol{D}_{t} = (1,1,1)}(u,v)=C_{\eta_{i}+V_{it-1},\,\,\eta_{i}+\alpha+V_{it-2}\mid\boldsymbol{D}_{t}=(1,0,1)}(u,v).
C_{\eta+V_{t},\,\,\eta+\alpha+V_{t-1}\mid\boldsymbol{D}_{t} = (1,1,1)}(u,v)=C_{\eta+V_{t-1},\,\,\eta+\alpha+V_{t-2}\mid\boldsymbol{D}_{t}=(1,0,1)}(u,v).
\end{equation} 
(The proof of Proposition \ref{Chap2_M1_prop_TWFE} is provided in the Supplementary Appendix Section A.5) %\ref{Chap2_Sec plausibility})
\end{proposition}  

{This sufficient condition, given in Equation \eqref{Chap2_M1_prop_TWFE Copula Eqn}, allows for serial correlation in the time-varying unobservables and permits their distributions to change over time for individuals who are treated in the last period $t$, %Due to this, it can be viewed as a relatively weaker condition than the stringent rank invariance assumption 
and is also similar to the condition outlined in \cite{callaway2021bounds} (See Proposition 2), under which his copula stability assumption would hold.}

\subsection{Testability of the GCE Assumptions}\label{Chap2_Subsec Testability}

The two GCE assumptions are the key assumptions employed in bounding the joint distribution and the target DTE parameters. In this section, we discuss the testability of these assumptions in practice. In particular, we consider a parametric and a non-parametric testing method when there are at least six periods of panels. Although our GCE Assumptions \ref{Chap2_ass3} and \ref{Chap2_ass3Model2} are not directly testable since $Y_{0t}$ is not observed for individuals in group $\boldsymbol{D}_{t} = (1,1,1)$, we can use historical observations when the two outcomes in these GCE assumptions were observed. Therefore, when multiple periods are available in the panel, it is possible to implement a test for these copula assumptions using historical observations.

Let us first introduce some additional notations as we now move to more than three time periods. Consider the simplest case with six time periods where  $ s \in \{t-5, t-4, t-3, t-2, t-1, t\}$. Recall that $ \boldsymbol{D}_{t} =(D_{t-2},\, D_{t-1},\, D_{t}) $ denotes the treatment pattern in the \textit{last three periods}. Let $ \boldsymbol{D}_{t-3} = (D_{t-5},\, D_{t-4},\, D_{t-3})$ denote the treatment pattern in the \textit{initial three periods}.

Recall that Assumption \ref{Chap2_ass3}, imposed under Model 1 assumes that the copula structure of $ (Y_{1t-1}, Y_{0t} )$ in the target group $ \boldsymbol{D}_{t} = (1,1,1) $, is same as that of $(Y_{1t-2}, Y_{0t-1} )$ in the copula recovery group $ \boldsymbol{D}_{t} = (1,0,1) $:
\begin{align*}
C^{\ast}_{Y_{1t-1},Y_{0t}|\boldsymbol{D}_{t} = (1, 1, 1)}(u, v) &= C_{Y_{1t-2},Y_{0t-1}|\boldsymbol{D}_{t} = (1, 0, 1)}(u, v).
\end{align*}
This can be tested using observations in the initial three time periods $ t-5$, $t-4$, and $ t-3$, by comparing the two identifiable dependences given below between the two subgroups: one with the pattern $ \boldsymbol{D}_{t-3} = (1,1,0)$ given $\boldsymbol{D}_{t} = (1,1,1) $, and the other with the pattern $ \boldsymbol{D}_{t-3} = (1,0,1)$ given $\boldsymbol{D}_{t} = (1,0,1) $, 
\begin{align}\label{test_firstGCE}
C_{Y_{1t-4},Y_{0t-3}|\boldsymbol{D}_{t-3} = (1,1,0),\, \boldsymbol{D}_{t} = (1, 1, 1)}(u, v) &= C_{Y_{1t-5},Y_{0t-4}|\boldsymbol{D}_{t-3} = (1,0,1),\, \boldsymbol{D}_{t} = (1, 0, 1)}(u, v).
\end{align}
Similarly, one could use the same procedure to test GCE Assumption \ref{Chap2_ass3Model2}, imposed under Model 2: 
\begin{align*}
C^{\ast}_{Y_{1t-1},Y_{0t}|\boldsymbol{D}_{t} = (1,1,1)}(u, v) &= C_{Y_{1t-1},Y_{0t}|\boldsymbol{D}_{t} = (1,1,0)}(u, v),
\end{align*}	
by considering the observed historical outcomes in the group with a treatment pattern $  \boldsymbol{D}_{t-3} = (1,1,0)$ and $  \boldsymbol{D}_{t} = (1,1,1)$, and the group with a treatment pattern $  \boldsymbol{D}_{t-3} = (1,1,0)$ and $  \boldsymbol{D}_{t} = (1,1,0)$,
\begin{align}\label{test_secondGCE}
C_{Y_{1t-4},Y_{0t-3}|\boldsymbol{D}_{t-3} = (1,1,0),\, \boldsymbol{D}_{t} = (1,1,1)}(u, v) &= C_{Y_{1t-4},Y_{0t-3}|\boldsymbol{D}_{t-3} = (1,1,0),\, \boldsymbol{D}_{t} = (1,1,0)}(u, v).
\end{align}	

To test Equations \eqref{test_firstGCE} and \eqref{test_secondGCE}, we can follow a similar procedure as used in \cite{callaway2021bounds}, which provides both a parametric and a non-parametric method.  Under the parametric method, we compute the dependence measures such as Spearman's rho or Kendall's tau of the two terms on both sides of Equation \eqref{test_firstGCE} (or \eqref{test_secondGCE} for Assumption \ref{Chap2_ass3Model2}), and check whether these pair of dependence measures are the same.%for the above-explained case in the earlier periods.
\footnote{However, as pointed out by \cite{callaway2021bounds}, the parametric testing method based on correlation measures, such as the Spearman’s rho or Kendall’s tau, might not be sufficient, because the same correlations do not necessarily imply their underlying copulas to be the same.} 
Alternatively, we can also use the non-parametric method introduced by \cite{remillard2009testing}, which tests whether the underlying two copulas on both sides of Equation \eqref{test_firstGCE} (or \eqref{test_secondGCE} for Assumption \ref{Chap2_ass3Model2}) are the same. Hence, will detect the violations of the GCE assumptions that the parametric test might not detect. 
We provide the testing results of  both parametric and non-parametric methods in our application section.

\section{Estimation and Inference } \label{Chap2_Estimation and Inference}

This section discusses the estimation and inference under both model setups for our group-wise DTE parameters. We follow similar approaches as in \cite{callaway2021bounds}.

First, we discuss the estimation process for Model 1. The proposed estimators for the DoTE and QoTE at period $ t $ for the individuals in group $ \boldsymbol{D}_{t}=(1,1,1) $ can be derived by substituting the first step estimators for terms derived in the lemmas into the expressions of the target parameters given in Theorems \ref{Theorem_DoTE} and \ref{Chap2_theorem_QoTE}. That is, 
\begin{align*}
&{\widehat{DoTE}}^{L}(\delta) = \dfrac{1}{n}\sum_{i=1}^{n}\dfrac{D_{it}}{\widehat{p}}\, \widehat{F}_{Y_{1 t}-Y_{0 t} \mid Y_{1t-1},\,\boldsymbol{D}_{t} =(1,1,1)}^{L}\left(\delta \mid Y_{1t-1}\right)  \\
&{\widehat{DoTE}}^{U}(\delta) = \dfrac{1}{n}\sum_{i=1}^{n}\dfrac{D_{it}}{\widehat{p}}\, \widehat{F}_{Y_{1 t}-Y_{0 t} \mid Y_{1t-1},\boldsymbol{D}_{t} =(1,1,1)}^{U}\left(\delta \mid Y_{1t-1}\right),
\end{align*}
where $ \widehat{p} =Pr\{\boldsymbol{D}_{t} =(1,1,1)\}$ denotes the proportion of individuals in group $ \boldsymbol{D}_{t} =(1,1,1) $, %, and $ D_{it} $ denotes the treatment status of the $ i^{th} $ individual in this group at period $ t $. 
and ${D_{it}}/{\widehat{p}}$ averages the values of $Y_{1t-1}$ over the individuals in group $ \boldsymbol{D}_{t} =(1,1,1) $. 

The estimators of $ {F}_{Y_{1 t}-Y_{0 t} \mid Y_{1t-1},\,\boldsymbol{D}_{t} =(1,1,1)}^{L}  $ and $ {F}_{Y_{1 t}-Y_{0 t} \mid Y_{1t-1},\,\boldsymbol{D}_{t} =(1,1,1)}^{U}  $ in Lemmas \ref{Chap2_lemma1} and \ref{Chap2_lemma_cond_TE} further rely on the preliminary estimators of $ {F}_{Y_{1 t} \mid Y_{1t-1},\boldsymbol{D}_{t} =(1,1,1)} $ and $ {F}_{Y_{0 t} \mid Y_{1t-1},\boldsymbol{D}_{t} =(1,1,1)} $. The distribution $ {F}_{Y_{1 t} \mid Y_{1t-1},\boldsymbol{D}_{t}=(1,1,1)} $ can be estimated directly using observed data, while 
the formula derived in Lemma \ref{Chap2_lemma1} can be used to estimate the counterfactual marginal distribution, $ {F}_{Y_{0 t} \mid Y_{1t-1},\boldsymbol{D}_{t} =(1,1,1)} $. As given in Lemma \ref{Chap2_lemma1}, the estimation of $ {F}_{Y_{0 t} \mid Y_{1t-1},\boldsymbol{D}_{t} =(1,1,1)} $ depends further on the estimators of $ {F}_{Y_{0 t-1} \mid Y_{1t-2},\boldsymbol{D}_{t} =(1,0,1)} $, $ {F}_{Y_{0 t-1} \mid \boldsymbol{D}_{t} =(1,0,1)} $, $ {F}_{Y_{0 t} \mid \boldsymbol{D}_{t} =(1,1,1)} $, $ {F}_{Y_{1 t-2} \mid \boldsymbol{D}_{t} =(1,0,1)} $, and $ {F}_{Y_{1 t-1} \mid\boldsymbol{D}_{t} =(1,1,1)} $. Observe that, except for the marginal distribution of $ {F}_{Y_{0 t} \mid \boldsymbol{D}_{t} =(1,1,1)} $, all the other four distributions can be estimated directly from the observed data. As mentioned in Assumption \ref{Chap2_ass2}, the estimation of $ {F}_{Y_{0 t} \mid \boldsymbol{D}_{t} =(1,1,1)} $ 
can be implemented using the Changes-in-changes (CiC) method:
\begin{equation*}
F_{Y_{0t}|\boldsymbol{D}_{t} = (1,1,1)}(y)=F_{Y_{1t-1}|\boldsymbol{D}_{t} = (1,1,1)}\left(F^{-1}_{Y_{1t-1}|\boldsymbol{D}_{t} = (1,1,0)}\left(F_{Y_{0t}|\boldsymbol{D}_{t} = (1,1,0)}(y)\right)\right),
\end{equation*}
which further relies on the estimators of $ F_{Y_{1t-1}|\boldsymbol{D}_{t} = (1,1,0)} $ and $ F_{Y_{0t}|\boldsymbol{D}_{t} = (1,1,0)} $.

In summary, to estimate the bounds on our target parameters under Model 1, we require the estimates of the following conditional and marginal distributions:
\begin{enumerate}
\item
$ {F}_{Y_{1 t}\mid Y_{1t-1},\,\boldsymbol{D}_{t} =(1,1,1)}  $ and $ {F}_{Y_{0 t} \mid Y_{1t-1},\,\boldsymbol{D}_{t} =(1,1,1)} $
%\item $ {F}_{Y_{0 t-1} \mid Y_{1t-2},\boldsymbol{D}_{t}=(1,0,1)} $ 
\item ${F}_{Y_{0 t-1} \mid \boldsymbol{D}_{t} =(1,0,1)} $,
%$ {F}_{Y_{0 t} \mid \boldsymbol{D}_{t} =(1,1,1)} $,
$ \, {F}_{Y_{1 t-2} \mid \boldsymbol{D}_{t} =(1,0,1)} $,
$ \, {F}_{Y_{1 t-1} \mid\boldsymbol{D}_{t} =(1,1,1)} $,
$ \, F_{Y_{1t-1}|\boldsymbol{D}_{t} = (1,1,0)} $, \, and 
$ \, F_{Y_{0t}|\boldsymbol{D}_{t} = (1,1,0)} $.
\end{enumerate}

Our study applies the distributional regression and the quantile regression approaches in the process of estimation \citep[similar to the work of][]{melly2015changes,callaway2021bounds,koenker2013distributional}. We obtain flexible parametric first-step estimators, using distributional regression for the conditional distributions $ {F}_{Y_{1 t}\mid Y_{1t-1},\,\boldsymbol{D}_{t} =(1,1,1)}  $ and $ {F}_{Y_{0 t} \mid Y_{1t-1},\,\boldsymbol{D}_{t} =(1,1,1)} $, and quantile regression for the marginal distributions listed above.

Next, we discuss the estimation process for Model 2. According to Lemma \ref{Chap2_lemma1Model2}, and the resulting expression to recover the counterfactual marginal of $ {F}_{Y_{0 t} \mid \boldsymbol{D}_{t} =(1,1,1)} $ under the CiC method (i.e., Equation \eqref{Chap2_CiC_marginal}), the following %first-step
estimators are required to estimate the bounds on the target parameters under Model 2: %They are, in summary, 
\begin{enumerate}
\item 
$ {F}_{Y_{1 t}\mid Y_{1t-1},\,\boldsymbol{D}_{t} =(1,1,1)}  $ and
$ {F}_{Y_{0 t} \mid Y_{1t-1},\,\boldsymbol{D}_{t} =(1,1,1)} $
%\item $ {F}_{Y_{0 t-1} \mid Y_{1t-2},\boldsymbol{D}_{t}=(1,0,1)} $ 

\item 
%$ {F}_{Y_{0 t} \mid \boldsymbol{D}_{t} =(1,1,1)} $,
$ {F}_{Y_{1 t-1} \mid\boldsymbol{D}_{t} =(1,1,1)} $,
$ F_{Y_{1t-1}|\boldsymbol{D}_{t} = (1,1,0)} $, and 
$ F_{Y_{0t}|\boldsymbol{D}_{t} = (1,1,0)} $
\end{enumerate}
These can be estimated by following similar procedures, as in Model 1.

For inference, the numerical bootstrap method proposed by \cite{hong2018numerical} can be applied. 
First, we simulate the limiting distributions of the first step estimators using the standard empirical bootstrap method. 
Denote the first step estimators obtained using the sample data by $ \hat{F}$ and the bootstrapped first step estimators by $ \hat{F}^{B}$. According to the numerical delta method proposed by \cite{hong2018numerical}, for  a tuning parameter $ \epsilon_{n} $ such that $ \epsilon_{n}\sqrt{n}\rightarrow \infty $ and  $ \epsilon_{n} \rightarrow 0  $ as $ n \rightarrow \infty $, we have, 
\begin{align}\label{Chap2_inference eqn}
\dfrac{\phi\left(\hat{F} - \epsilon_{n}\sqrt{n}\left(\hat{F}^{B} - \hat{F}\right) \right) - \phi\left(\hat{F}\right)}{\epsilon_{n}} & \rightsquigarrow \mathcal{J},
\end{align}
where $ \phi $ is a mapping from a distribution function to the bounds of the parameters of interest, and $ \mathcal{J} $ is the limiting distribution for the estimators of the lower and upper bounds. 
The confidence intervals for the bounds of our target parameters, DoTE and QoTE, can be constructed by simulating the term on the LHS of the Equation \eqref{Chap2_inference eqn} for a larger number of times. Let $ \phi^{L} $ and $ \phi^{U} $ be the maps from distribution functions to the lower and upper bounds of our target DTE parameters, respectively. Then, the ($ 1 - \alpha $) confidence interval for the lower  bound is $ \phi^{L}(\hat{F}) - \frac{\hat{c}^{L}_{1-\alpha}}{\sqrt{n}} $, where $ \hat{c}^{L}_{1-\alpha} $ is the $ (1-\alpha)^{th} $ percentile of the term on the LHS of the Equation \eqref{Chap2_inference eqn}. The same approach can be applied to obtain confidence intervals for the  upper bounds of our target parameters.

\section[Numerical Illustrations and Simulations]{Numerical Illustrations and Monte Carlo Simulations} \label{Chap2_Numerical Illustration and Sim}

\subsection{Numerical Illustrations }\label{Chap2_Sec Numerical Illustration}

This section illustrates the tightness of the true DTE bounds obtained using our proposed method, 
utilizing a data generating process (DGP) similar to that used in \cite{callaway2021bounds}.  %(given in the section SB in the Appendix, in his study). 
As in previous sections, the group $ \boldsymbol{D}_{t} =(1,1,1) $ is used as an example to perform the numerical illustrations. As the bounds on the QoTE can be obtained by simply taking the inverse of the DoTE bounds, we present the numerical illustrations only for our target parameter DoTE given the target group $ \boldsymbol{D}_{t} =(1,1,1) $, at period $ t $. Suppose the following joint distributions hold in the DGP:
\begin{align*}
\begin{pmatrix}
Y_{1t} \\
Y_{1t-1}
\end{pmatrix}\Big| \boldsymbol{D}_{t} =(1,1,1) &\sim N\left(\begin{pmatrix}0\\0\end{pmatrix},\begin{pmatrix}1&\rho_1\\\rho_1&1\end{pmatrix}\right)\\
&=C(F_{Y_{1t}| \boldsymbol{D}_{t} =(1,1,1)}(Y_{1t}),F_{Y_{1t-1}| \boldsymbol{D}_{t} =(1,1,1)}(Y_{1t-1}),\,\rho_1),\\\\
\begin{pmatrix}
Y_{0t} \\
Y_{1t-1}
\end{pmatrix}\Big| \boldsymbol{D}_{t} =(1,1,1)&\sim N\left(\begin{pmatrix}0\\0\end{pmatrix},\begin{pmatrix}1&\rho_0\\\rho_0&1\end{pmatrix}\right)\\
&=C(F_{Y_{0t}| \boldsymbol{D}_{t} =(1,1,1)}(Y_{0t}),F_{Y_{1t-1}| \boldsymbol{D}_{t} =(1,1,1)}(Y_{1t-1}),\,\rho_0),
\end{align*}
where $C(\cdot,\cdot,\rho_{j})$ is the normal copula with dependence parameter $\rho_{j}$ for $ j \in \{0,1\} $.

Given these  joint distributions, we have %the following conditional distributions given the previous outcomes can be identified.
\begin{align}\label{DGP_numer_conditional1}
&Y_{1t}| \boldsymbol{D}_{t} =(1,1,1), Y_{1t-1}=v'\sim N(\rho_1v',(1-\rho^2_1)),\\\label{DGP_numer_conditional2}
&Y_{0t}| \boldsymbol{D}_{t} =(1,1,1), Y_{1t-1}=v'\sim N(\rho_0 v',(1-\rho^2_0)).
\end{align}
Next, the DoTE bounds are computed by substituting the two distributions in Equations \eqref{DGP_numer_conditional1} and \eqref{DGP_numer_conditional2} into the formulas provided in Lemma \ref{Chap2_lemma_cond_TE}. These results are then applied to Theorem \ref{Theorem_DoTE} to compute the lower and upper bounds on the DoTE for group $ \boldsymbol{D}_{t} =(1,1,1) $. To demonstrate the performance of our methods under different dependence degrees, we set $ \rho_1=0 $ and $ \rho_0 \in \{0, 0.3, 0.5, 0.9\} $, and compare these DoTE bounds with the WD bounds, which do not incorporate any dependence information.\footnote{Where $ \rho_0 $ is our main dependence parameter,  which denotes the dependence between $ Y_{0t} $ and $ Y_{1t-1} $ for group $\boldsymbol{D}_{t} =(1,1,1)$.}

\begin{figure}[H]
\caption{\centering Numerical Illustration of the DoTE Bounds for Group $\boldsymbol{D}_{t} =(1,1,1)$} 
\label{fig nume bounds}
\begin{subfigure}[b]{0.5\linewidth}
\centering
\includegraphics[width=0.7\linewidth]{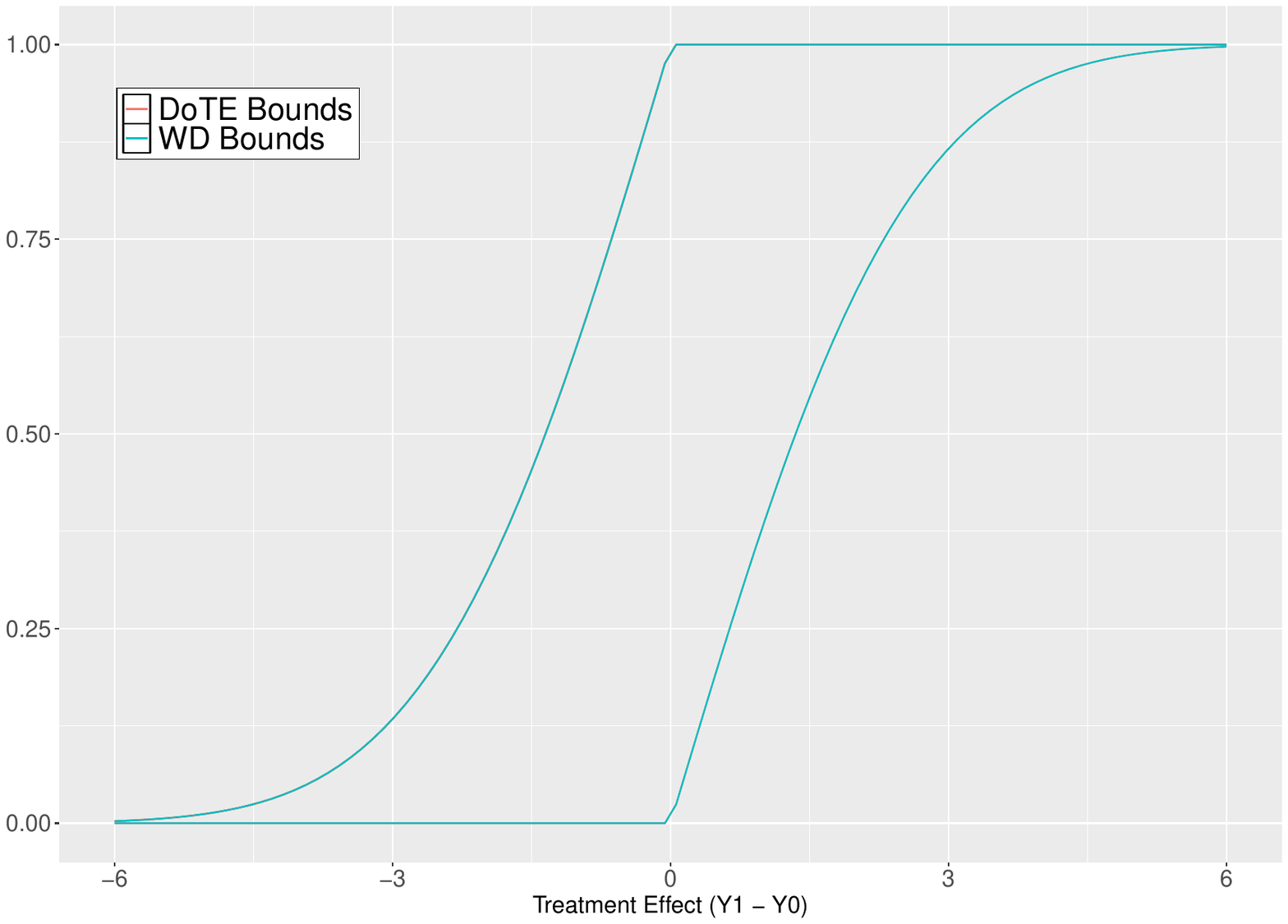}
\caption{$ \rho_{0} $ = 0 }
\label{fig1:a}
\vspace{4ex}
\end{subfigure}
\begin{subfigure}[b]{0.5\linewidth}
\centering
\includegraphics[width=0.7\linewidth]{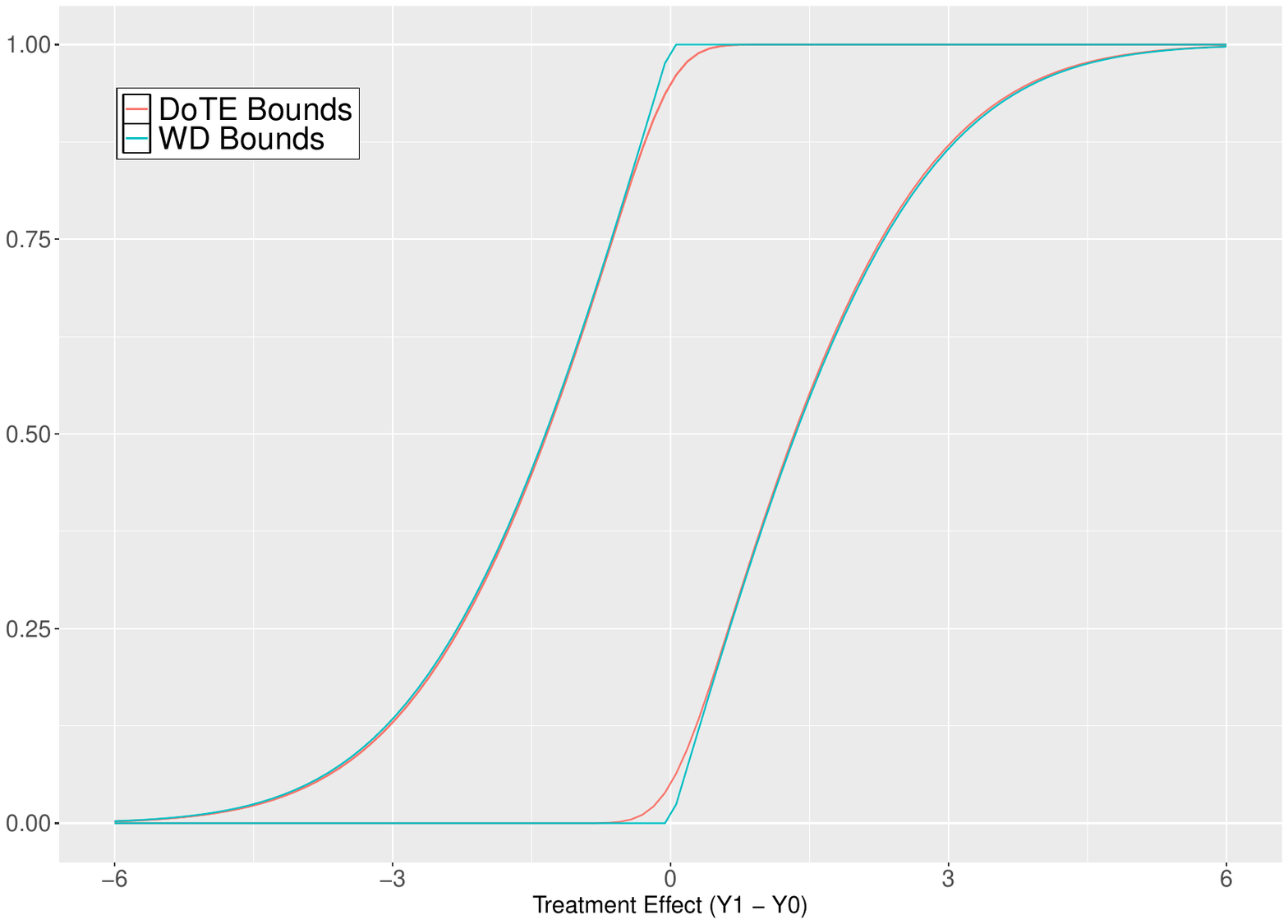}
\caption{$ \rho_{0} $ = 0.3 }
\label{fig1:b}
\vspace{4ex}
\end{subfigure}
\begin{subfigure}[b]{0.5\linewidth}
\centering
\includegraphics[width=0.7\linewidth]{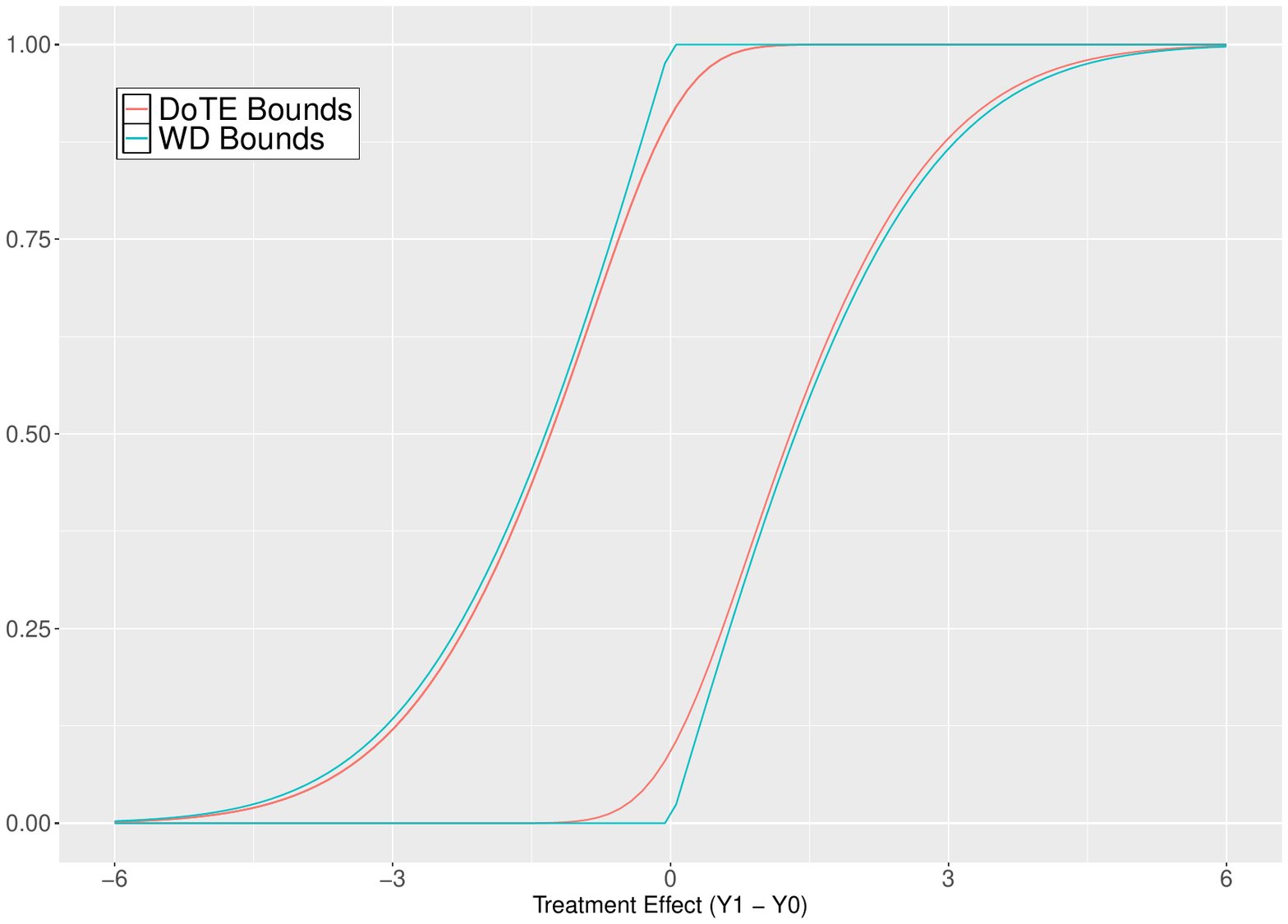}
\caption{$ \rho_{0} $ = 0.5}
\label{fig1.c}
\end{subfigure}
\begin{subfigure}[b]{0.5\linewidth}
\centering
\includegraphics[width=0.7\linewidth]{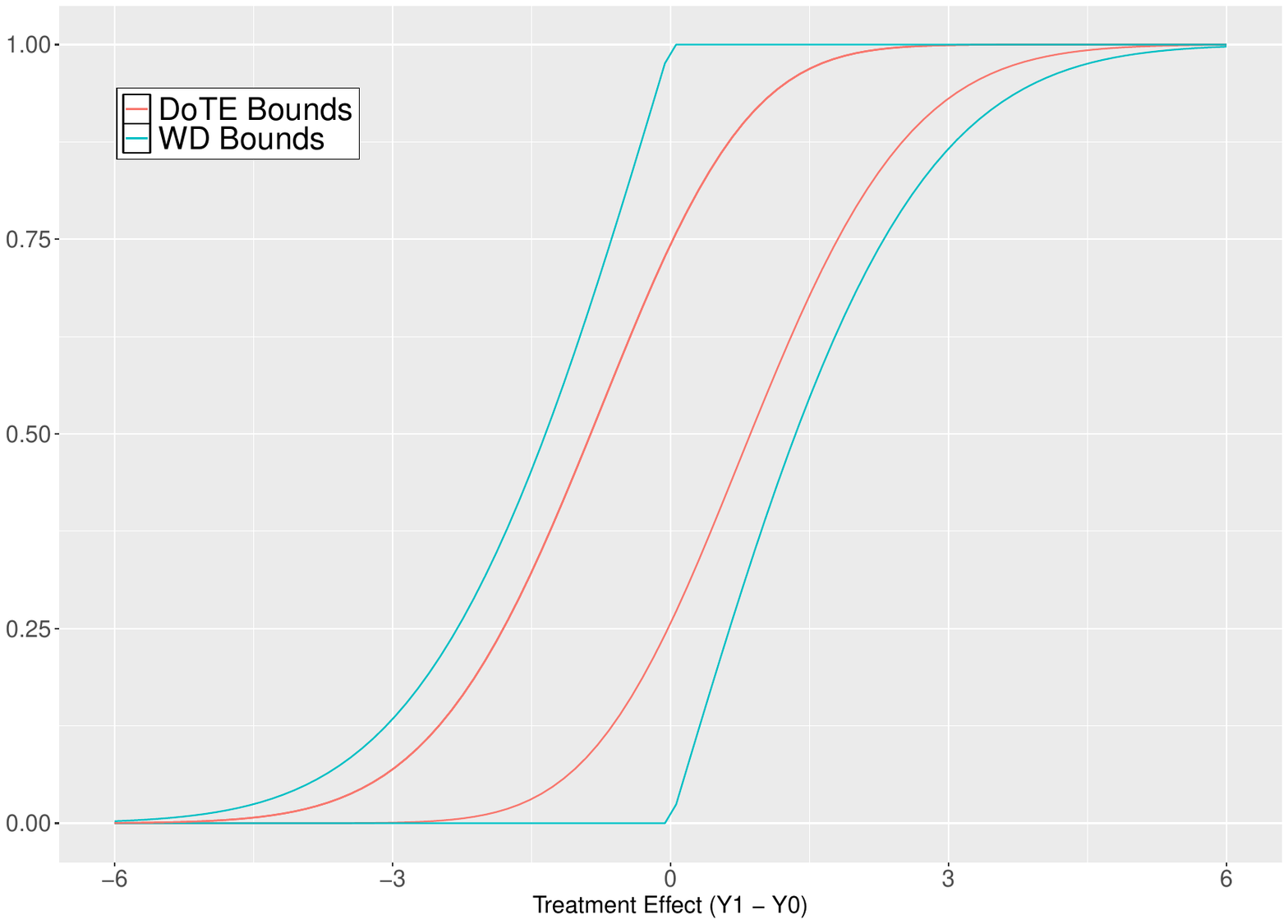}
\caption{$ \rho_{0} $ = 0.9}
\label{fig1:d}
\end{subfigure}

\textit{Notes: The figure plots the bounds on the DoTE, calculated using Theorem \ref{Theorem_DoTE} by considering the dependence between the potential outcomes (red) and using the Williamson and Downs (WD) bounds without assuming any restrictions on the dependence (blue). 
}
\end{figure}

These resulting bounds are presented in Figure \ref{fig nume bounds}. 
We can see that our proposed DoTE bounds plotted in Figure \ref{fig nume bounds} across different values of $ \rho_{0} $ are always inside the WD bounds, and they become tighter for larger values of $ \rho_{0} $. 
This result implies that for the group $ \boldsymbol{D}_{t} =(1,1,1) $, the bounds of the DTE parameters become tighter when there is a stronger dependence between $ Y_{0t} $ and $ Y_{1t-1} $. For example, under Assumption \ref{Chap2_ass3}, this will be true when the dependence between $ Y_{0t-1} $ and $ Y_{1t-2} $ for the group $ \boldsymbol{D}_{t} =(1,0,1) $, is stronger. 

\subsection{Monte Carlo Simulations}\label{Chap2_Sec_Simulations Main Text}

In this section, we conduct a Monte Carlo simulation study to investigate the finite sample performance of the proposed estimators under the two alternative GCE assumptions. We simulate the bounds of our target DTE parameters, the DoTE and QoTE, using a DGP that satisfies both key GCE assumptions (I) and (II) imposed under Models 1 and 2 and other model assumptions necessary to identify the DTEs. The primary objectives of this simulation study are to investigate the finite sample performance of our methods and to demonstrate the tightness of the DTE bounds when accounting for the dependence structure of the potential outcomes.

We consider a simple DGP design, where the potential outcomes, $$\boldsymbol{Y_{it}} = (Y_{1t}, Y_{0t}, Y_{1t-1}, Y_{0t-1}, Y_{1t-2},Y_{0t-2} )^{T} \equiv \boldsymbol{Y}_{t}|\boldsymbol{D}_{t},$$ which denotes both observed and unobserved treated and untreated potential outcomes of an individual in group $\boldsymbol{D}_{t} =(1,1,1)$, are generated by a joint normal distribution. 

The simulation DGP used here defines $\boldsymbol{Y_{it}}$ as the \textit{Kronecker product} of a time effect, denoted by the random vector $\boldsymbol{\theta_{t}}=(\theta_{1t}, \theta_{0t}, \theta_{1t-1}, \theta_{0t-1}, \theta_{1t-2}, \theta_{0t-2} )^{T}$, and a random vector  $\boldsymbol{\eta_{g}}=(\eta_{1}, \eta_{2}, \eta_{3}, \eta_{4}, \eta_{5}, \eta_{6}, \eta_{7}, \eta_{8})^{T}$ for $g=\{1,2,\dots, 8\}$, which is a group effect defined by the treatment patterns  $\boldsymbol{D}_{t}$.  Assuming both random vectors $\boldsymbol{\theta_{t}}$ and $\boldsymbol{\eta_{g}}$ to be independent, we assume, 
\begin{align} \label{simulation DGP}
\boldsymbol{Y_{it}} \sim N\bigg(\boldsymbol{\mu}=\left(\boldsymbol{\mu_{\eta_{g}}} \otimes \boldsymbol{\mu_{\theta_{t}}}\right), \, \boldsymbol{\Sigma}=\left(\boldsymbol{\Sigma_{\eta_{g}}} \otimes \boldsymbol{\Sigma_{\theta_{t}}}\right)\bigg).
\end{align}
The mean vector $\boldsymbol{\mu}$ is obtained as the Kronecker product of the mean vectors $ \boldsymbol{\mu_{\eta_{g}}}  $ and $ \boldsymbol{\mu_{\theta_{t}}} $, of the random variables $  \boldsymbol{\eta_{g}} $ and $ \boldsymbol{\theta_{t}} $, respectively, where,
\begin{align}\label{simulation d111 eta}
\boldsymbol{\mu_{\eta_{g}}} = (\mu_{\eta_{1}}, \mu_{\eta_{2}}, \cdots, \mu_{\eta_{8}})^{T} & \text{ and }\quad 
\boldsymbol{\mu_{\theta_{t}}} =(\mu_{\theta_{1t}},
\mu_{\theta_{0t}},
\mu_{\theta_{1t-1}},
\mu_{\theta_{0t-1}},
\mu_{\theta_{1t-2}},
\mu_{\theta_{0t-2}})^{T},
\end{align}
and the correlation matrix $\boldsymbol{\Sigma}$ is obtained as the Kronecker product of the correlation matrices, $ \boldsymbol{\Sigma_{\eta_{g}}}  $ and $ \boldsymbol{\Sigma_{\theta_{t}}} $ of the random variables $  \boldsymbol{\eta_{g}} $ and $ \boldsymbol{\theta_{t}} $, respectively, where,
\begin{align}\label{simulation d111 varetag}
\boldsymbol{\Sigma_{\eta_{g}}} = \begin{pmatrix}
1 & \rho_{\eta_{1}\eta_{2}} &  \cdots & \rho_{\eta_{1}\eta_{8}} \\
\rho_{\eta_{2}\eta_{1}} & 1  & \cdots & \rho_{\eta_{2}\eta_{8}}\\
\vdots   & \vdots & \ddots & \vdots \\
\rho_{\eta_{8}\eta_{1}} & \rho_{\eta_{8}\eta_{2}} & \cdots  & 1\end{pmatrix}\end{align}
and,
\begin{align}\label{simulation d111 yt}
\boldsymbol{\Sigma_{\theta_{t}}}=\begin{pmatrix}
1 & \rho_{\theta_{1t}.\theta_{0t}}& \rho_{\theta_{1t}.\theta_{1t-1}} & \rho_{\theta_{1t}.\theta_{0t-1}} & \rho_{\theta_{1t}.\theta_{1t-2}} & \rho_{\theta_{1t}.\theta_{0t-2}}\\
\rho_{\theta_{0t}.\theta_{1t}} & 1 & \rho_{\theta_{0t}.\theta_{1t-1}} & \rho_{\theta_{0t}.\theta_{0t-1}} & \rho_{\theta_{0t}.\theta_{1t-2}} & \rho_{\theta_{0t}.\theta_{0t-2}}\\
\rho_{\theta_{1t-1}.\theta_{1t}} &  \rho_{\theta_{1t-1}.\theta_{0t}}  & 1 &  \rho_{\theta_{1t-1}.\theta_{0t-1}} & \rho_{\theta_{1t-1}.\theta_{1t-2}} & \rho_{\theta_{1t-1}.\theta_{0t-2}}\\
\rho_{\theta_{0t-1}.\theta_{1t}}  & \rho_{\theta_{0t-1}.\theta_{0t}} & \rho_{\theta_{0t-1}.\theta_{1t-1}} & 1 & \rho_{\theta_{0t-1}.\theta_{1t-2}} & \rho_{\theta_{0t-1}.\theta_{0t-2}}  \\
\rho_{\theta_{1t-2}.\theta_{1t}} & \rho_{\theta_{1t-2}.\theta_{0t}} &  \rho_{\theta_{1t-2}.\theta_{1t-1}} & \rho_{\theta_{1t-2}.\theta_{0t-1}}& 1 &  \rho_{\theta_{1t-2}.\theta_{0t-2}}   \\
\rho_{\theta_{0t-2}.\theta_{1t}}& \rho_{\theta_{0t-2}.\theta_{0t}} & \rho_{\theta_{0t-2}.\theta_{1t-1}} &  \rho_{\theta_{0t-2}.\theta_{0t-1}} & \rho_{\theta_{0t-2}.\theta_{1t-2}} & 1\end{pmatrix}.\end{align}

This DGP serves as the foundational framework for generating data used in our Monte Carlo simulation study. It can be applied to both models, Model 1 and Model 2, which rely on the GCE assumptions \ref{Chap2_ass3} and \ref{Chap2_ass3Model2}, respectively. We illustrate the Monte Carlo simulations only for the case discussed in the main text, where $ \boldsymbol{D}_{t} =(1,1,1) $. 

We consider different DGP designs using various correlation matrices by allowing the main dependence measure $\rho_{\textasteriskcentered}$, which is the correlation between the potential outcomes $Y_{1t-1}$ and $Y_{0t}$ given the target group $\boldsymbol{D}_{t}$, to vary as $\rho_{\textasteriskcentered} \in \{0, 0.6, 0.9\}$—in order to investigate the tightness of the bounds with the strength of the dependence of the potential outcomes. More details of this simulation design are given in the Supplementary Appendix under Section C. %\ref{Section simulation}.

To test the improvement of the DoTE and the QoTE bounds with the sample size, for each design, we consider three different sample sizes $N$: 100, 500, and 1000 ($N$ is the number of observations in each treatment group).\footnote{To estimate DTEs for this target group $\boldsymbol{D}_{t}=(1,1,1)$, we require information from the two treatment groups $\boldsymbol{D}_{t}=(1,1,0)$, and $\boldsymbol{D}_{t}=(1,0,1)$ under both GCE assumptions.}  In order to keep the computation time reasonable, we set the number of Monte Carlo replications to 1000, and we draw 500 bootstrap replications in each of these Monte Carlo replications. We estimate the target distributional treatment effects parameters, the DoTE and QoTE, and their 95\% (point-wise) confidence regions using the numerical bootstrapping method discussed in Section \ref{Chap2_Estimation and Inference}. 
We present simulation results only for DoTE, as the QoTE bounds are simply the inverse of these DoTE bounds. 

\begin{figure}[H]
\begin{center}
\caption{\centering Simulation Results: DoTE Bounds for Target Group $  \boldsymbol{D}_{t}=(1,1,1)$ under \textbf{Model 1}}\label{fig main:D sim1 M1}
\begin{subfigure}[b]{0.33\textwidth}	
\centering
\caption[]{\centering{N=100 and $ \rho_{\ast}=0 $}}
\includegraphics[height=5.5cm, width=5cm]{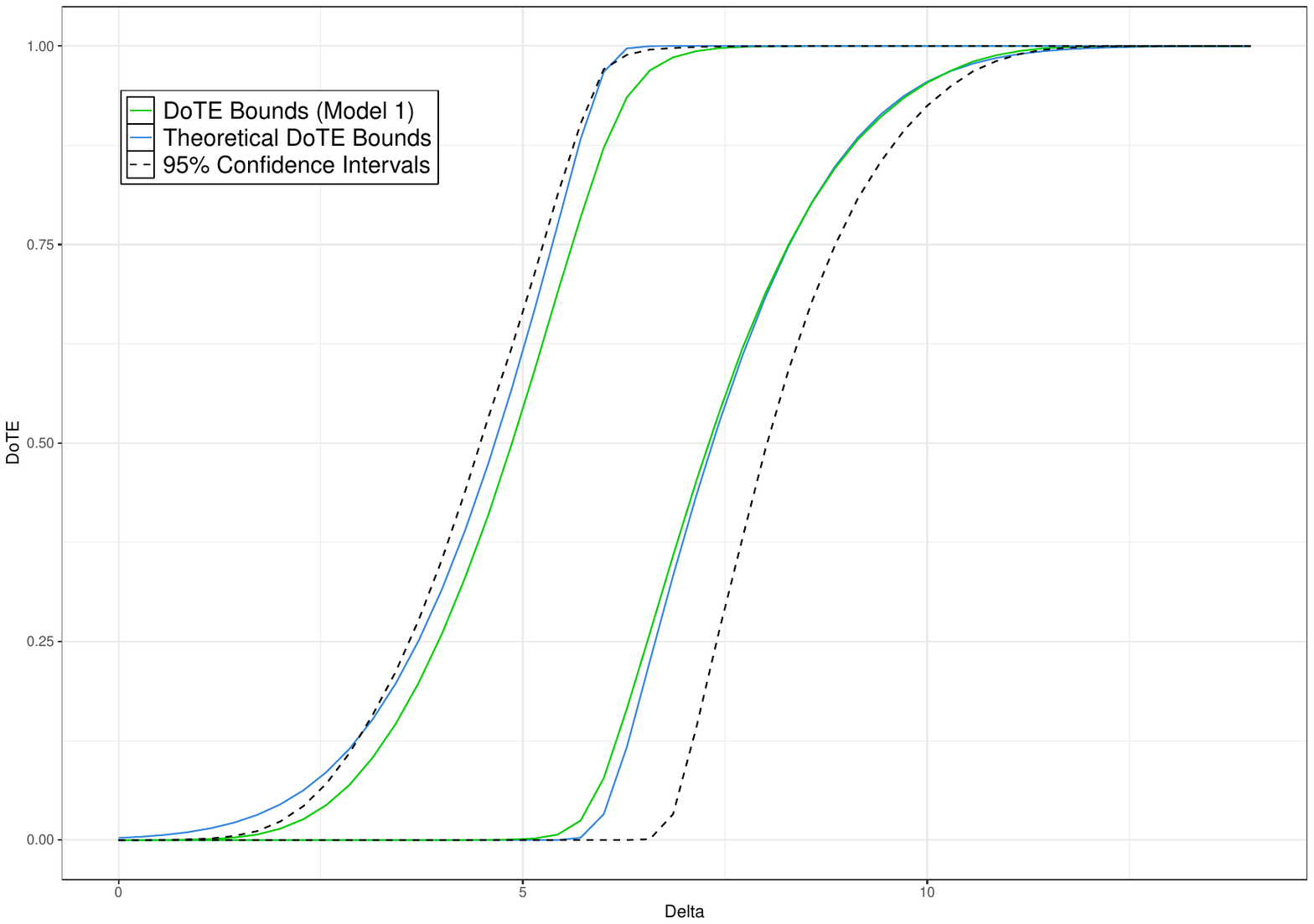}
\label{fig1:Dsim1 M1}
\end{subfigure}%
\begin{subfigure}[b]{0.33\textwidth}
\centering
\caption[]{\centering{N=100 and $ \rho_{\ast}=0.6 $}}
\includegraphics[height=5.5cm, width=5cm]{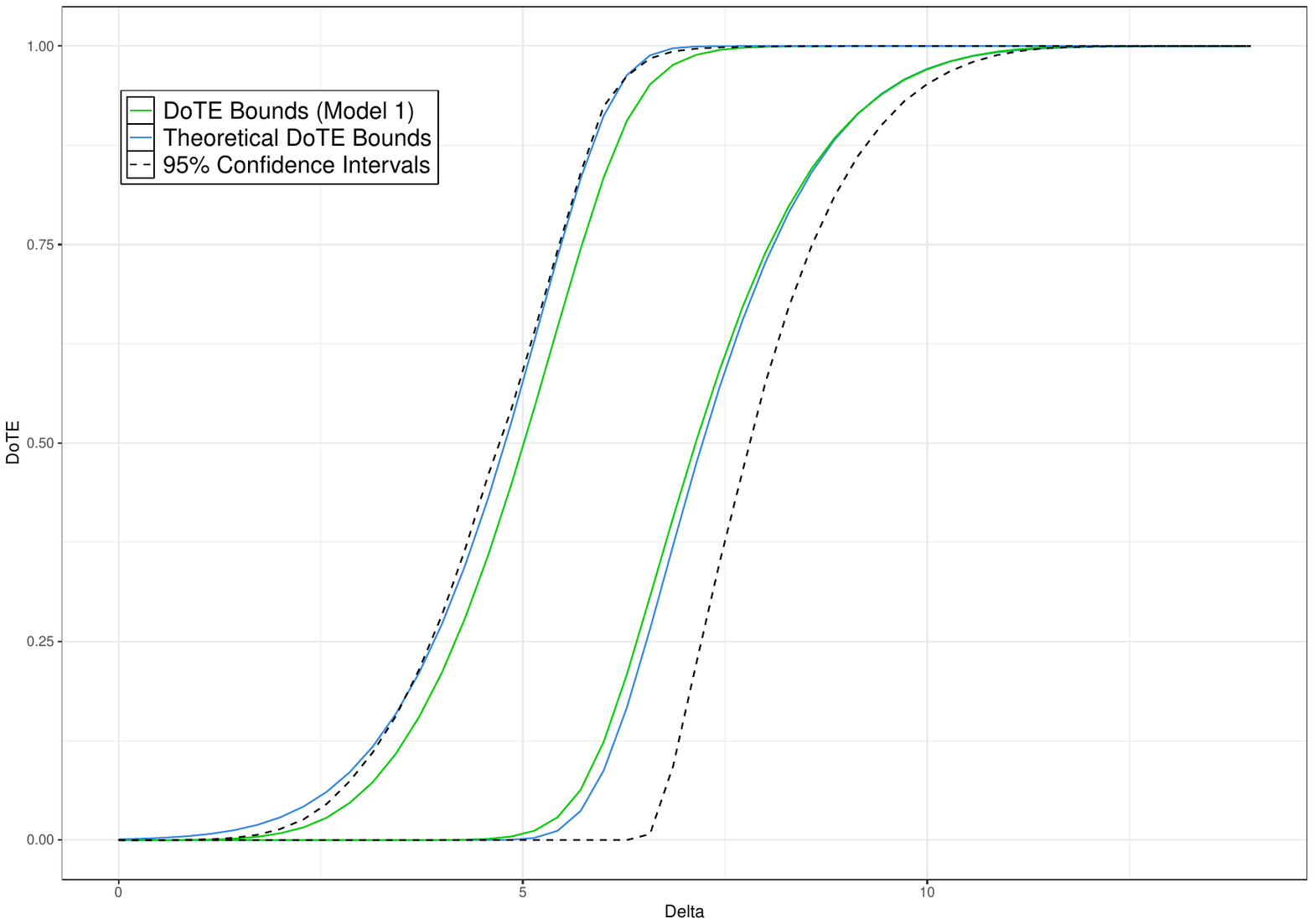}

\label{fig2:Dsim1 M1}
\end{subfigure}
\begin{subfigure}[b]{0.33\textwidth}
\centering
\caption[]{\centering{N=100 and $ \rho_{\ast}=0.9 $}}
\includegraphics[height=5.5cm, width=5cm]{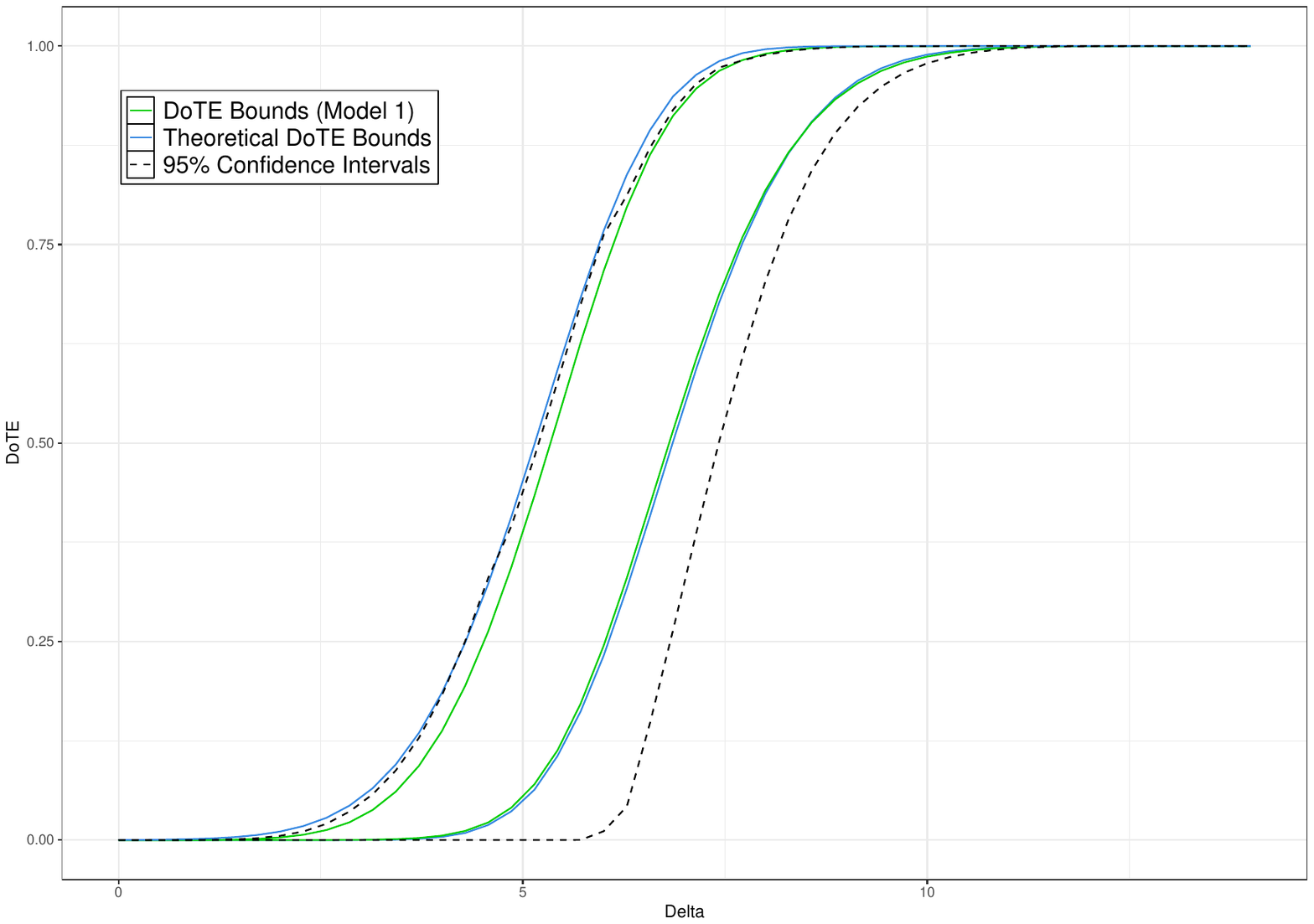}

\label{fig3:Dsim1 M1}
\end{subfigure}
\begin{subfigure}[b]{0.33\textwidth}	
\centering
\caption[]{\centering{N=500 and $ \rho_{\ast}=0 $}}
\includegraphics[height=5.5cm, width=5cm]{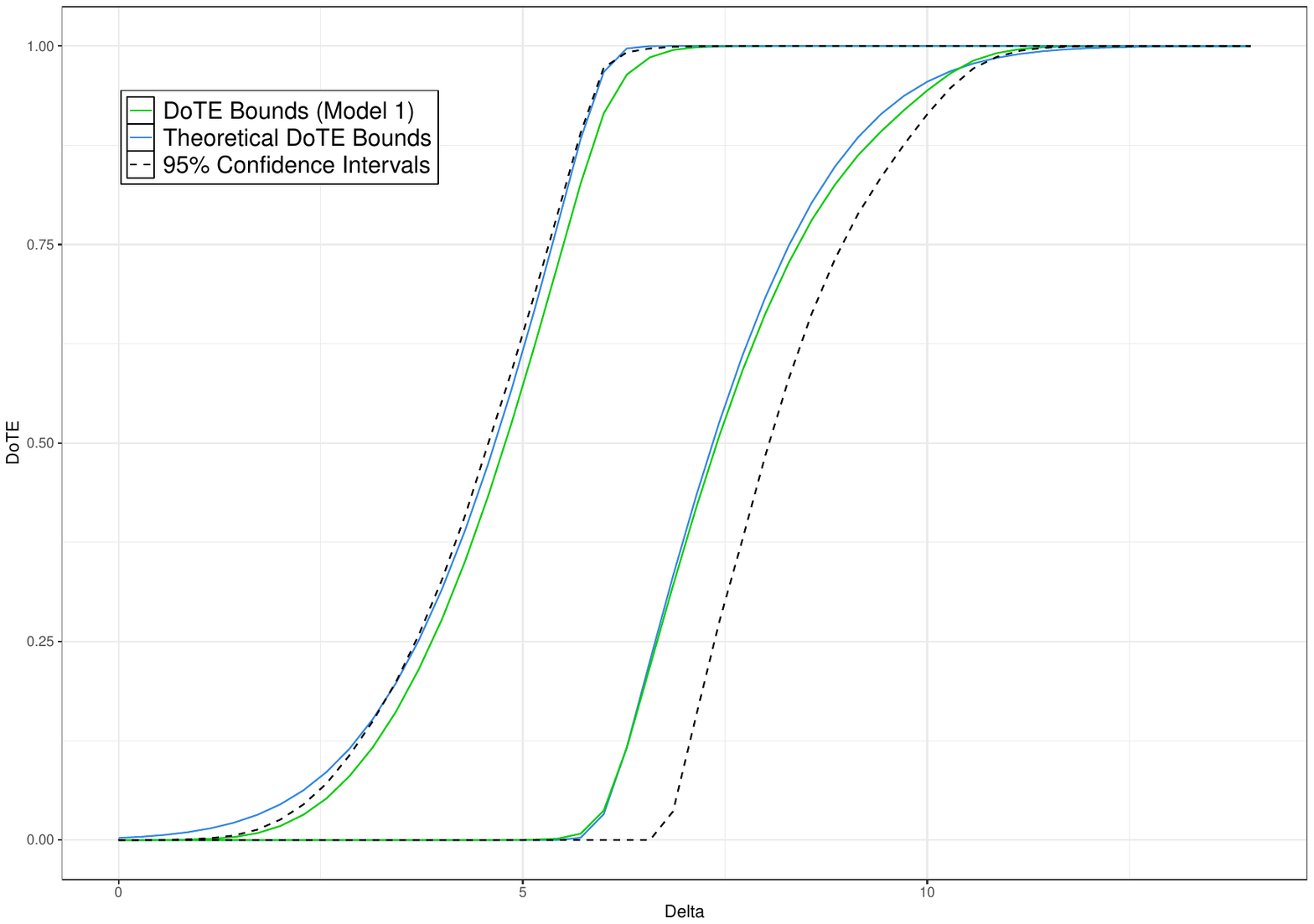}

\label{fig4:Dsim1 M1}
\end{subfigure}%
\begin{subfigure}[b]{0.33\textwidth}
\centering
\caption[]{\centering{N=500 and $ \rho_{\ast}=0.6 $}}
\includegraphics[height=5.5cm, width=5cm]{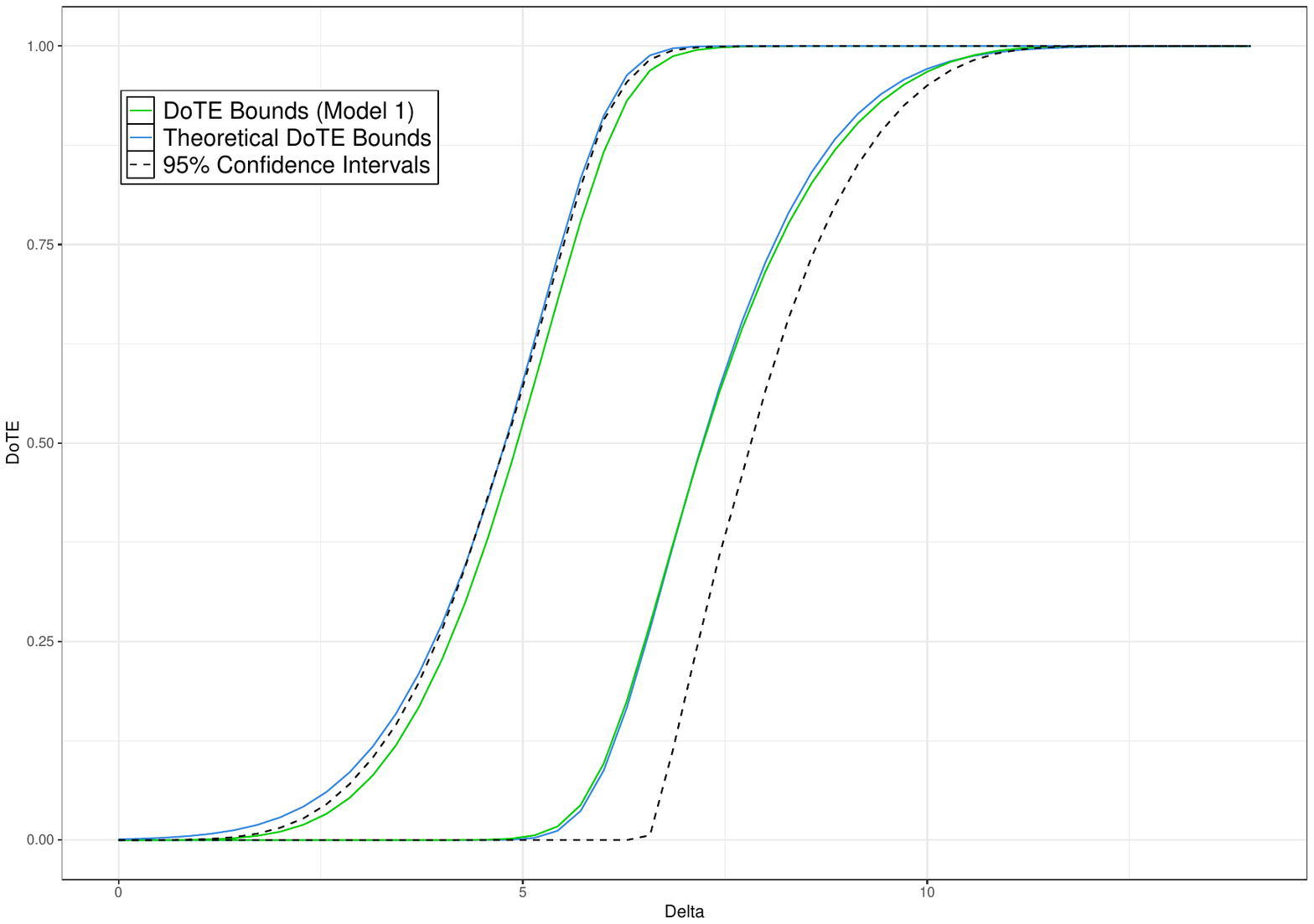}

\label{fig5:Dsim1 M1}
\end{subfigure}
\begin{subfigure}[b]{0.33\textwidth}
\centering
\caption[]{\centering{N=500 and $ \rho_{\ast}=0.9 $}}
\includegraphics[height=5.5cm, width=5cm]{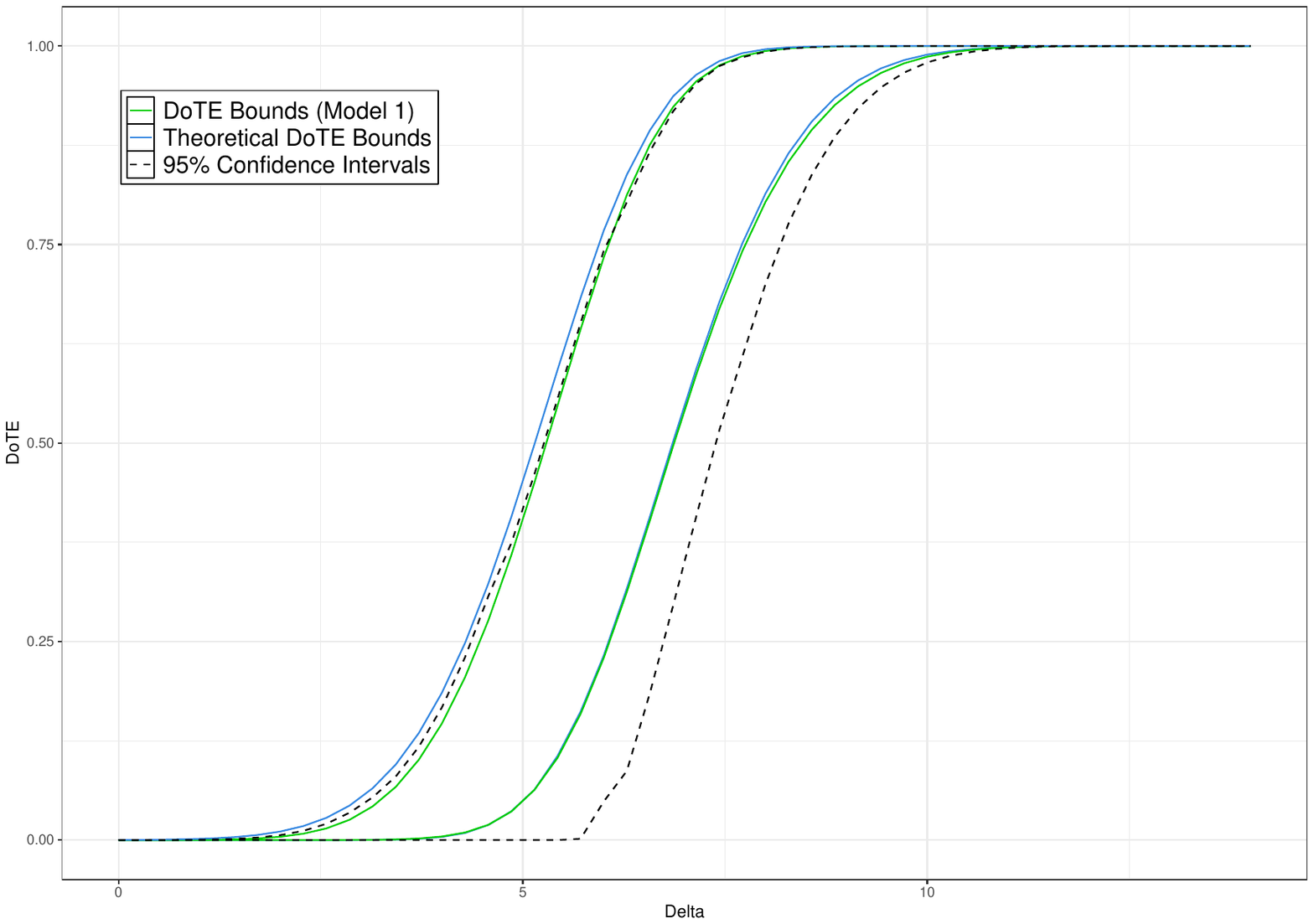}

\label{fig6:Dsim1 M1}
\end{subfigure}
\begin{subfigure}[b]{0.33\textwidth}	
\centering
\caption[]{\centering{N=1000 and $ \rho_{\ast}=0 $}}
\includegraphics[height=5.5cm, width=5cm]{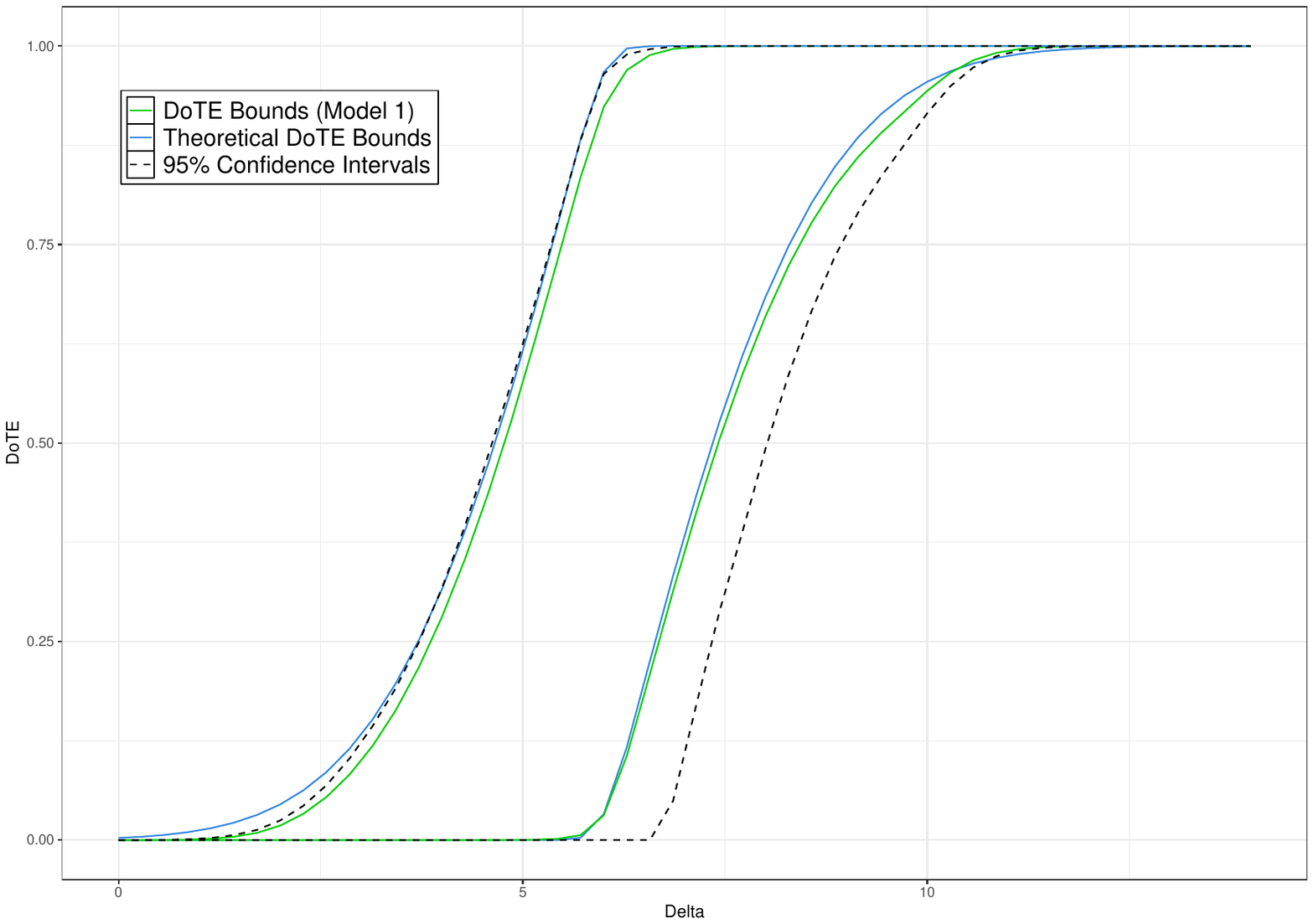}

\label{fig7:Dsim1 M1}
\end{subfigure}%
\begin{subfigure}[b]{0.33\textwidth}
\centering
\caption[]{\centering{N=1000 and $ \rho_{\ast}=0.6 $}}
\includegraphics[height=5.5cm, width=5cm]{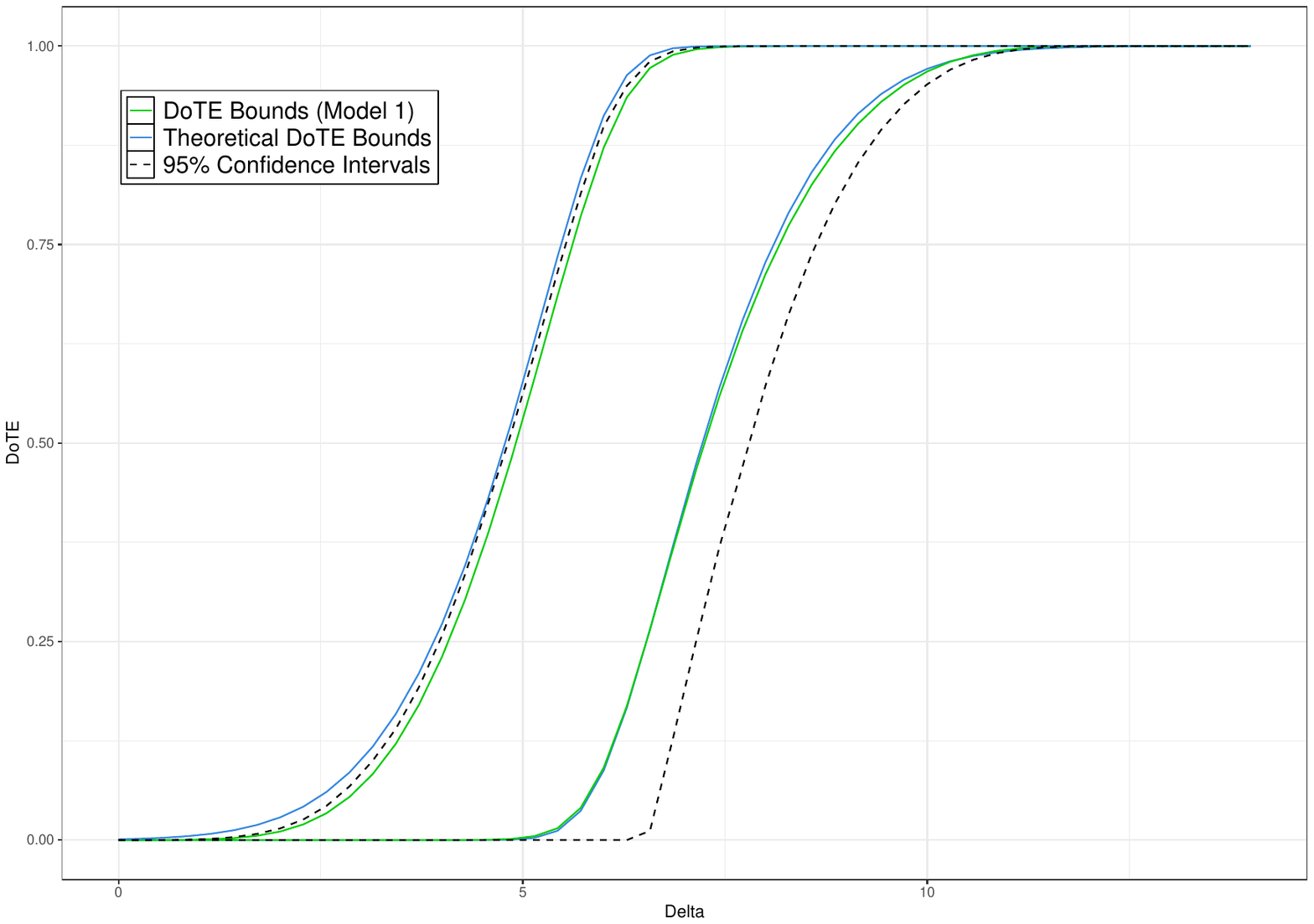}

\label{fig8:Dsim1 M1}
\end{subfigure}
\begin{subfigure}[b]{0.33\textwidth}
\centering
\caption[]{\centering{N=1000 and $ \rho_{\ast}=0.9 $}}
\includegraphics[height=5.5cm, width=5cm]{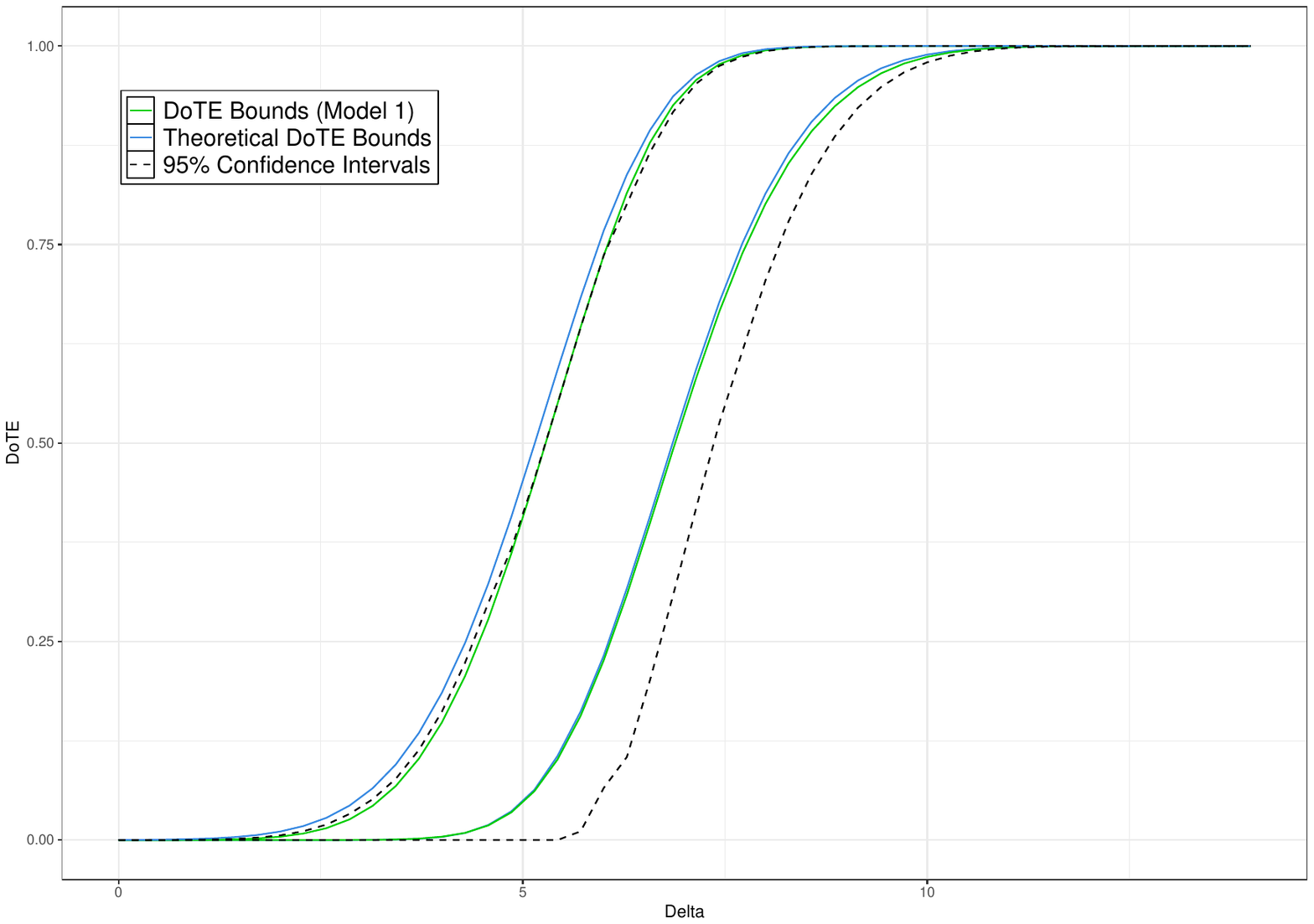}

\label{fig9:Dsim1 M1}
\end{subfigure}
\end{center} 
\textit{Notes: This figure presents the results of the Monte Carlo experiments derived under Model 1. The sub-figures provide the plots of the theoretical bounds of the DoTE (blue), the DoTE bounds derived under Model 1 using our first GCE Assumption \ref{Chap2_ass3} (green), and the 95\% bootstrapped confidence intervals constructed using the numerical bootstrap method (dotted curves), for the target group $  \boldsymbol{D}_{t}=(1,1,1)$. Each sub-figure given under each row plots the estimates of the DoTE lower and upper bounds across different values of the dependence parameter, $\rho_{\textasteriskcentered} \in \{0, 0.6, 0.9\}$ which is the correlation between $ Y_{1t-1} $ and $ Y_{0t} $ for the group $ \boldsymbol{D}_{t}=(1,1,1)$, for the sample sizes, $ N=100 $, $ N=500 $, and $ N=1000 $. These results are obtained after performing 1000 Monte Carlo simulations.}
\end{figure}

The resulting bounds on the DoTE under Model 1 with Assumption \ref{Chap2_ass3} are presented in Figure \ref{fig main:D sim1 M1}. In each sub-figure, we have plotted the estimated DoTE lower and upper bounds under Model 1, the theoretical DoTE lower and upper bounds, and their 95\% confidence intervals. Furthermore, these sub-figures given under each row plots the estimates of the DoTE lower and upper bounds across different values of the dependence parameter $\rho_{\textasteriskcentered} \in \{0, 0.6, 0.9\}$, for the sample sizes $ N=100 $, $ N=500 $, and $ N=1000 $.

The key insights of this Figure \ref{fig main:D sim1 M1} can be summarised as follows. First, by examining the columns that present results across different sample sizes with the same correlation value, they show an improvement in the overall performance of the estimates of both the upper and lower bounds of the DoTE as the sample size increases. Second, when considering the rows that include results across different correlation values with fixed sample sizes, they show that with stronger dependence between the potential outcomes $ Y_{1t-1} $ and $ Y_{0t} $ of group $ \boldsymbol{D}_{t}=(1,1,1)$ (or when the correlation value $\rho_{\textasteriskcentered}$ is increasing), the bounds on the DoTE tend to be tighter.

Other than these, we can observe a finite sample bias in the estimates of the DoTE lower and upper bounds. This bias arises due to two main reasons: (i) sampling variation and (ii) the application of supremum and infimum operators when deriving the DoTE lower and upper bounds (see Lemma \ref{Chap2_lemma_cond_TE}). In the literature, \cite{chernozhukov2013intersection} refer to such bounds, which involve an infimum or a supremum operator in their bounding functions as intersection bounds. They state that these types of bounds may create complexities in the estimation and inference. Further, they have argued that (and also in the related discussions of \cite{callaway2021bounds}, \cite{flores2018average}, and \cite{manski1998monotone}), the sample estimators of these intersection bounds may experience a significant bias due to the convex nature of the maximum functions and concave nature of the minimum functions associated with it. Hence, this observed biasedness of our DoTE estimators is natural and expected in finite samples, as our bounds also belong to this class of intersection bounds. However, overall, the biasedness of these estimators of the DoTE lower and upper bounds diminishes with reasonable sample sizes. In future, we aim to address this issue using an appropriate de-biasing method, such as the CLR method of \cite{chernozhukov2013intersection}.

Next, we discuss the Monte Carlo simulation results derived under Model 2 using our second alternative GCE assumption, \ref{Chap2_ass3Model2}. Since our DGP satisfies both Model 1 and Model 2 GCE assumptions, Model 2 also yields similar DoTE estimates to those obtained under Model 1. Thus, the results and interpretations for the DoTE lower and upper bounds derived under Model 2 are analogous to those under Model 1.\footnote{Further, using these DoTE estimates one could obtain the bounds on the QoTE by simply inverting the DoTE lower and upper bounds, for both Models 1 and 2. Similar results hold for the QoTE lower and upper bounds, as observed for the DoTE bounds.} Although the GCE assumptions \ref{Chap2_ass3} and \ref{Chap2_ass3Model2} are not subsets of one another, in scenarios where both assumptions are satisfied, the bounds derived under each model are expected to provide similar results, as in this case. We use this evidence to justify our key alternative GCE assumptions in the application; when comparing the bounds obtained under Models 1 and 2, and as well as when comparing our bounds with \cite{callaway2021bounds} (Section \ref{Chap2_Sec Appl2 BC&Model2}).

In summary, our simulation results show that both Models 1 and 2 work well in reasonable sample sizes, and lead to substantially tighter bounds with stronger or larger values of the main dependence parameter, $\rho_{\textasteriskcentered}$. %Details of the simulation results are given in the Appendix.

\section{An Application: The Impact of Exercise on BMI} \label{Chap2_Section Application}

In this section, we apply our proposed methods to examine the heterogeneous treatment effects of physical activity (exercising) on body weight, using a panel dataset from the Household, Income and Labour Dynamics in Australia (HILDA) Survey (\cite{HILDA_2015}). 
The HILDA survey follows more than 17,000 Australians annually, starting from the year 2001 and over the course of their lifetime. It offers rich information on a variety of aspects, such as family background, socioeconomic conditions, and health-related lifestyle behaviours. 
We illustrate this application for both Model 1 and Model 2, together with the parametric and non-parametric tests for the key GCE assumptions. Additionally, we presented results of DTEs bounds, conditioned on the previous year’s BMI values and controlled for covariates, to provide further insight into treatment effect heterogeneity. 
Furthermore, by computing the DTE bounds for the group considered by \cite{callaway2021bounds} (i.e., $\boldsymbol{D}_{t} =(0,0,1)$), we compare the results obtained through our method with those derived from his method utilizing the copula stability assumption (CSA), for this specific group.

The outcome variable in this study is the body weight measured by the \textit{body mass index} (BMI). The BMI is defined as a person's body weight in kilograms divided by the square of their height in meters ($ kg/m^{2} $), 
which is the most widely used and internationally recognized standard to classify the body weights of adults. 
Our (binary) treatment variable is, exercising ($D_s =1$) or not exercising ($D_s=0$). 
This is defined from the responses to a questionnaire about \textit{physical activity} in HILDA\textminus ``In general, how often do you participate in moderate or intensive physical activity for at least 30 minutes?" The responses can be any of the six categories: (1) Not at all; (2) Less than once a
week; (3) One to two times a week; (4) Three times a week; (5) More than 3 times a week; and (6) Every day. We define those who have responded (1), (2), or (3), i.e., exercising less than three times a week, as $D_s = 0$ (untreated), and those who have responded (4), (5), or (6), i.e., exercising three or more times as $D_s = 1$ (treated).

The \cite{HILDA_2015} consists of 217,917 observations from the years 2001 to 2015 (unbalanced panel). 
Our analysis uses the yearly data from 2006 to 2015 that provides complete information about BMI. 
Our target parameters are the DoTE and QoTE in the last period (i.e., 2015), for the treatment group $ \boldsymbol{D}_{t} =(1,1,1) $ where individuals continued exercising in all three periods, 2013, 2014, and 2015. 
After extracting information from the three periods that we consider (i.e., 2013, 2014, and 2015) from the panel, we obtain a balanced panel by excluding individuals who do not provide adequate information.
Furthermore, we restrict our analysis to individuals whose BMI is over $ 17.5 \, kg/m^2$ and under $ 42 \, kg/m^2$, 
leading to a sample of 10,755 unique individuals.

\subsection{Effects of Exercising on BMI under Model 1} \label{Chap2_Subsec app2model1}

Under Model 1, to derive bounds on the DoTE and QoTE for the target group $\boldsymbol{D}_{t} = (1,1,1)$, the panel dataset should include information on this group, the copula recovery group $\boldsymbol{D}_{t} = (1,0,1)$, and the group $\boldsymbol{D}_{t} = (1,1,0)$ that is required by the Change-in-changes method to recover the counterfactual marginal  $ {F}_{Y_{0 t} \mid \boldsymbol{D}_{t} =(1,1,1)} $. In our sample, out of 10,755 individuals, there are 3,353 individuals in group $\boldsymbol{D}_{t} = (1,1,1)$, 636 individuals in the copula recovery group $\boldsymbol{D}_{t} = (1,0,1)$, and 785 individuals in group $\boldsymbol{D}_{t} = (1,1,0)$, in the years 2013, 2014, and 2015. We use these observations to obtain the results in our study under Model 1.

\begin{figure}[H]
\begin{center}
\caption{\centering Effects of Exercising on BMI: QoTE Bounds under \textbf{Model 1} for $ \boldsymbol{D}_{t} = (1,1,1)  $}\label{Chap2_fig main: App2 M1}
\includegraphics[width=0.65\linewidth]{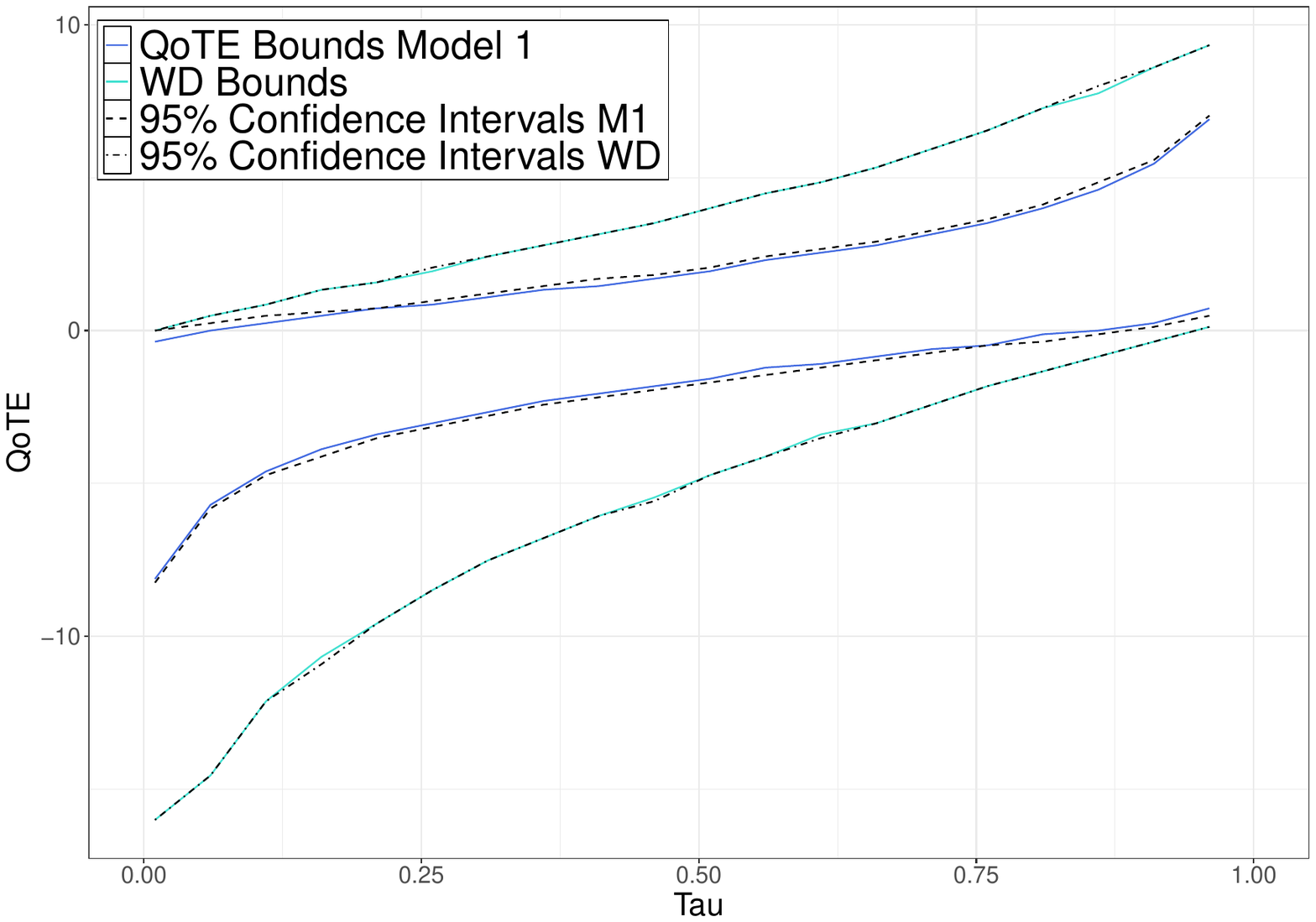}	
\end{center}
\textit{Notes: The figure provides bounds on the QoTE for group $ \boldsymbol{D}_{t} = (1,1,1) $ under the WD method \citep{williamson1990probabilistic}  and under Model 1 with the first alternative GCE Assumption \ref{Chap2_ass3}. The dotted curves denote the 95\% confidence intervals for the lower and upper bounds obtained using the numerical bootstrap method as discussed in Section \ref{Chap2_Estimation and Inference}.}
\end{figure}

The average treatment effects alone may not be sufficient to get a clear picture of the underlying distribution of the  effect of physical activity on BMI.  Hence, by applying the method developed in this study, we examine the individual-level heterogeneous treatment effects of exercising on BMI by estimating the \textit{distributional} treatment effects for group $ \boldsymbol{D}_{t} = (1,1,1)  $. In Figure \ref{Chap2_fig main: App2 M1}, we present the estimates of the QoTE parameter for group $ \boldsymbol{D}_{t} = (1,1,1)  $, obtained by taking the inverse of the DoTE bounds. 
This Figure \ref{Chap2_fig main: App2 M1} plots the QoTE bounds obtained using the \cite{williamson1990probabilistic} (WD) method with no assumptions on the dependence  between the potential outcome distributions (light blue), the QoTE  bounds proposed in this study obtained under Model 1 (blue), and  their 95\% confidence intervals for the estimated lower and upper bounds using the numerical bootstrap method discussed in Section \ref{Chap2_Estimation and Inference}.  

This figure highlights several results. Firstly, both methods imply that the effect of performing exercises on BMI is not the same for all individuals, as the curves do not appear to be a constant across the {interior} quantiles. For example, the upper bound estimate on the $10^{th}$ percentile obtained using the WD bounds indicates a 0.73 $ kg/m^{2} $ increase in BMI, while the lower bound on the $90^{th}$ percentile shows a 0.48 $ kg/m^{2} $ reduction in BMI when they exercise. 
Even though, these WD bounds indicate that treatment effects are heterogeneous, due to the width of these bounds they tend to be relatively less informative.

{Nevertheless, the QoTE bounds obtained using our method are much tighter than the WD bounds across all quantiles. For instance, at the $10^{th}$ percentile, its estimated effects of exercising on BMI is between a reduction of 4.73 $ kg/m^{2} $ and a gain of 0.24 $kg/m^{2} $ in BMIs, suggesting that exercising is beneficial for certain individuals than the average impacts might suggest.} Further, these results indicate that around 6.74\% of individuals would definitely reduce their BMIs when they exercise compared to if they had not exercised. However, when considering the higher percentiles we can see that, around 12.38\% of individuals would definitely increase their BMIs when they exercise than if they had not exercised. {These estimates imply that exercising can be more beneficial for some individuals in reducing their BMIs while it might not be for others \textminus findings that are not evident from average treatment effects.}

\subsection{Effects of Exercising on BMI under Model 2} \label{Chap2_Subsec app2model2}

In this section, we obtain the bounds on the DTEs for the same group $ \boldsymbol{D}_{t} = (1,1,1) $ under Model 2. 
In this model, we need the data from group $ \boldsymbol{D}_{t} = (1,1,0) $ to identify both the counterfactual marginal $ {F}_{Y_{0 t} \mid \boldsymbol{D}_{t} =(1,1,1)} $ using the CiC method and the unknown copula using Assumption \ref{Chap2_ass3Model2} for the group $ \boldsymbol{D}_{t} = (1,1,1) $.

\begin{figure}[H]
\begin{center}\caption{\centering Effects of Exercising on BMI: QoTE Bounds under \textbf{Model 2} for $ \boldsymbol{D}_{t} = (1,1,1)  $}\label{Chap2_fig main: App2 M2}
\includegraphics[width=0.65\linewidth]{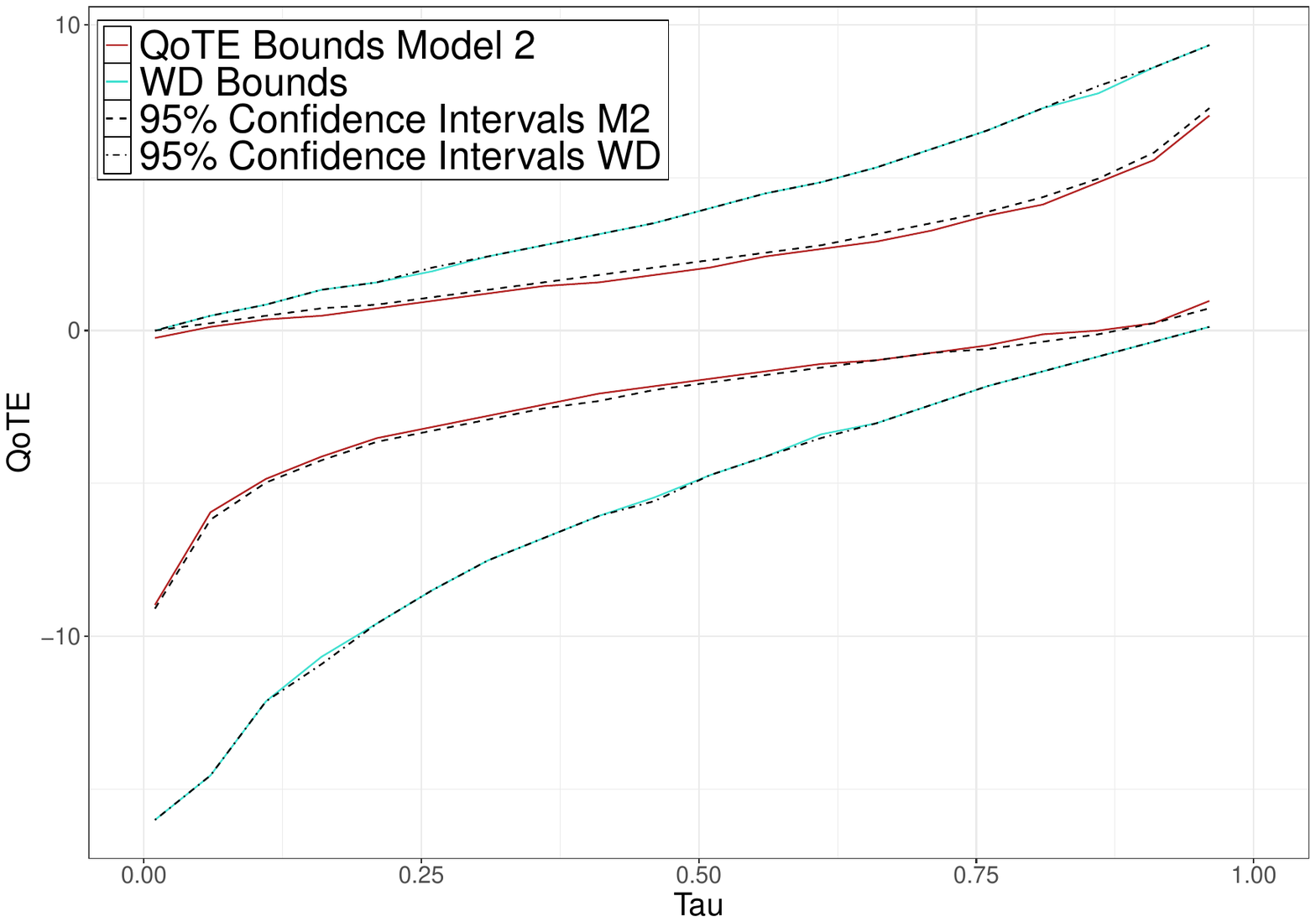}
\end{center}
\textit{Notes: The figure provides bounds on the QoTE for group $ \boldsymbol{D}_{t} = (1,1,1) $ under the WD method \citep{williamson1990probabilistic}  and under Model 2 with the second alternative GCE Assumption \ref{Chap2_ass3Model2}. The dotted curves denote the 95\% confidence intervals for the lower and upper bounds obtained using the numerical bootstrap method as discussed in Section \ref{Chap2_Estimation and Inference}.}
\end{figure}

The estimates of the QoTE bounds under Model 2, the WD bounds, and the 95\% confidence intervals for the lower and upper bounds using the numerical bootstrap method are presented in Figure \ref{Chap2_fig main: App2 M2}. The estimated QoTE bounds using WD method  are the same as those plotted in Figure \ref{Chap2_fig main: App2 M1} \textminus wider than our proposed bounds and less informative. %, and hence, interpretations of the results using these bounds may not be informative and meaningful.
The QoTE bounds derived under Model 2 are tighter than that obtained from WD bounds, and again indicate heterogeneous treatment effects of exercising on BMI, as in Model 1. 

\begin{figure}[H]
\begin{center}
\caption{\centering{Effects of Exercising on BMI: QoTE Bounds from \textbf{Model 1} and \textbf{Model 2}}}\label{Chap2_fig: App2 M1&M2}
\includegraphics[width=0.65\linewidth]{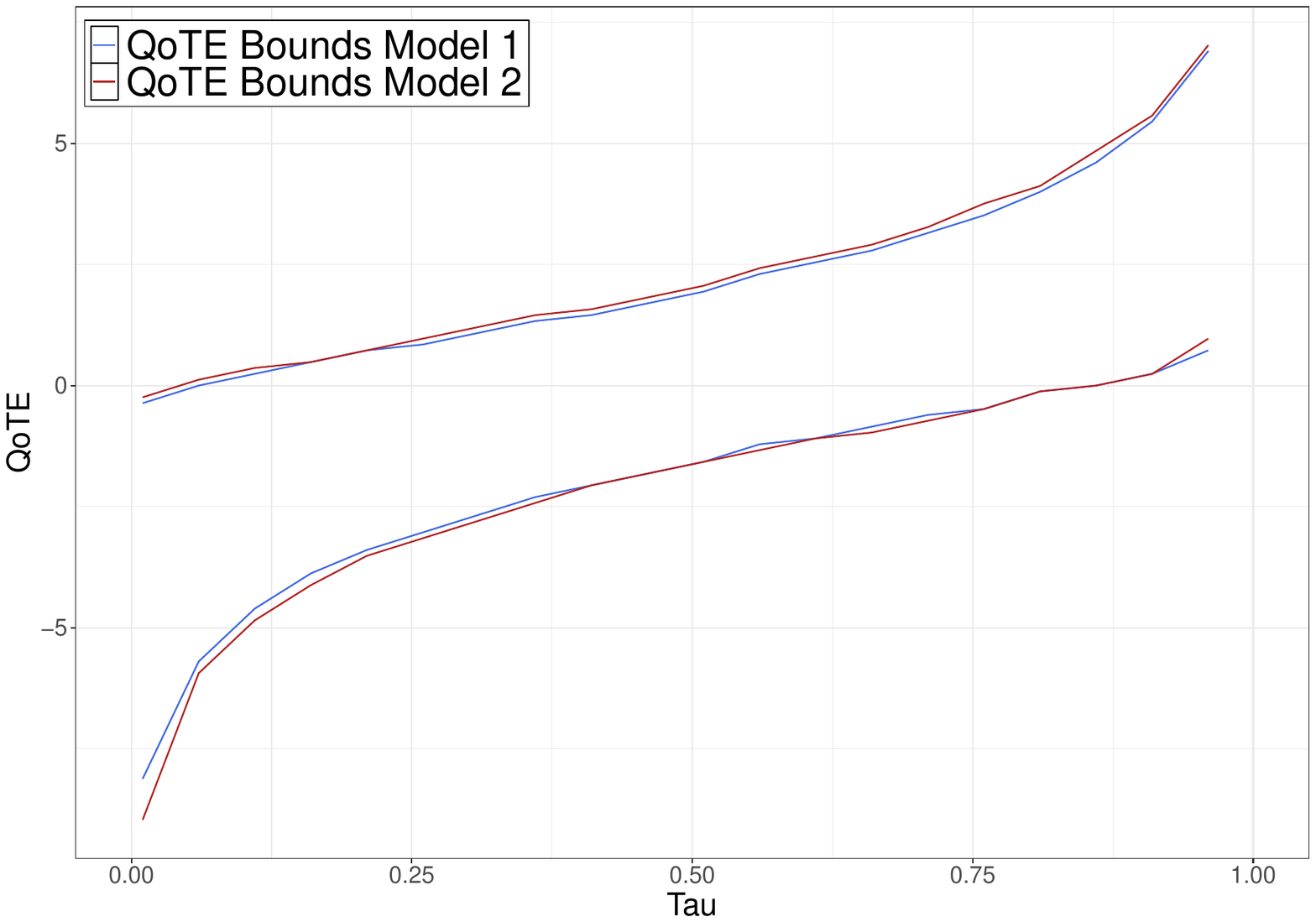}
\end{center}
\textit{Notes: This figure plots the QoTE bounds for $ \boldsymbol{D}_{t} =(1,1,1) $, derived under Model 1 using GCE assumption \ref{Chap2_ass3} (blue), and Model 2 using GCE assumption \ref{Chap2_ass3Model2} (red).}
\end{figure}

Figure \ref{Chap2_fig: App2 M1&M2} presents the QoTE bounds derived under both Model 1 and Model 2. For this particular application, we can see that the bounds obtained under both alternative GCE assumptions are similar, as they do not make much difference in the overall estimated results. Notably, the key finding here is that the bounds obtained under both alternative assumptions have improved the QoTE bounds more than in the case with WD bounds, providing meaningful results compared to the WD method.

\subsection{Testing GCE Assumptions}{\label{Chap2_Sec Appl2 Testing}}

In this section, we test whether our two key GCE assumptions hold in the exercising example. As explained in Section \ref{Chap2_Subsec Testability}, the test is carried out using data in the earlier periods (i.e., from 2006 to 2012), assuming that the GCE assumptions also hold in those periods.\footnote{As we have considered the last three year time window (i.e., 2013, 2014, and 2015) to evaluate the effects of exercising on BMIs, the previous years from 2006 to 2012 are used to conduct the test.}

Recall that the two GCE assumptions use information from different copula recovery groups when identifying the unknown copula for group $ \boldsymbol{D_{t}}=(1,1,1) $. Assumption \ref{Chap2_ass3} uses group $ \boldsymbol{D_{t}}=(1,0,1) $ and Assumption \ref{Chap2_ass3Model2} uses group $ \boldsymbol{D_{t}}=(1,1,0) $. To implement the test, we consider all the \textit{three-year time windows} from 2006 to 2012. The first column in Table \ref{Table pretest kendalM} provides all possible three-year time windows. The next two columns in Table \ref{Table pretest kendalM} present the  Kendall's Tau values (dependence measures) of these observed pair of dependences given in \eqref{test_firstGCE} and \eqref{test_secondGCE} for the target and copula recovery groups, in each three period time window. The last two columns present the p-values for the parametric and non-parametric tests introduced in Section \ref{Chap2_Subsec Testability}. In particular, the parametric test checks whether the Kendall's Tau of the target and copula recovery groups are the same, and the non-parametric two copula equality test adopted from \cite{remillard2009testing} tests whether the two underlying copulas are the same.
\footnote{The results for the non-parametric test are obtained using the ``TwoCop" R package developed by \cite{remillard2009testing} (see https://cran.r-project.org/src/contrib/Archive/TwoCop/).}

\begin{table}[h!]
\caption{\centering Kendall's Tau Values and p-values for the Parametric and Non-parametric Tests} \label{Table pretest kendalM} \subcaption{\textbf{Model 1}}\label{Chap2_Table pretest kendalM1}
\begin{center}
\begin{tabular}{c|cccc}
\hline\hline
\textbf{Year} & \begin{tabular}[c]{@{}c@{}}\textbf{Kendall's Tau}\\ $\boldsymbol{D}_{t}=(1,1,1)$\end{tabular} & \begin{tabular}[c]{@{}c@{}}\textbf{Kendall's Tau}\\ $\boldsymbol{D}_{t}=(1,0,1)$\end{tabular} & \begin{tabular}[c]{@{}c@{}}\textbf{Parametric}\\ \textbf{p-value}\end{tabular} & \begin{tabular}[c]{@{}c@{}}\textbf{Non-parametric}\\ \textbf{p-value}\end{tabular} \\ 
\hline
(2006, 2007, 2008) & 0.759                                                                        & 0.806                                                                           & 0.651                                                                & 0.092                                                                                           \\ 
(2007, 2008, 2009) & 0.790                                                                       & 0.807                                                                           & 0.826                                                                & 0.148                                                                                          \\ 
(2008, 2009, 2010) & 0.781                                                                         & 0.667                                                                         & 0.343                                                                & 0.161                                                                                            \\ 
(2009, 2010, 2011) & 0.799                                                                          & 0.754                                                                           & 0.600                                                                 & 0.064                                                                                           \\
(2010, 2011, 2012) & 0.786                                                                         & 0.718                                                                           & 0.505                                                               & 0.217                                                                                            \\ \hline\hline
\end{tabular}
\smallskip
\subcaption{\textbf{Model 2}} \label{Table pretest kendalM2}
\begin{tabular}{c|cccc}
\hline\hline
\textbf{Year} & \begin{tabular}[c]{@{}c@{}}\textbf{Kendall's Tau}\\ $\boldsymbol{D}_{t}=(1,1,1)$\end{tabular} & \begin{tabular}[c]{@{}c@{}}\textbf{Kendall's Tau}\\ $\boldsymbol{D}_{t}=(1,1,0)$\end{tabular} & \begin{tabular}[c]{@{}c@{}}\textbf{Parametric}\\ \textbf{p-value}\end{tabular} & \begin{tabular}[c]{@{}c@{}}\textbf{Non-parametric}\\ \textbf{p-value}\end{tabular} \\
\hline
(2006, 2007, 2008) & 0.759                                                                         & 0.760                                                                     & 0.996                                                                 & 0.170                                                                     \\ 
(2007, 2008, 2009) & 0.790                                                                        & 0.787                                                                        & 0.980                                                                & 0.160                                                                     \\ 
(2008, 2009, 2010) & 0.781                                                                          & 0.699                                                                      & 0.357                                                                 & 0.194                                                                     \\ 
(2009, 2010, 2011)  & 0.799                                                                        & 0.843                                                                       & 0.472                                                                 & 0.065                                                                     \\ 
(2010, 2011, 2012) & 0.786                                                                         & 0.780                                                                         & 0.936                                                                & 0.263                                                                   \\ \hline\hline
\end{tabular}
\end{center}
\bigskip

\textit{Notes: This table presents the Kendall's Tau for the target group $\boldsymbol{D}_{t}=(1,1,1)$ and the recovery group, and the p-values for both parametric and non-parametric tests, for both models 1 and 2 in each three year time window. }
\end{table}

According to the Kendall's Tau values and the p-values of the parametric test, for both Model 1 and Model 2, we fail to reject that the Kendall's Taus are the same for both treated and copula recovery groups in each three-year time window. 
In addition, the  non-parametric test further conclude that we fail to reject the GCE assumption in  both Models 1 and 2  in this application. 

\subsection{DTE Bounds Conditional on Previous BMI Values}\label{Chap2_Sec Appl2 withYt-1}

The previous section provides the results of the unconditional QoTE bounds, which are obtained by averaging the conditional QoTE bounds given the outcomes in the previous period (see Lemma \ref{Chap2_lemma_cond_TE} and Theorems \ref{Theorem_DoTE} and \ref{Chap2_theorem_QoTE}). However, our methodology also allows for the estimation of heterogeneous DTE bounds for individuals with different starting BMIs in the previous period. For example, this allows for the illustration of (heterogeneous) DTEs for those with low BMIs and those with high BMIs in the past.

\begin{figure}[h!]
\caption{\centering Conditional QoTE Bounds of Exercising on BMI Given the Previous BMI Values under Model 1 for  $\boldsymbol{D}_{t} =(1,1,1)$} %when Considering the Dependence Between the Treated and Untreated Potential Outcomes}%
\label{Chap2_fig QoTE with Y'}
\begin{subfigure}[b]{0.5\linewidth}
\centering
\includegraphics[width=1\linewidth]{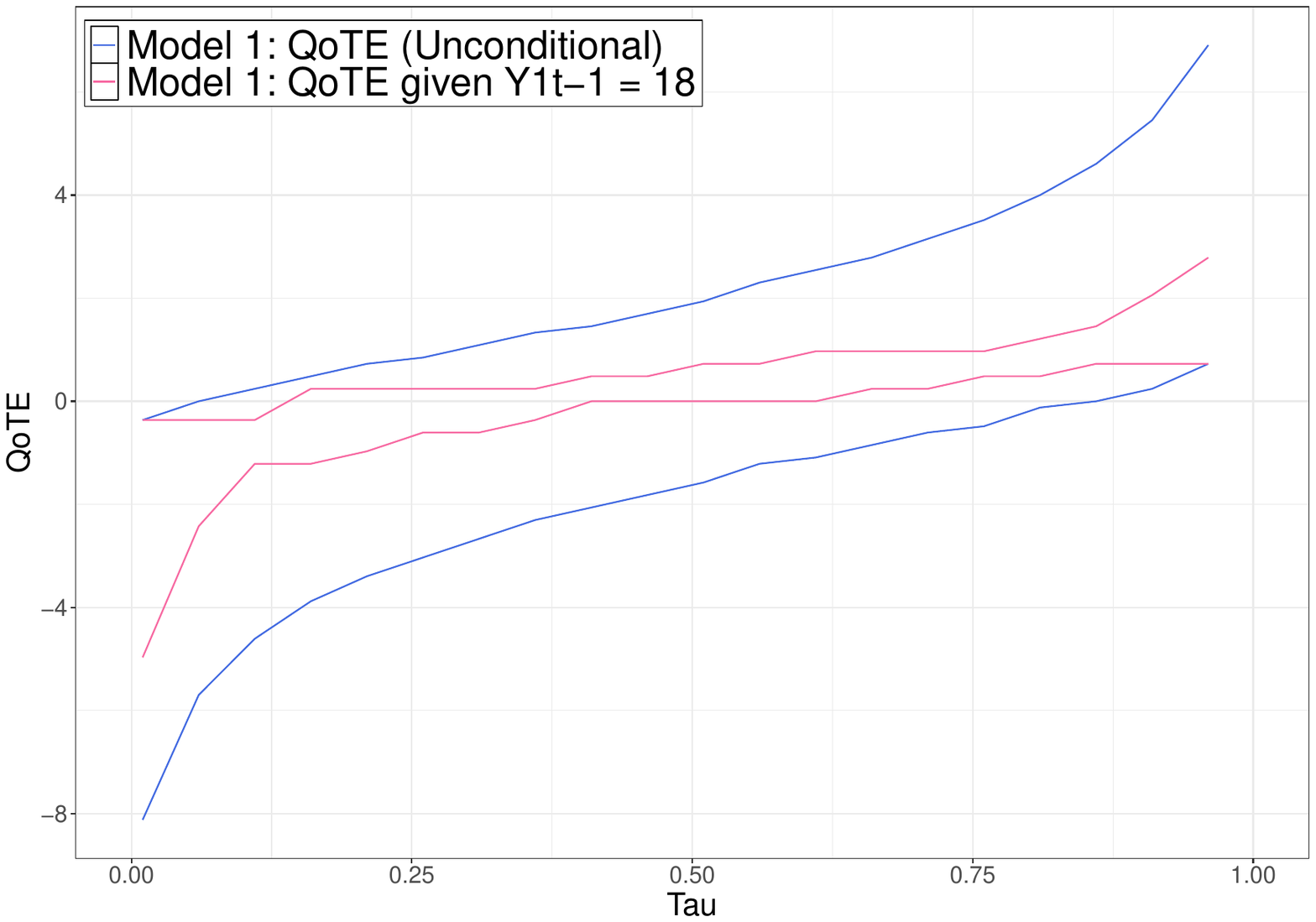}
\caption{Previous BMI Value: $ Y_{1t-1}  = 18 \, kg/m^{2}$ }
\label{fig1Y':a}
\vspace{4ex}
\end{subfigure}
\begin{subfigure}[b]{0.5\linewidth}
\centering
\includegraphics[width=1\linewidth]{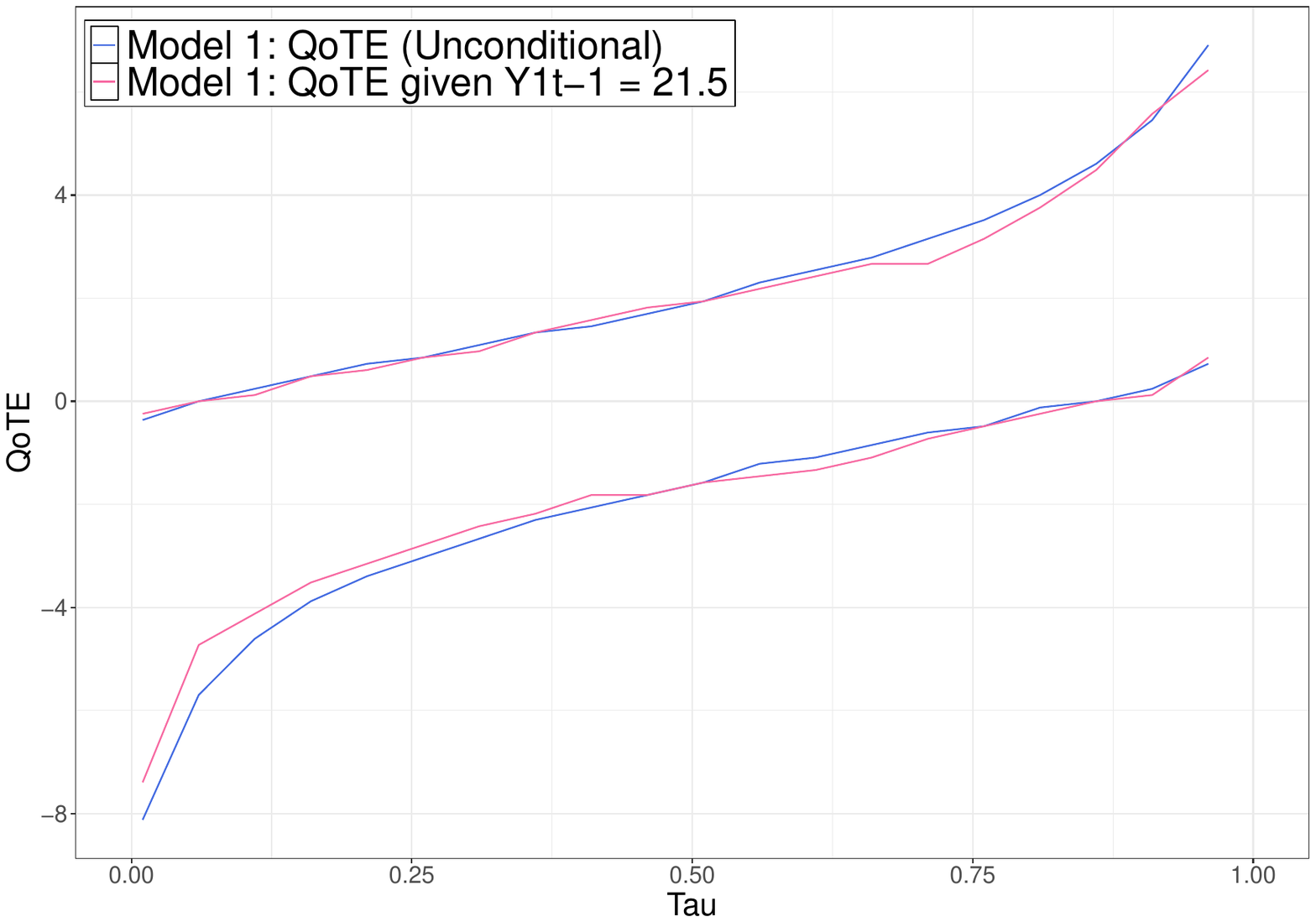}
\caption{Previous BMI Value: $ Y_{1t-1}  = 21.5 \, kg/m^{2}$}
\label{fig1Y':b}
\vspace{4ex}
\end{subfigure}
\begin{subfigure}[b]{0.5\linewidth}
\centering
\includegraphics[width=1\linewidth]{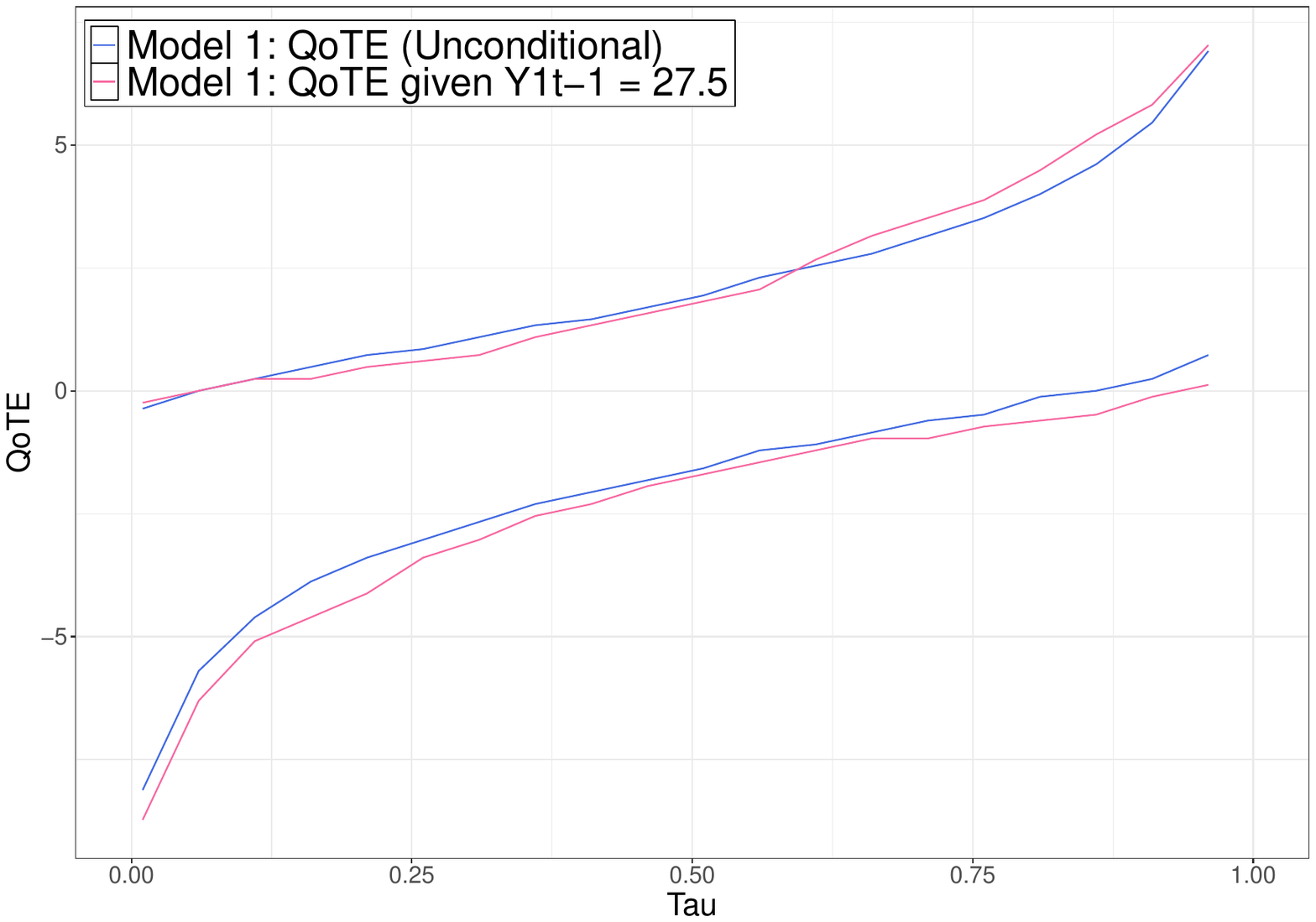}
\caption{Previous BMI Value: $ Y_{1t-1}  = 27.5 \, kg/m^{2}$}
\label{fig1Y':c}
\end{subfigure}
\begin{subfigure}[b]{0.5\linewidth}
\centering
\includegraphics[width=1\linewidth]{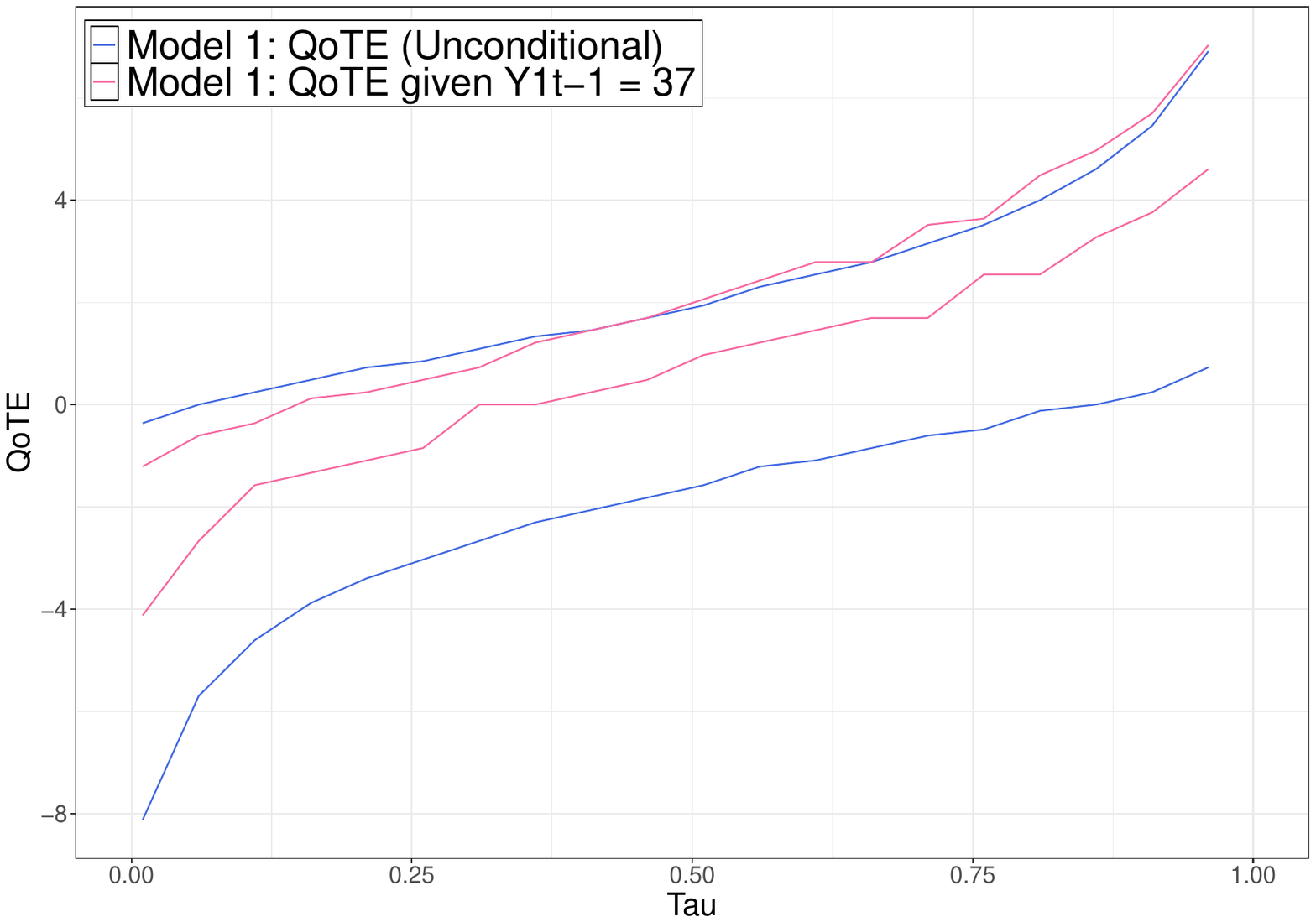}
\caption{Previous BMI Value: $ Y_{1t-1}  = 37 \, kg/m^{2}$}
\label{fig1Y':d}
\end{subfigure}

\textit{Notes: Each sub-figure plots the (unconditional) QoTE bounds at 2015 for the target group $\boldsymbol{D}_{t}=(1,1,1)$, derived under Model 1 (blue) together with their QoTE bounds conditional on the previous years' BMI values, where we consider only the set values, $Y_{t-1} = \{18 \, kg/m^{2}, \, 21.5\, kg/m^{2}, \, 27.5\, kg/m^{2}, \, 37\, kg/m^{2}\}$}.

\end{figure}

In this section, we present the results of the conditional QoTE bounds derived under Model 1 for the target group $\boldsymbol{D}_{t} = (1,1,1)$ by conditioning on four different past BMI values in the previous period: $Y_{t-1} = \{18 \, kg/m^{2}, \, 21.5\, kg/m^{2}, \, 27.5\, kg/m^{2}, \, 37\, kg/m^{2}\}$, each representing a different BMI classification group.\footnote{Our analysis adheres to the World Health Organization (WHO) BMI classification guidelines. Specifically, BMI $< 18.5\, kg/m^{2}$ is categorized as underweight, BMI $18.5\, kg/m^{2} - 24.9\, kg/m^{2}$ as normal weight, BMI $25.0\, kg/m^{2} - 29.9\, kg/m^{2}$ as overweight, and BMI $\geq 30.0\, kg/m^{2}$ as obese (https://www.who.int/europe/news-room/fact-sheets/item/a-healthy-lifestyle---who-recommendations).} The conditional QoTE bounds for each of these previous BMI values are plotted in sub-figures \ref{fig1Y':a} - \ref{fig1Y':d}, together with the unconditional QoTE bounds derived by averaging over all $Y_{t-1}$ (in blue).\footnote{These unconditional QoTE bounds are the same as those plotted in Figure \ref{Chap2_fig main: App2 M1}.} These conditional QoTE bounds, shown in each sub-figure for different previous BMI values, appear distinct from one another, providing greater insight into treatment effect heterogeneity. Therefore, this developed approach offers a significant advantage in studying heterogeneous treatment effects not only at an individual or group level but also based on previous outcomes, accommodating for highly heterogeneous treatment effects.

\subsection{Effects of Exercising on BMI Conditional on Covariates}{\label{Chap2_Sec Appl2 withCov}}

So far, we have focused on the QoTE bounds without using the information of covariates. 
In this section, we aim to examine our proposed bounding methods by controlling and leveraging the availability of covariates. 

\begin{figure}[h!]
\begin{center}
\caption{\centering QoTE Bounds for Different Age-groups under Model 1}\label{Chap2_fig covariates}
\begin{subfigure}[b]{0.33\textwidth}	
\centering
\caption{\centering{Young Adults (Males vs. Females)}}
\includegraphics[width=1\linewidth]{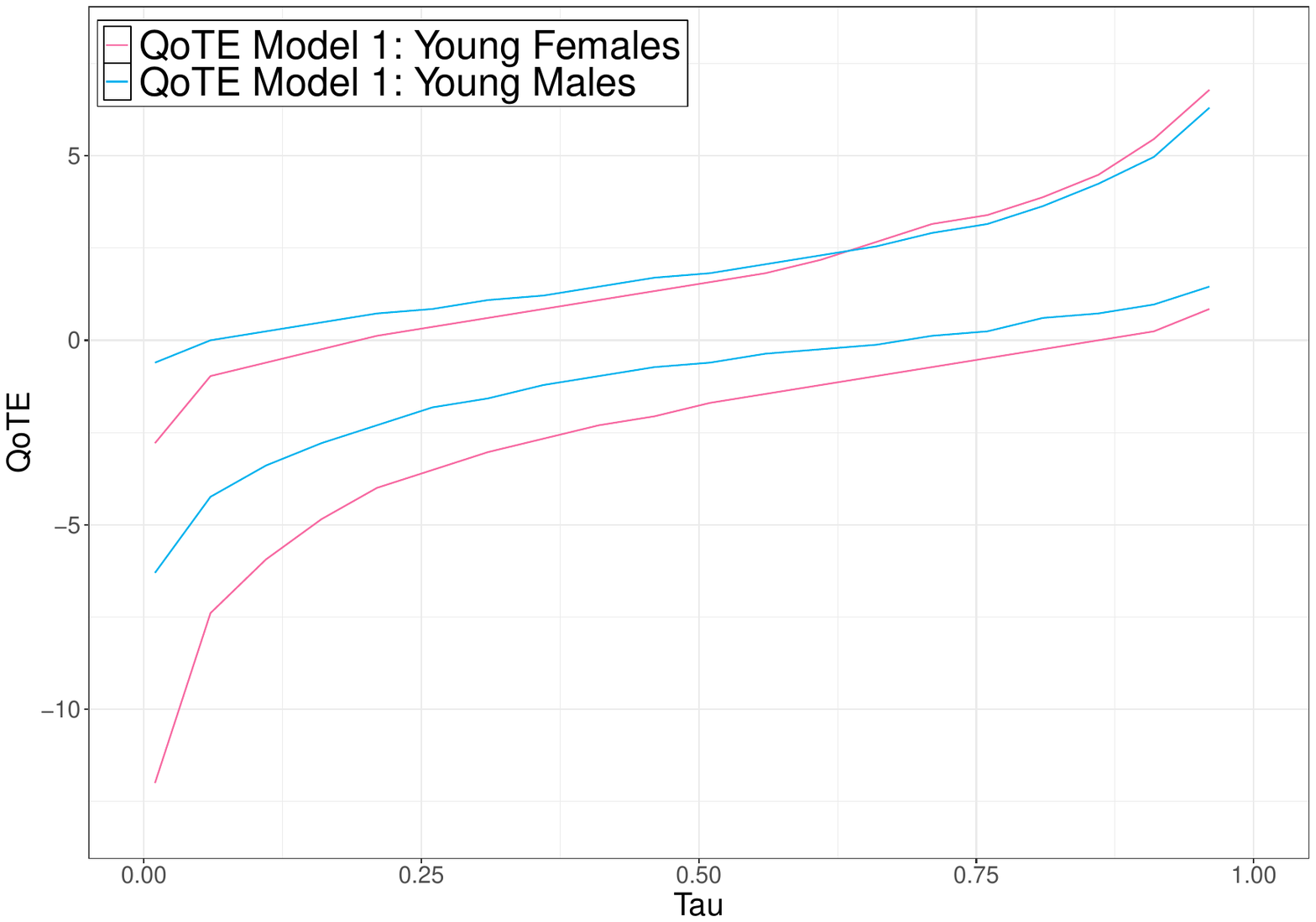}
\label{fig4:covariates}
\end{subfigure}
\begin{subfigure}[b]{0.33\textwidth}
\centering
\caption{\centering{Middle-aged Adults (Males vs. Females)}}
\includegraphics[width=1\linewidth]{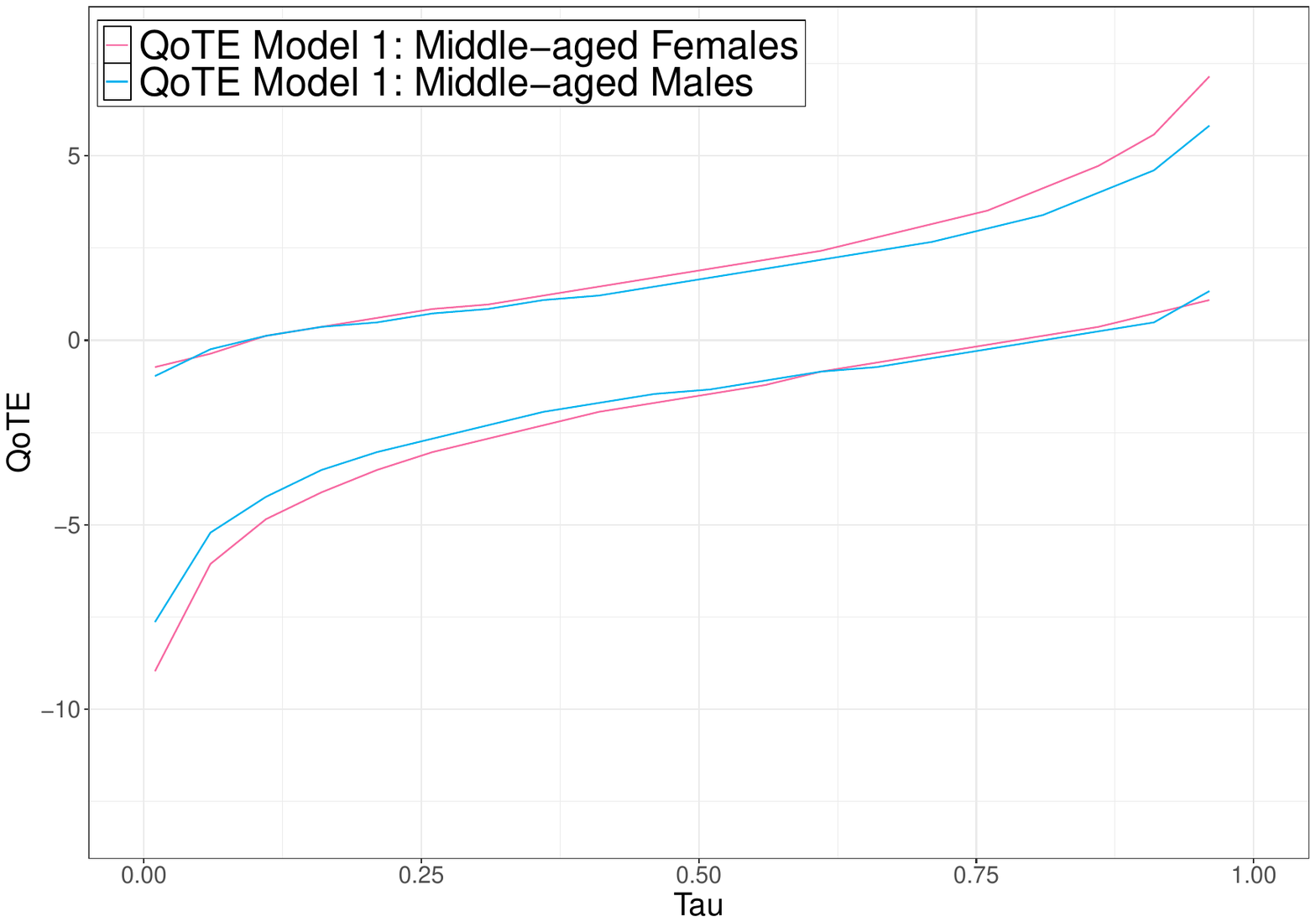}
\label{fig3:covariates}
\end{subfigure}%
\begin{subfigure}[b]{0.33\textwidth}
\centering
\caption{\centering{Older Adults (Males vs. Females)}}
\includegraphics[width=1\linewidth]{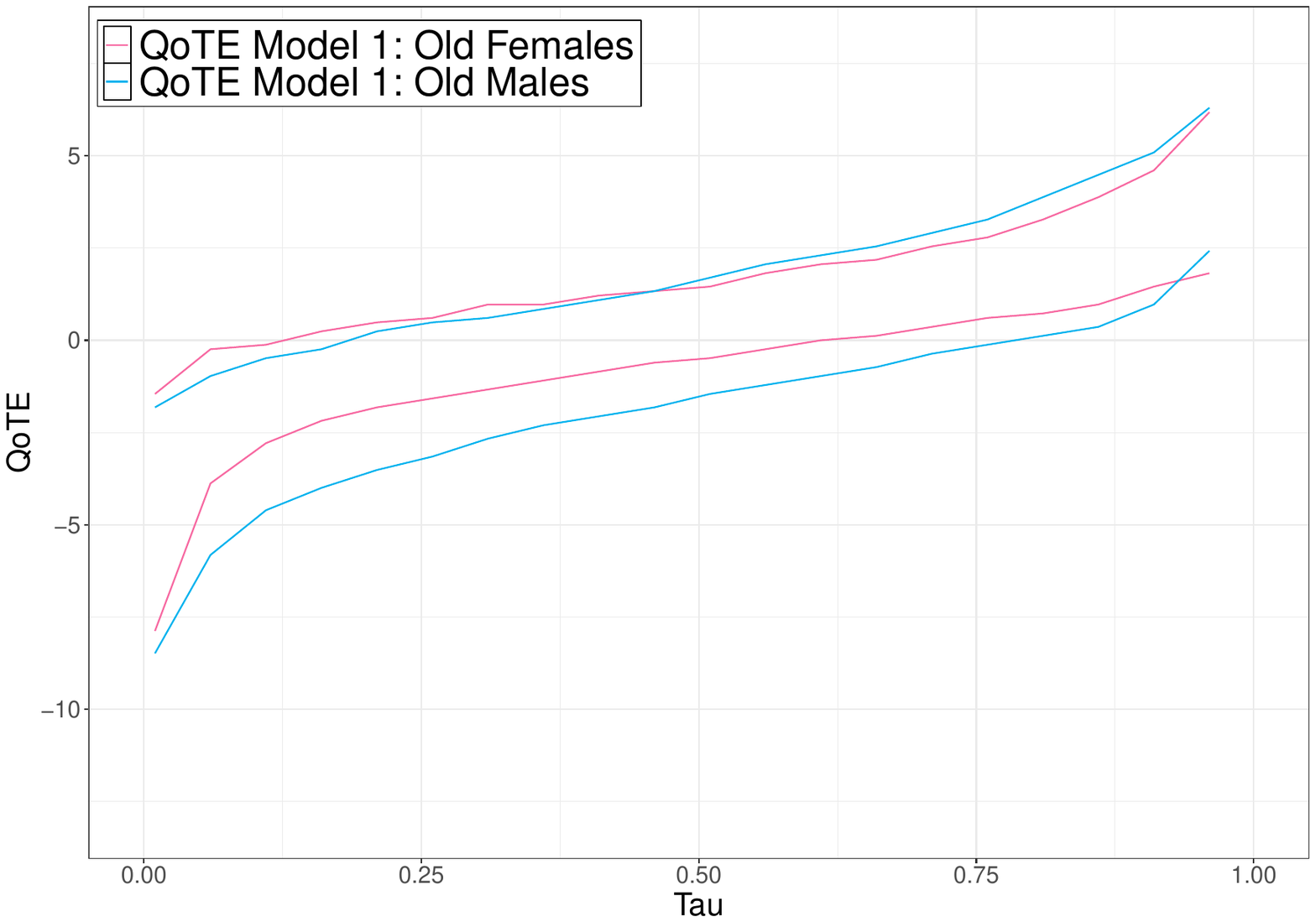}
\label{fig5:covariates}
\end{subfigure}
\end{center} 
\textit{Notes: This figure presents the QoTE bounds derived under under Model 1 by conditioning on covariates age and gender. The three sub-figures \ref{fig3:covariates}, \ref{fig4:covariates}, and \ref{fig5:covariates}, compare the QoTE bounds for males (blue) and females (pink) across the specified age groups: young, middle-aged, and older.}
\end{figure}

We focus on the target group $\boldsymbol{D}_{t}=(1,1,1)$ and Model 1. We compare the QoTE bounds by considering a simple case where we condition on covariates gender (i.e., male and female) and age (i.e., young, middle aged, and older)\footnote{The age categories are based on standard age classifications: young (ages 18 - 34), middle-aged (ages 35 - 64), and older (ages 65 and above)}. %(see \textcolor{red}{Table \ref{Table summary}}). 
The resulting bounds for each subpopulation are illustrated in Figure \ref{Chap2_fig covariates}. The sub-figures \ref{fig4:covariates}, \ref{fig3:covariates}, and \ref{fig5:covariates} plot the QoTE bounds for the three age groups: young, middle-aged, and old, respectively. In each sub-figure, we compare the bounds for males and females within the specified age group. 

The main takeaways can be outlined as follows. Incorporating covariates through Copula Assumption \ref{Chap2_ass3} provides more insights than the analysis without covariates. For example, consider the sub-figure \ref{fig4:covariates} which highlights the treatment effect heterogeneity between young males and females. According to the QoTE estimates, it is interesting to note that overall, the reduction in BMI associated with physical activity is greater for young females than for young males. Specifically, approximately 20\% of young females are likely to experience a definitive reduction in BMI through exercise, whereas only around 9\% of young males experience a reduction in BMI through exercise. In contrast, older males show a greater reduction in BMI compared to older females. However, the effects of physical activity on BMI appear to be similar for middle-aged females and males. 
Consequently, these results further underscore the importance of including covariates in the derivation of the DTE bounds together with our copula assumptions as they provide more insights about the treatment effect heterogeneity.\footnote{Existing research has also shown that including covariates often leads to tighter DTE bounds \citep[see, for example][]{fan2009partial,fan2010sharp,firpo2019partial,callaway2021bounds}. For instance, these studies illustrate how the integration of relevant covariates can improve the DTE bounds. We are still in the process of examining how the integration of covariates affects the DTE bounds under our GCE assumptions.}

\subsection{A Comparison with \cite{callaway2021bounds} Bounds for Group $ \mathbf{\textit{D}}_{t} =(0,0,1) $}{\label{Chap2_Sec Appl2 BC&Model2}}

Whilst Callaway's copula stability assumption (CSA) cannot be applied to group $ \boldsymbol{D}_{t} = (1,1,1) $, and only be applicable for group $ \boldsymbol{D}_{t} = (0,0,1) $, our work delivers a flexible approach that can be applied to all eight possible treatment patterns observed over a three-period timeframe. In this subsection, we derive the DTE bounds for this group $ \boldsymbol{D}_{t} = (0,0,1) $ considered by \cite{callaway2021bounds} and justify our GCE assumptions by comparing the DTE bounds using CSA with those derived from our Model 2 with assumption GCE (II). We illustrate this using the same previous empirical example and show that our results do not differ significantly from the \cite{callaway2021bounds} bounds.

\subsubsection{Method Proposed by \cite{callaway2021bounds} for the Group $\mathbf{\textit{D}}_{t} =(0,0,1) $}\label{Chap2_Subsec CSA}

Callaway's approach requires at least three periods of panels, assuming that all individuals are untreated in the first two periods and can either choose to be treated or remain untreated only in the last period $t$. Depending on this setup, the treatment pattern vector $\boldsymbol{D}_{t}$ can only take two possible patterns, namely, $\boldsymbol{D}_{t} =(0,0,1) $ (target group) and $ \boldsymbol{D}_{t}=(0,0,0) $ (untreated group). Then, he provides a method to obtain tighter bounds on the DTEs for the target group $\boldsymbol{D}_{t} =(0,0,1) $, using the dependence assumption: copula stability assumption (CSA). The CSA restricts the dependence between the untreated potential outcomes over time for the individuals in group $ \boldsymbol{D}_{t} =(0,0,1) $, as follows.

\paragraph{\textit{Copula Stability Assumption (CSA):}} \label{Chap2_ass CSA}  \textit{For all $ (u, v) \in [0, 1]^{2} $} and individuals with treatment pattern $ \boldsymbol{D}_{t} =(0,0,1) $, it's assumed that,
\begin{align}\label{Eqn CSA}
C^{\ast}_{Y_{0t-1}, Y_{0t}|\boldsymbol{D}_{t} = (0,0,1)}(u, v) &= C_{Y_{0t-2}, Y_{0t-1}|\boldsymbol{D}_{t} = (0,0,1)}(u, v),
\end{align}
\textit{where the superscript ${\ast} $ denotes the ``unobserved" component.}

The CSA assumes that the two dependences or the copulas between the untreated potential outcomes in periods $ t-2 $ and $ t-1 $, and in periods $ t-1 $ and $ t $ to be the same for individuals in the target group, $\boldsymbol{D}_{t} =(0,0,1) $. In other words, it assumes the dependence between untreated potential outcomes to be invariant over time for this treatment group. 

Observe that the CSA given in Equation \eqref{Eqn CSA} recovers the unobserved copula term (left), using information from the same group $\boldsymbol{D}_{t} =(0,0,1) $ at different points in time (right). This is one of the key advantages of limiting his analysis to a setting where all individuals are untreated in the initial two periods and either be treated or untreated only in the last period. 
This setup, naturally enables to recover the unknown copula of the lagged potential outcomes for an individual by leveraging their \textit{own} observed outcomes from the two pre-treatment periods, $t-2$ and $t-1$.
%	\end{remark}
Even though \cite{callaway2021bounds} has discussed his approach only for the case $\boldsymbol{D}_{t} =(0,0,1) $, analogous assumptions can be made when identifying DTE bounds for $\boldsymbol{D}_{t} =(1,1,0) $. Unfortunately, his approach fails to identify bounds on the DTEs for all the other six possible treatment patterns, such as $\boldsymbol{D}_{t} =(1,1,1) $ or $\boldsymbol{D}_{t} =(0,0,0) $.

Since we consider more general treatment patterns and focus on the identification of  DTEs for all eight treatment patterns, for patterns such as $\boldsymbol{D}_{t} =(1,1,1) $, it limits the possibilities that can be considered more naturally as CSA, when recovering the unknown copula. In other words, even though the two alternative GCE assumptions that are imposed in our study to recover the unobserved dependence are plausible and sensible, they might not be more natural as the CSA except for the two cases where $\boldsymbol{D}_{t} =(0,0,1) $ and $\boldsymbol{D}_{t} =(1,1,0) $. For example, the identification of DTEs for group $ \boldsymbol{D}_{t}= (1,1,1)$ is more difficult compared to Callaway's case, $ \boldsymbol{D}_{t}= (0,0,1)$. The primary reason for this is that the individuals in group $ \boldsymbol{D}_{t}= (1,1,1)$ have continued to be treated throughout the three periods. Hence, we cannot observe their untreated potential outcomes, even at different points in time, as in $\boldsymbol{D}_{t}=(0,0,1)$. This limitation leads to exploring information from another similar treatment group as the target group $ \boldsymbol{D}_{t}= (1,1,1)$, to recover the target unknown dependence and, consequently, the DTEs (see Supplementary Appendix Section D, %\ref{Sec other groups}, 
which provides the GCE assumptions assumed under Models 1 and 2 for the remaining seven cases).

It's interesting to note that the logic employed to recover the unknown copula under Model 1 (GCE assumption \ref*{Chap2_ass3}) for group $\boldsymbol{D}_{t} =(0,0,1) $, has led to the same conditions imposed under the CSA,  proposed by \cite{callaway2021bounds} (See Supplementary Appendix Section D). %\ref{Sec other groups}. 
However, the logic employed under the second alternative GCE assumption introduced under Model 2 for $\boldsymbol{D}_{t} =(0,0,1) $, leads to conditions different from the CSA. Therefore, we focus on the bounds on the DTEs for the target group $\boldsymbol{D}_{t} =(0,0,1) $, obtained under Model 2 using the second alternative GCE assumption. This is discussed in the following subsection, where we justify these results by comparing them with the DTE bounds derived using Callaway's method under the CSA.

\subsubsection{Method Proposed in the Current Study for the Group $\mathbf{\textit{D}}_{t} =(0,0,1) $} \label{Chap2_Subsec Dt(001) DA(II)}

The method developed in the current study utilizes the second alternative GCE assumption for the case $\boldsymbol{D}_{t} =(0,0,1) $ as given below (Equation \eqref{Chap2_Eq_d(0,0,1)DA(II)}), which uses an analogous technique as in Assumption \ref{Chap2_ass3Model2} imposed under Model 2 for the treated group $\boldsymbol{D}_{t} = (1, 1, 1)$. 
\begin{assumptionp}{\ref*{Chap2_ass3Model2}}\label{d(0,0,1)DA(II)} For all $ (u, v) \in [0, 1]^{2} $ and for individuals with the treatment pattern $ \boldsymbol{D}_{t} = (0,0,1) $, we assume: 
\begin{align}
C^{\ast}_{Y_{ {0}t-1},Y_{0t}|\boldsymbol{D}_{t} = ({0},  {0}, 1)}(u, v) &= C_{Y_{{0}t-1},Y_{0t}|\boldsymbol{D}_{t} = ({0}, {0}, 0)}(u, v) \label{Chap2_Eq_d(0,0,1)DA(II)}
\end{align}
where the superscript ${\ast} $ denotes the ``unobserved" component.
\end{assumptionp}
For this target group $\boldsymbol{D}_{t} =(0,0,1) $, under Assumption \ref*{d(0,0,1)DA(II)} we require the information of its untreated group at period $ t $: $ \boldsymbol{D}_{t} = (0, 0, 0) $, to identify the unobserved copula term on the left of Equation \eqref{Chap2_Eq_d(0,0,1)DA(II)}. %, and also when identifying the counterfactual marginal $F_{Y_{0t}|\boldsymbol{D}_{t} = (0, 0, 1)}$ under the CiC method, for this target group.
Using this assumption together with panel data, we estimate the bounds of our target parameters following a similar approach discussed in Section \ref{Subsec Model setup 2} for the case $\boldsymbol{D}_{t} = (1, 1, 1)$ under Model 2. 

However, as discussed earlier, the CSA can be viewed as a natural and intuitive technique compared to Assumption \ref*{d(0,0,1)DA(II)} for this group $\boldsymbol{D}_{t} =(0,0,1) $. The primary reason is that CSA utilizes its own group's information to estimate this same unknown copula (left), while Assumption \ref*{d(0,0,1)DA(II)} for $\boldsymbol{D}_{t} =(0,0,1) $, requires another group's information for the same purpose. Hence, we justify our Assumption \ref*{d(0,0,1)DA(II)} by making comparisons with the bounds obtained by invoking this assumption and the CSA, for the target group  $ \boldsymbol{D}_{t} = (0, 0, 1) $, using the same empirical illustration.

\begin{figure}[H]
\caption{\centering{Effects  of Exercising on BMI for Group $\boldsymbol{D}_{t} =(0,0,1) $: QoTE Bounds under CSA of \cite{callaway2021bounds} and \textbf{Model 2} with Assumption \ref*{d(0,0,1)DA(II)}}}\label{Chap2_fig main: App2 D(001)BC&M2}
\begin{center}
\includegraphics[width=0.7\linewidth]{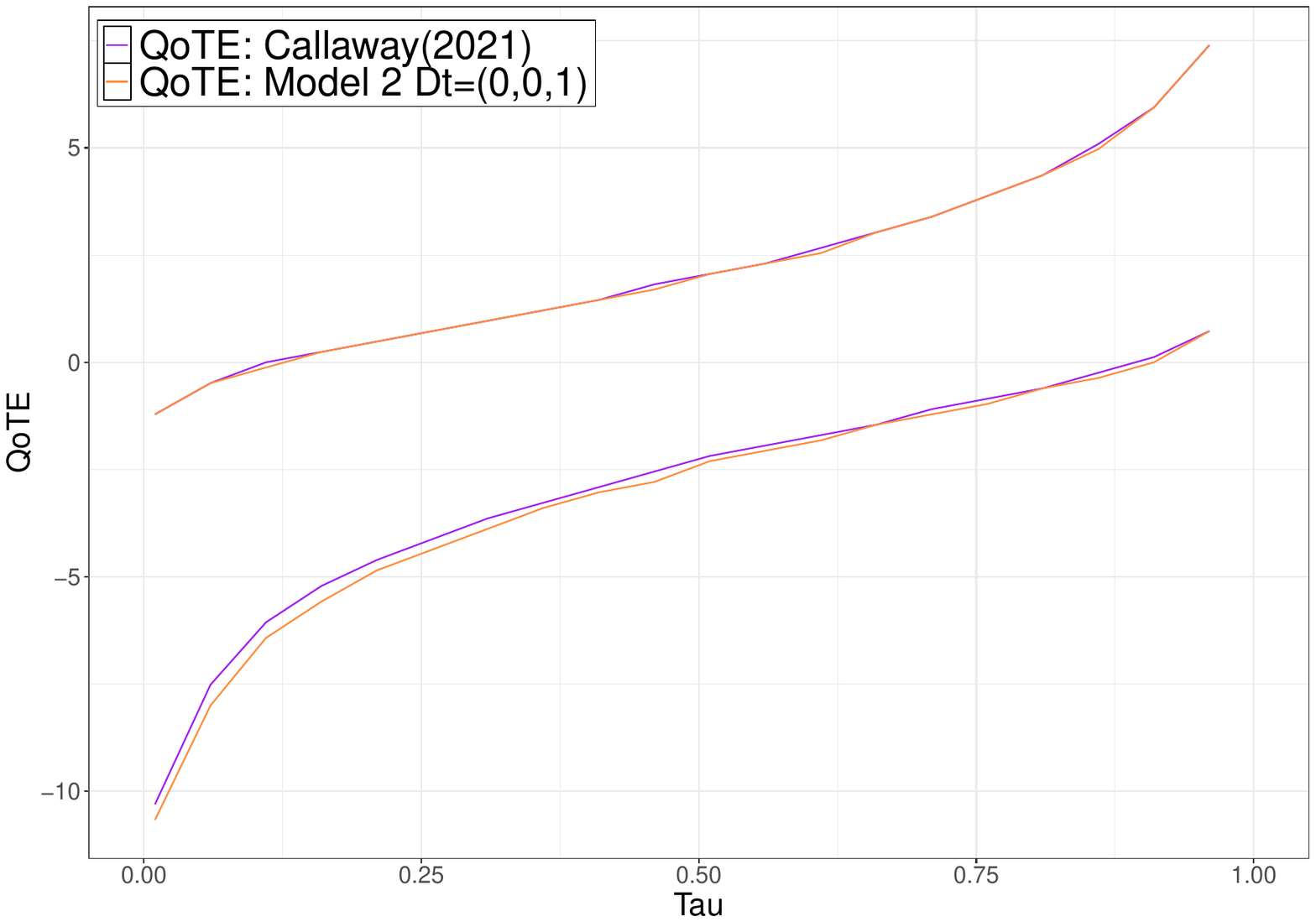}
\end{center}
\textit{Notes: This figure plots the QoTE bounds for group $\boldsymbol{D}_{t} =(0,0,1) $, obtained using the method proposed in \cite{callaway2021bounds} under CSA (purple) with the bounds derived under our Model 2 using the second alternative GCE Assumption \ref*{d(0,0,1)DA(II)} (orange)}.
\end{figure}

We compute the estimates of the QoTE bounds considering the same three time periods, 2013, 2014, and 2015, and the same final sample (balanced panel) drawn from the HILDA survey, which comprises 10,755 unique individuals. Among these individuals, 626 belong to the target group $ \boldsymbol{D}_{t} = (0, 0, 1) $, and 3,141 belong to the untreated group $ \boldsymbol{D}_{t} = (0, 0, 0) $. Exploiting these observations, we compute the estimates of the QoTE bounds for group $\mathbf{\textit{D}}_{t} =(0,0,1) $ using, (i) the Callaway's method with the dependence assumption; CSA, and (ii) the approach developed in the current study under Model 2 with GCE Assumption \ref*{d(0,0,1)DA(II)}. These QoTE estimates obtained under each method are plotted in Figure \ref{Chap2_fig main: App2 D(001)BC&M2}. Even though CSA is more natural and intuitive compared to our second alternative GCE Assumption \ref*{d(0,0,1)DA(II)} for group $ \boldsymbol{D}_{t} = (0, 0, 1) $, in this application, both methods provide similar DTE bounds with no huge differences.

\section{Conclusions} \label{Chap2_Sec Conclusion}

This study introduces two alternative methods to derive tighter bounds on the (group-wise) distributional treatment effects parameters, which depend on the joint distribution of the potential outcomes, when there is access to a rich panel with at least three periods. Both methods accommodate general treatment patterns by allowing individuals to switch between the treated and untreated states unrestrictedly over time.
This setup can be viewed as a commonly encountered setting in empirical research. The resulting bounds in this study depend on the following key ingredients: (i) having access to at least three periods of panels, 
and (ii) the two alternative GCE assumptions on the copula structure; Assumption \ref{Chap2_ass3} and Assumption \ref{Chap2_ass3Model2} introduced under Model 1 and Model 2, respectively. The latter are the key assumptions in this study, where both utilize information from past observations when estimating the unknown estimates. Depending on the economic setting and the information available in the panel dataset, one could use \textit{either} of these GCE assumptions to recover the unknown copula and consequently obtain tighter bounds on the DTEs. Unlike existing methods that derive bounds on the DTEs only for the treated or one specific group of individuals, our approach delivers the flexibility to partially identify DTEs for both treated and untreated subpopulations (i.e., for all eight possible treatment patterns). Hence, it provides an opportunity to identify DTEs for the entire population.
Furthermore, the application (exercising on body weight) that we used to illustrate our developed approaches affirms the need for the identification of DTEs rather than limiting our focus only on the ATEs, which precludes information on the treatment effect heterogeneity. In addition to that, it also demonstrates the improvement of the identification power of the DTE bounds when considering the copula structure between the potential outcomes, compared with the existing methods, which use no assumptions on this copula.

\bibliography{refpaper2, refpaperM3}

\end{document}